\documentclass[a4paper,11pt]{article}
\usepackage{jcappub} 
\usepackage{lineno}

\usepackage{aas_macros}
\usepackage{subcaption}  

\usepackage[dvipsnames]{xcolor}

\newcommand{\eqn}[1]{equation~(\ref{#1})}
\newcommand{\eqns}[2]{equations~(\ref{#1}) and (\ref{#2})}

\newcommand{\secn}[1]{Section~\ref{#1}}
\newcommand{\fig}[1]{Figure~\ref{#1}}
\newcommand{\figs}[2]{Figures~\ref{#1} and \ref{#2}}

\newcommand{\tab}[1]{Table~\ref{#1}}
\newcommand{\app}[1]{Appendix~\ref{#1}}
\newcommand{\angstrom}{\text{\normalfont\AA}}

\newcommand{\Msun}{\rm{M_\odot}}
\newcommand{\LUV}{L_{\rm UV}}
\newcommand{\MUV}{M_{\rm UV}}
\newcommand{\Mcrit}{M_{\rm crit}}
\newcommand{\kappaUV} {\mathcal{K}_{{\rm UV}}}
\newcommand{\kappaUVfid} {\mathcal{K}_{{\rm UV,fid}}}
\newcommand{\etaGamma} {\eta_{\gamma\ast}}
\newcommand{\etaGammafid} {\eta_{{\rm \gamma\ast,fid}}}

\defcitealias{Chakraborty2024}{CC24}

\arxivnumber{2503.07590} 
\title{Probing Reionization-Era Galaxies with JWST UV Luminosity Functions and Large-Scale Clustering}

\author[1]{Anirban Chakraborty\note{Corresponding author.}}
\author{and Tirthankar Roy Choudhury}
\affiliation{National Centre for Radio Astrophysics, Tata Institute of Fundamental Research, \\Pune University Campus, Ganeshkhind, Pune 411007, India.}

\emailAdd{anirban@ncra.tifr.res.in; anirban.chakraborty096@gmail.com; tirth@ncra.tifr.res.in}

\abstract{
The James Webb Space Telescope (JWST) has transformed our understanding of early galaxy formation, providing an unprecedented view of the first billion years of cosmic history. These observations offer a crucial opportunity to probe the interplay between galaxy formation and reionization, placing stringent constraints on theoretical models. In this work, we build upon our previously developed semi-analytical framework that self-consistently models the evolving UV luminosity function (UVLF) of galaxies and the global reionization history while incorporating the effects of radiative feedback.
Comparing our predictions with JWST and HST data, we identify a fundamental tension: models that match the UVLF fail to reproduce the observed evolution of galaxy clustering (bias) with redshift, and vice versa. To resolve this, we introduce a redshift- and mass-dependent duty cycle linked to the duration of star formation. This duty cycle increases towards higher redshifts, requiring either an enhanced production of UV radiation or increased star formation efficiency at $z>10$ to match the JWST UVLFs, but declines at lower redshifts ($5 < z \leq 9$) and towards low-mass halos to remain consistent with the bias and HST UVLF measurements. Reconciling theory with observations requires the characteristic star formation timescale to be longer in massive halos, and to decrease with redshift at fixed halo mass, evolving from $\approx 85$ Myr at $z = 6$ to $\approx 45$ Myr at $z = 14$ for $10^{10} M_\odot$ halos. Finally, our extended model, assuming a halo mass-dependent escaping ionizing efficiency ($\varepsilon_{\rm esc} \equiv f_{\rm esc} \times \xi_{\rm ion}$), whose population-averaged value gradually increases with redshift and corresponds to $\langle f_{\rm esc} \rangle \approx 15\%$ at $z=5$ for a fixed value of $\xi_{\rm ion} = 10^{25.23}~\mathrm{erg^{-1}\,Hz}$ across all galaxies, produces a reionization history consistent with current constraints.
These findings underscore the importance of jointly constraining high-redshift galaxy models using both UVLF and bias statistics to accurately interpret JWST data and refine our understanding of early cosmic evolution.}


\begin{document}
\maketitle
\flushbottom

\section{Introduction}
\label{sec:intro}

Understanding the formation and evolution of galaxies in the early Universe is one of the most fundamental areas of research in modern cosmology. In the hierarchical model of structure formation, dark matter halos serve as the cradles where galaxies form and evolve. As a result, the statistical and physical properties of galaxies are expected to be closely tied to those of their parent dark matter halos. Unraveling this relationship, commonly referred to as the `galaxy-halo connection', can offer critical insights into the astrophysical processes that govern star formation within dark matter halos (see \cite{Wechsler2018} for a review).  

Over the past decade, the ultraviolet luminosity function (UVLF) has emerged as an important observable for understanding the statistical properties of galaxies. Deep imaging surveys with space- and ground-based facilities, such as the Subaru Telescope, Hubble Space Telescope (HST), and more recently, James Webb Space Telescopes (JWST), have enabled measurements of UVLFs at high redshifts ($z \geq $ 6) and even out to redshifts as high as $z \approx 25$ \cite{Bouwens2015, Bouwens2017, Atek2018, Ono2018, Bowler2020, harikane2022, Donnan2023, Harikane2023, Bouwens2023, McLeod2024, Donnan2024, Whitler2025, PerezGonzalez2025}. These measurements have been widely used to understand how high redshift galaxies populate dark matter halos \cite{Trenti2010, Tacchella2013, Dayal14, Mason2015, Sun&Furlanetto2016, Tacchella2018, Park++2019, Ferrara2023, Wang2024}. However, one-point statistics such as the UVLFs would have been sufficient for constraining the galaxy-halo connection at high redshifts if the mapping between the observed light from galaxies and their host dark-matter halo properties were strictly one-to-one. Instead, the complex interplay of different baryonic processes within dark matter halos gives rise to a complex relationship between galaxy properties and their host halos, allowing for multiple ways to populate galaxies inside dark matter halos while still producing the same number density of galaxies as a function of luminosity \cite{Ren2018, Ren2019, Munoz2023, Mirocha2023, Gelli2024}. Therefore, to distinguish between widely different models of galaxy-halo connection, that are otherwise tuned to reproduce the observed UVLFs, one needs to consider other higher-order summary statistics of galaxies.  In this regard, measuring the clustering of galaxies detected in large-scale surveys can prove to be very useful \cite{Lee2009, Weinberger2019, Mirocha2020, Munoz2023, Sun2024}. An important quantity obtained from these two-point correlation studies is the galaxy bias, which quantifies the extra clustering of galaxies compared to the underlying dark matter distribution in the Universe. Since the clustering strength of dark matter halos is known to depend on their mass \cite{Mo1996, ST99, Jenkins2001, Tinker2010}, measurements of the galaxy bias provide a means to infer the masses of the halos hosting these galaxies, thereby placing tighter constraints on how high-redshift galaxies populate dark matter halos \cite{Park2016, Hatfield2018, Harikane2016}.

While clustering studies have been widely conducted at lower redshifts to investigate the host halo properties of Lyman-break galaxies (e.g., \cite{Jose2013,Bielby2013,Park2016}), such analyses at higher redshifts have been considerably more challenging due to the lack of a statistical sample of high-$z$ galaxies, with only a handful of studies available for $z > 6$ \cite{Barone-Nugent2014,Hatfield2018,harikane2022,Qiu2018,DalmassoHST, Shuntov2025, Paquereau2025}. However, exploiting the increased depth, sensitivity, and wide-field coverage of the instruments onboard JWST, it has recently been possible to measure the angular clustering of galaxies during the first billion years of the Universe, out to redshifts of $z \approx 11$ \cite{DalmassoJWST}. Therefore, it is essential to check whether the wide variety of galaxy-halo connection models that had been proposed to explain the overabundance of UV-bright galaxies \cite{Dekel2023,Renzini2023,Mirocha2023,Shen2023,Pallottini2023,Inayoshi2022,Harikane2023,Ferrara2023,Chakraborty2024,Hutter2024} seen in JWST observations, remain consistent with these latest observations of galaxy two-point statistics. In this paper, our primary goal is, therefore, to obtain insights into the astrophysical properties of high-redshift galaxies by comparing the self-consistently coupled theoretical model of high-$z$ galaxy formation, evolution, and cosmic reionization, introduced in our previous work \cite{Chakraborty2024} (hereafter, \citetalias{Chakraborty2024}), against the most recent and updated JWST UVLF and clustering measurements as well as constraints on the progress of reionization. In this work, the cosmological parameters are taken to be $\Omega_m$ = 0.308, $\Omega_{\Lambda}$ = 0.692, $\Omega_b$ = 0.0482, $h$ = 0.678, $\sigma_8$ = 0.829 and $n_s$ = 0.961 \cite{Planck2014}.

The paper is organized as follows: In \secn{sec:theory_model_baseline}, we describe the details of the theoretical model from our previous work (\citetalias{Chakraborty2024}). \secn{sec:data_and_methods} describes the various observational datasets used in this work and the Bayesian formalism used for parameter estimation.  We discuss the results obtained from comparing our earlier model to the observational datasets presently available in \secn{sec:results_baseline}. In \secn{sec:duty_cycle}, we discuss some modifications to this model that can help in explaining all the different observables simultaneously. Finally, we conclude with a summary of our main results in \secn{sec:conclusion}.

\section{Theoretical Formalism: The Baseline Model}
\label{sec:theory_model_baseline}
In this section, we describe the theoretical framework for modelling the star formation and ionizing properties of galaxies at high redshifts and calculating the different global high-redshift galaxy and reionization observables. 

In \citetalias{Chakraborty2024}, we presented a semi-analytical framework for modeling the astrophysical properties of high-redshift galaxies. This model calculates the evolving galaxy UV luminosity function across a wide range of redshifts and simultaneously tracks the evolution of the neutral hydrogen fraction in the intergalactic medium with time. While computing various galaxy observables, it self-consistently accounts for the effects of reionization feedback that suppresses star formation in low-mass galaxies. We briefly summarize here the main features of the model and refer interested readers to \citetalias{Chakraborty2024} for more details. We will refer to this model as the \textbf{baseline} model.

In this model, each dark matter halo is assumed to host only one galaxy\footnote{In reality, some halos may host multiple galaxies, including satellites. However, at the redshifts relevant to this work, the satellite fraction is expected to be very small (e.g., Paquereau et al. 2025 \cite{Paquereau2025} found the satellite fraction to be less than $ 5\%$ in galaxies with $\log_{10}(M_\ast/M_\odot) \geq 8.5$ at $z \gtrsim 5$) and the halo occupation distribution models populate only very massive halos ($M_h \gtrsim 10^{12.6} ~M_\odot$; Shuntov et al. 2025 \cite{Shuntov2025}) with satellites, making this a reasonable assumption.}, whose properties are primarily determined by the mass of the halo that hosts it. For instance, the star-formation rate $\dot{M}_*$ of a galaxy residing within a halo of mass $M_h$ is calculated as
\begin{equation}
\label{eq:SFR}
\dot{M}_*(M_h,z) = \dfrac{M_\ast(M_h,z)}{t_\ast(M_h,z)} =\dfrac{f_*(M_h,z)~f_{\rm gas}(M_h)~ \bigg(\dfrac{\Omega_b}{\Omega_m}\bigg) M_h}{t_\ast(M_h,z)},
\end{equation}
In the equation above, $f_*(M_h,z)$ denotes the star-formation efficiency (i.e., the fraction of baryons within halos that are converted into stars), $f_{\rm gas}(M_h)$ represents the gas fraction retained inside a halo after photo-heating/photo-evaporation due to the rising ionizing UV background (UVB), and $t_\ast(M_h,z)$ is the average star formation time scale of a halo of mass $M_h$ at redshift $z$. For halos forming in already ionized regions, the gas fraction is assumed to be $f_{\rm gas}(M_h) =  2^{-\Mcrit/M_h}$, wherein the parameter $\Mcrit$ represents the characteristic mass of halos that are capable of retaining 50 percent of their gas reservoir. This simple parameterization, based on results of 1D spherically symmetric collapse simulations investigating the impact of the UVB on baryon condensation in dark matter halos \cite{Sobacchi&Mesinger_1Dsims_2013}, indicates that the baryonic content is increasingly depleted in lower-mass halos, while halos that are much more massive than the critical mass $\Mcrit$ remain largely unaffected \cite{Choudhury&Dayal2019}. Although the critical halo mass $\Mcrit$ could, in principle, vary with redshift and the intensity of the ionizing UV background, we assume it to be redshift-independent in this work, given that current reionization observations struggle to disentangle its potential evolution from that of other uncertain factors such as the production and escape of ionizing photons \cite{Hutter2021, Trebitsch2022, Chakraborty2024}. In the future, precise measurements of the faint-end UV luminosity function across multiple redshifts (e.g., $z=6-8$) will be essential for constraining any redshift dependence of $\Mcrit$ \cite{Choudhury&Dayal2019, Hutter2021}.  We set $f_{\rm gas}(M_h)$ as unity for halos located in neutral regions, where radiative feedback is absent.

As a result, the monochromatic rest-frame UV luminosity ($\LUV$), which is directly linked to the star-formation rate (SFR) through a constant conversion factor $\kappaUV$ \footnote{This conversion factor $\kappaUV$ is defined as $\LUV = \dot{M}_*(M_h,z) / \kappaUV$ and depends on the star formation history as well as the assumed properties of the stellar population (age, IMF, binarity, metallicity).}, depends on the extent to which a galaxy is affected by radiative feedback due to reionization. The UV luminosity of galaxies in regions where radiative feedback is absent can be written as  - 
\begin{equation}
\label{eq:LUV_nofb}
L^{\rm nofb}_{{\rm UV}}(M_h,z) = \dfrac{\dot{M}^{\rm nofb}_*(M_h,z)}{\mathcal {K}_{{\rm UV}}} = \dfrac{\varepsilon_{{\rm *,UV}}(M_h,z)}{\mathcal {K}_{{\rm UV,fid}}}~\bigg(\dfrac{\Omega_b}{\Omega_m}\bigg) M_h
\end{equation} 
In contrast, for halos located in ionized regions, where the feedback arising from the ionizing UV background can be appreciable, the UV luminosity is suppressed and given by \cite{Sobacchi&Mesinger_1Dsims_2013} :
\begin{equation}
\label{eq:LUV_fb}
L^{\rm fb}_{{\rm UV}}(M_h,z) = 2^{-\Mcrit/M_h}~L^{\rm nofb}_{{\rm UV}} = 2^{-\Mcrit/M_h}~\dfrac{\varepsilon_{{\rm *,UV}}(M_h,z)}{\mathcal {K}_{{\rm UV,fid}}}~\bigg(\dfrac{\Omega_b}{\Omega_m}\bigg) M_h
\end{equation} 

Here, the parameter $\varepsilon_{{\rm *,UV}}(M_h,z)$ denotes the UV efficiency of a halo of mass $M_h$ at redshift $z$. This quantity encapsulates several key astrophysical factors, including the star-formation efficiency, the characteristic timescale of star formation, and the amount of non-ionizing UV luminosity produced per unit of star formation, and is defined as:
\begin{equation}
\label{eq:epsilonstarUV}
\varepsilon_{{\rm *,UV}}(M_h,z) \equiv \dfrac{f_{*}(M_h,z)}{t_{*}(M_h,z)}~\dfrac{\mathcal {K}_{\rm UV,fid}}{\mathcal {K}_{\rm UV}}.
\end{equation}

Motivated by both observational inferences and simulations that indicate a strong correlation between stellar and halo masses in galaxies \cite{Harikane2016, Ma2018, Behroozi19, Zhu2020, Stefanon2021, Kannan2022, DiCesare2023, Shuntov2025_SHMR, Paquereau2025}, we parameterize the star-formation efficiency as $
f_{\ast}(M_h) = f_{*,10} \, \left(M_h/10^{10} M_\odot\right)^{\alpha_\ast}$,
and assume the star-formation timescale \footnote{Here, $t_H(z) = H^{-1}(z)$ denotes the Hubble time at redshift $z$ and $c_\ast$ is a dimensionless constant.} to be given by  $t_\ast(M_h,z) = c_\ast \, t_H(z)$  in this model. We note that, while internal feedback processes — such as supernova- and stellar-driven outflows or radiation pressure from stars within the galaxy — are not explicitly modeled in our framework, their impact on regulating the efficiency of star-formation is partially captured through the mass-dependence of $f_\ast(M_h)$ (e.g., see Furlanetto et al. (2017) \cite{Furlanetto2017}).
Under these assumptions, the mass- and redshift-dependent UV efficiency can be expressed as  
\begin{equation}
\label{eq:epsilonstarUV_reExpressed}
\varepsilon_{{\rm *,UV}}(M_h,z)  = \dfrac{\varepsilon_{{\rm *10,UV}}}{t_H(z)} \left(\dfrac{M_h}{10^{10} M_\odot}\right)^{\alpha_\ast} ,
\end{equation}
where the normalization is defined as  

\begin{equation}
\label{eq:epsilonstarUV_normalisation}
\varepsilon_{{\rm *10,UV}}\equiv \dfrac{f_{*,10}}{c_{*}} \, \dfrac{\mathcal{K}_{\rm UV,fid}}{\mathcal{K}_{\rm UV}} .
\end{equation}

The rest-frame UV luminosities obtained from this model are finally converted to an absolute UV magnitude (in the AB system) using the relation \cite{Oke_ABmag, Oke&Gunn_ABmag} - 
\begin{equation}
{\rm log_{10}}\left(\frac{L_{\rm UV}}{{\rm erg \ s^{-1} \ Hz^{-1}}} \right) = 0.4 \times (51.6 - M_{\rm UV}). 
\end{equation} 

At a given redshift $z$, the globally averaged UV luminosity function (${\rm \Phi^{total}_{UV}}$) is thereafter obtained by appropriately combining the feedback-affected UV luminosity function (${\rm \Phi^{fb}_{UV}}$) from ionized regions and the feedback-unaffected UV luminosity function (${\rm \Phi^{nofb}_{UV}}$) from neutral regions : 
\begin{align}
\label{eqn:lumfunc_full}
\Phi^{\rm total}_{\rm UV} (\MUV, z) &= Q_{\rm HII}(z)~{ \Phi^{\rm fb}_{\rm UV}} + [1-Q_{\rm HII}(z)]~{\rm \Phi^{\rm nofb}_{\rm UV}}
\nonumber \\
&= Q_{\rm HII}(z) \frac{{\rm d}n}{{\rm d}M_h} \left|\frac{{\rm d}M_h}{{\rm d}{L^{\rm fb}_{\rm UV}}}\right|~\left|\frac{{\rm d}{ L^{\rm fb}_{\rm UV}}}{{\rm d}{ M_{\rm UV}}} \right| 
+ \big[1 - Q_{\rm HII}(z)\big] \frac{{\rm d}n}{{\rm d}M_h} \left|\frac{{\rm d}M_h}{{\rm d}{L^{\rm nofb}_{\rm UV}}}\right|~\left|\frac{{\rm d}{ L^{\rm nofb}_{\rm UV}}}{{\rm d}{M_{\rm UV}}}\right|,
\end{align}
where  $Q_{\rm HII}(z)$ is the global volume-averaged ionization fraction at redshift $z$ and ${\rm d}n / {\rm d}M_h$ is dark matter halo mass function. We adopt the fitting formula from Jenkins et al. (2001) \cite{Jenkins2001} for ${\rm d}n / {\rm d}M_h$. It is worth emphasizing that the probability of a galaxy residing in an ionized region will also be influenced by the clustering of galaxies owing to their highly biased nature and is in fact expected to be larger than the global ionized volume fraction at that epoch, as assumed in deriving \eqn{eqn:lumfunc_full}. Since such an effect is challenging to model self-consistently within a semi-analytical framework like ours and is more reliably captured through numerical simulations, we neglect it when calculating various globally averaged quantities in this work.

We self-consistently compute the evolution of the globally averaged ionization fraction $Q_{\rm HII}$, which is required for determining the UVLF (see \eqn{eqn:lumfunc_full}), from the model assuming star-forming galaxies to be the primary sources of ionizing photons at high redshifts. While JWST has uncovered relatively large numbers of low- and moderate-luminosity active galactic nuclei (AGN) candidates at $z > 5$, studies so far have differed on their contribution to reionization: some suggest AGN could be the main drivers \citep{Madau2024, Grazian2024}, whereas others find their role to be subdominant compared to star-forming galaxies and contributing only $\approx$ 17–23\% of the total share of the ionizing budget \citep{Dayal2025, Asthana2024, Jiang2025}. As a result, we do not include AGN as ionizing sources in our model.

To carry out this calculation, one requires information about the intrinsic ionizing photon production rate per unit comoving volume within a galaxy as well as the fraction of these photons that escape the galaxy and reach the IGM. We model the intrinsic photon production rate in a halo in terms of its star formation rate and the number of ionizing photons emitted per unit mass of stars formed ($\etaGamma$). The escape fraction of ionizing photons, however, remains poorly understood, both from observational studies of low-redshift analogues (e.g., \cite{Nakajima2020, Izotov2021, Flury2022, Begley2022, Saxena2022, Naidu2022, Chisholm2022, Pahl2023}) and from state-of-the-art numerical simulations \cite{Kimm2014, Paardekooper2015, Kimm2017, Trebitsch2017, Lewis2020, Rosdahl2022, Yeh2023}. While theoretical simulations indicate that the escape fraction, $f_\mathrm{esc}$, is connected to global halo properties such as its mass, the exact nature of this dependence remains uncertain. Some studies find $f_\mathrm{esc}$ to increase with halo mass \cite{Sharma2016, Naidu2020}, while others report a decrease (e.g., \cite{Kimm2014, Paardekooper2015, Xu2016, Lewis2020, Mutch2023}), and still others suggest a peak at intermediate masses (e.g., \cite{Ma2020, Rosdahl2022, Yeh2023, Kostyuk2023}). Motivated by these results, the escape fraction of hydrogen ionizing photons, $f_{\rm esc}$, is also taken to be dependent on the halo mass in our model. Under these assumptions, the comoving number density of ionizing photons per unit time contributed by galaxies in neutral regions, unaffected by radiative feedback, $\dot{n}^{\rm nofb}_{\rm ion}$, is given by
\begin{equation}
\dot{n}^{\rm nofb}_{\rm ion} (z) =\eta_{\gamma*,{\rm fid}}  \int_{M_{\rm cool}(z)}^{\infty} \varepsilon_{{\rm esc}}(M'_h,z)~~\varepsilon_{{\rm *,UV}}(M'_h,z) ~\left(\frac{\Omega_b}{\Omega_m}\right) ~~ M'_h ~~ \frac{dn}{dM_h}(M'_h,z) ~~ dM'_h
\label{eq:niondot_nofb}
\end{equation}
while for galaxies in ionized regions, where radiative feedback suppresses star formation in low-mass halos, the corresponding rate becomes
\begin{equation}
\dot{n}^{\rm fb}_{\rm ion} (z) = \eta_{\gamma*,{\rm fid}} \int_{M_{\rm cool}(z)}^{\infty} ~2^{-\Mcrit/M'_h}~~\varepsilon_{{\rm esc}}(M'_h,z)~~\varepsilon_{{\rm *,UV}}(M'_h,z) ~\left(\frac{\Omega_b}{\Omega_m}\right) ~~ M'_h ~~ \frac{dn}{dM_h}(M'_h,z) ~~ dM'_h
\label{eq:niondot_fb}
\end{equation}

Here, $\varepsilon_{{\rm esc}}(M_h,z)$ denotes the escaping ionizing efficiency and is defined as - 
\begin{equation}
\label{eqn:espilon_escape}
\varepsilon_{{\rm esc}}(M_h,z) \equiv \dfrac{\kappaUV}{\kappaUVfid} ~ \dfrac{\etaGamma}{\etaGammafid}~f_{{\rm esc}}(M_h,z).
\end{equation}

In the interest of simplicity, we model the ionizing escape fraction as a power-law function of halo mass, assuming it to be independent of redshift at fixed halo mass: $f_{\rm esc}(M_h,z) = f_{\rm esc,10} \left(M_h/10^{10} M_\odot\right)^{\alpha_{\rm esc}}$. However, in this case, we emphasize that the population-averaged escape fraction $\langle f_\mathrm{esc}\rangle$ still evolves with redshift as a consequence of the evolving halo mass function. Therefore, the escaping ionizing efficiency introduced in \eqref{eqn:espilon_escape} can be written as  
\begin{equation}
\label{eqn:espilon_escape_reExpressed}
\varepsilon_{\rm esc}(M_h,z) = \varepsilon_{{\rm esc,10}} \left(\frac{M_h}{10^{10} M_\odot}\right)^{\alpha_{\rm esc}},
\end{equation}
where the normalization is defined as  
\begin{equation}
\label{eqn:espilon_escape_normalisation}
\varepsilon_{{\rm esc,10}} \equiv \dfrac{\kappaUV}{\kappaUVfid} \, \dfrac{\etaGamma}{\etaGammafid}\, f_{{\rm esc,10}}.
\end{equation}

Note that the product of $\kappaUV$ and $\etaGamma$ equates to the ionizing photon production efficiency $\xi_{\rm ion}$ \cite{Chakraborty2024}, which quantifies the production rate of ionizing photons per unit UV luminosity at 1500 $\angstrom$, and thus, allows \eqn{eqn:espilon_escape_normalisation} to also be interpreted as : 
\begin{equation}
\label{eqn:espilon_escape_normalisation_ver2}
\varepsilon_{{\rm esc,10}} \equiv \left(\dfrac{\xi_{\rm ion}}{\xi_{\rm ion,fid}}\right)~f_{{\rm esc, 10}}.
\end{equation}

In all these calculations, we assume that only halos with masses above the threshold $M_{\rm cool}(z)$, where atomic cooling becomes effective (i.e., $T_{\rm vir} \geq 10^4$ K), contribute ionizing photons.

The total comoving  number density of ionizing photons that leaks into the IGM per unit time  at a given redshift $z$ is therefore calculated as \cite{Dayal2017, Choudhury&Dayal2019} -
\begin{equation}
\label{eq:niondot_total}
\dot{n}_{\rm ion}(z) = Q_{\rm HII}(z)~\dot{n}^{\rm fb}_{\rm ion}(z) + [1- Q_{\rm HII}(z)]~\dot{n}^{\rm nofb}_{\rm ion}(z) 
\end{equation}

We adopt a fiducial value of  $\mathcal {K}_{{\rm UV,fid}} = 1.15485 \times 10^{-28} ~{\rm \mathrm{M}_{{\odot }}}\ {\rm yr}^{-1}/$ $({\rm erg}~{\rm s}^{-1}~{\rm Hz}^{-1})$  and $\eta_{\gamma*,{\rm fid}}$ = $4.62175 \times 10^{60}$ photons per M$_\odot$ in all our calculations. These values were obtained using {\tt{ STARBURST99} v7.0.1}\footnote{https://www.stsci.edu/science/starburst99/docs/default.htm}\cite{Starburst99} for a stellar population with a Salpeter IMF  (0.1 - 100 $\Msun$) and metallicity $Z = 0.001 ( = 0.05~Z_\odot)$ at an age of 100 Myr, assuming continuous star formation. The assumed fiducial values for $\mathcal {K}_{{\rm UV}}$ and $\eta_{\gamma*}$ correspond to an ionizing photon production efficiency $\log_{10} \big[\xi_{\rm ion,fid}/({\rm erg}^{-1}\ {\rm Hz}) \big] \approx 25.23$.  This value is consistent with the canonical ionizing photon production efficiency ($\xi_{\mathrm{ion}} \simeq 10^{25.2} \, \mathrm{erg^{-1} ~Hz}$) commonly assumed in reionization studies before the launch of JWST \cite{Robertson2013, Robertson2015, Bouwens2016_xi_ion}. In contrast, the initial JWST-based measurements of $\xi_{\mathrm{ion}}$, obtained from studies based on strong emission-line galaxies, reported significantly higher efficiencies of $\xi_{\mathrm{ion}} \simeq 10^{25.8}-10^{26.0} \, \mathrm{~erg^{-1} ~ Hz}$ \cite{Atek2024_Spectroscopy, Simmonds2024}. However, more recent analyses with the JWST have revised these estimates downward, yielding values closer to the pre-JWST benchmark \cite{Simmonds2024_JADES_xiION, Pahl2025_xiION, Begley2025_xiION, Llerena2025_xiION}.

As a caveat, we note that treating the properties of high-redshift galaxies as deterministic, one-to-one functions of their host halo mass (apart from whether or not it is affected by radiative feedback) is a simplifying assumption of our model. In reality, the relationship between galaxy properties and their host dark matter halos is expected to show scatter, introducing stochasticity into the ``mean'' relations (e.g., $\varepsilon_{*,\mathrm{UV}}$ – $M_h$, $\varepsilon_{\mathrm{esc}}$ – $M_h$) discussed above. A comprehensive analysis of how this scatter affects various globally averaged observables \cite{Nikolic2024} is beyond the scope of this paper and will be pursued in future work.

In \citetalias{Chakraborty2024}, the parameters $\log_{10}(\varepsilon_{\rm *10,UV})$ and $\alpha_*$ that determine the UV efficiency of halos (see \eqn{eq:epsilonstarUV_reExpressed}) were modeled as redshift-dependent, evolving according to an empirically motivated tanh-based parameterization (see their Appendix A for details). Although a potential temporal offset between the transitions of these two parameters was also considered in that work, it was found to be negligible. Consequently, in this study, we make the simplifying assumption that the transition redshift and the redshift width for this tanh evolution are identical for both parameters, that is, $z_\alpha =z_\varepsilon = z_\ast$ and $\Delta z_\varepsilon = \Delta z_\alpha = \Delta z_\ast$. Therefore, the respective redshift evolution is now given by   

\begin{equation}
\log_{10}(\varepsilon_{\rm *10,UV}) = \ell_{\varepsilon,0} + \dfrac{\ell_{\varepsilon, \mathrm{jump}}}{2} \tanh\left(\dfrac{z-z_\ast}{\Delta z_\ast}\right),
\end{equation}
and
\begin{equation}
\alpha_\ast = \alpha_0 + \dfrac{\alpha_\mathrm{jump}}{2} \tanh\left(\dfrac{z-z_\ast}{\Delta z_\ast}\right).
\end{equation}
In this formulation, the parameter $\log_{10} \varepsilon_{*10,\mathrm{UV}} $ ($\alpha_\ast$) asymptotes to $\ell_{\varepsilon,0} - \ell_{\varepsilon, \mathrm{jump}} / 2$  ($\alpha_{0} - \alpha_{\mathrm{jump}} / 2$) at low redshifts and to $\ell_{\varepsilon,0} + \ell_{\varepsilon, \mathrm{jump}} / 2$ ($\alpha_{0} + \alpha_{\mathrm{jump}} / 2$) at high redshifts, with the transition between these values occurring at a characteristic redshift $z_\ast$ over a range $\Delta z_\ast$.

Once the global reionization history $Q_{\rm HII}(z)$ is obtained, the Thomson scattering optical depth of the CMB photons for that particular model is computed as
\begin{equation}
\label{eq:tauCMB}
\tau_{\rm el} \equiv \tau (z_{\rm LSS}) = \sigma _T \bar{n}_{H}c \int _0^{z_{\rm LSS}} \frac{\mathrm{d}z^{\prime }}{H(z^{\prime })} ~ (1 + z^{\prime })^2 ~ \chi _{\mathrm{He}}(z^{\prime }) ~ Q_{\mathrm{HII}}(z^{\prime }),
\end{equation}
where $z_{\rm LSS}$ is the redshift of last scattering,  $\bar{n}_{\rm H} $ is the current mean comoving number density of hydrogen, and $\sigma_T$ is the Thomson scattering cross-section. In practice, the integral in \eqn{eq:tauCMB} does not pick up any contributions from redshifts before the onset of reionization $z_{\text{start}}$, as $Q_{\mathrm{HII}}(z > z_{\text{start}}) = 0$.

Besides the observables discussed so far, one can also compute the effective number-weighted linear bias, $b^{\rm eff}_{\rm gal}$, of galaxies at a redshift $z$ from the model:
\begin{equation}
\label{eqn:effective_bias_definition}
b^{\rm eff}_{\rm gal}(z) = \dfrac{\displaystyle \int _{M_{\rm UV,min}}^{M_{\rm UV,max}} {\rm d}\MUV~ \{Q_{\rm HII}(z)~{ b^{\rm fb}_{\rm gal}(\MUV,z)~\Phi^{\rm fb}_{\rm UV}} + [1-Q_{\rm HII}(z)]~b^{\rm nofb}_{\rm gal}(\MUV,z)~{\rm \Phi^{\rm nofb}_{\rm UV}} \} }{ \displaystyle \int _{M_{\rm UV,min}}^{M_{\rm UV,max}} {{\rm d} \MUV}~ \{Q_{\rm HII}(z)~{ \Phi^{\rm fb}_{\rm UV}} + [1-Q_{\rm HII}(z)]~{\rm \Phi^{\rm nofb}_{\rm UV}} \}},
\end{equation}
where $b^{\rm fb}_{\rm gal}(\MUV,z)$ and $b^{\rm nofb}_{\rm gal}(\MUV,z)$ represents the linear bias of galaxies, with absolute magnitude $\MUV$ at redshift $z$, residing in ionized and neutral regions respectively and is calculated from the linear halo bias $b_{\rm halo}$, as follows
\begin{align}
\label{eqn:bgal_definition}
b^{\rm fb}_{\rm gal}(\MUV,z) &= b_{\rm halo}\left(M_h (\MUV)\Big|_{\rm fb} \, ,z\right)
\nonumber \\
b^{\rm nofb}_{\rm gal}(\MUV,z) &= b_{\rm halo}\left(M_h (\MUV)\Big|_{\rm nofb} \, ,z\right).
\end{align}

While calculating the effective bias, we take $M_{\rm UV,max}$ to be same as that adopted in the observational study under consideration and $M_{\rm UV,min}$ = $-21$ \footnote{This is because our model does not account for the effects of active galactic nuclei (AGN) feedback or dust attenuation that are likely to affect the brighter galaxies. If dust attenuation is also significant within faint galaxies, then the inferred UV efficiency parameter $\varepsilon_\mathrm{*10,UV}$ should be interpreted as incorporating both the impact of dust obscuration and the processes governing the production of non-ionizing UV emission in these galaxies.}. We use the fitting formula of Tinker et al. (2010) \cite{Tinker2010} to compute the linear halo bias $b_{\rm halo}$.

To summarize, the \textbf{baseline} model of high-redshift star-forming galaxies has \textbf{\emph{nine}} free parameters - $ \boldsymbol \theta = \{ \ell_{\varepsilon,0} - \ell_{\varepsilon, \mathrm{jump}} / 2 ~,~ \alpha_{0} - \alpha_{\mathrm{jump}} / 2 ~,~  \ell_{\varepsilon,0} + \ell_{\varepsilon, \mathrm{jump}} / 2  ~,~ \alpha_{0} + \alpha_{\mathrm{jump}} / 2 ~,~  z_\ast ~,~ \Delta z_\ast ~,~ \newline \alpha_{esc} ~,~  \log_{10}~(\varepsilon_{\mathrm{esc,10}}) ~,~ \log_{10}(M_{\mathrm{crit}}/M_\odot) \}$. 

\section{Observational Datasets and Likelihood Analysis}
\label{sec:data_and_methods}

We utilize several available observational datasets to constrain the theoretical model described in the previous section through a Bayesian analysis. In this section, we briefly summarize them and also describe the Bayesian formalism used to constrain the free parameters of our model.

\begin{enumerate}
    \item \textbf{Thomson scattering optical depth of CMB photons: } For our analysis, we use the latest measurement of $\tau_{\rm el} = 0.054 \pm 0.007$ reported by the Planck collaboration \citep{Planck2020}.

    \item  \textbf{Global Reionization History: } We utilize measurements of the globally averaged neutral hydrogen fraction ($Q_\mathrm{HI} = 1 - Q_\mathrm{HII} $) in the IGM  at different redshifts derived from Lyman-$\alpha$ absorption studies of distant quasars and galaxies, similar to our previous work \cite{Chakraborty2024}. It is essential to remember that all these derived constraints are however model-dependent.
    
    \item \textbf{Galaxy UV Luminosity Functions:} We use measurements of the galaxy UV luminosity functions $\Phi_{\rm UV} (\MUV, z)$ at nine redshifts spanning the range: 5 $\leq z \leq$ 15, obtained from various surveys conducted with the HST \cite{Bouwens2021} and the JWST \cite{Donnan2023, Harikane2023, Bouwens2023, McLeod2024}. In addition to the datasets used in \citetalias{Chakraborty2024}, we have included new JWST measurements at $z \geq 9$ \cite{Donnan2024}  and the measurements from HST at $z = 5$ in this present work. As our theoretical model does not incorporate the effects of feedback from AGN activity or the significant dust attenuation present in bright galaxies, we consider only the observational data points with $M_{\rm UV} \geq -21$ from these studies in our analysis \cite{Mauerhofer&Dayal2023}.

    \item  \textbf{Galaxy Bias: }  We use the most recent estimates of the galaxy bias, $b_{\rm gal}(z)$, obtained from JWST observations. While it would have been more appropriate to fit the projected correlation function directly using our model of galaxy–halo connection, we take a simpler approach in this work and model only the large-scale bias, derived from a power-law fit or halo occupation distribution (HOD) modeling of the angular correlation function (ACF). A more detailed analysis using the ACF measurements directly will be undertaken in future work. 
    
    In this work, we incorporate results from two independent analyses: 
    
    (a) Dalmasso et al. (2024) \cite{DalmassoJWST}, who measure the galaxy bias from a power-law fit to the angular two–point correlation function (2PCF) of a photometric sample of Lyman-break galaxies spanning the redshift range $5.5 \leq z \leq 10.6$, and 
    
    (b) Shuntov et al. (2025) \cite{Shuntov2025}, who constrain the galaxy bias at mean redshifts $\langle z \rangle = 5.4$ and $\langle z \rangle = 7.3$  by jointly fitting the measurements of the UV luminosity functions and the 2PCF of spectroscopically selected H$\alpha$ and [O III] emitters in the GOODS–North and GOODS–South fields under a HOD framework. 
    
    For consistency, when computing the effective galaxy bias from the theoretical model, we impose the same magnitude cuts ($M_{\rm UV,max}$) as adopted in the respective studies. The analysis of Dalmasso et al. (2024) used a redshift-dependent threshold, enabling them to probe comparatively fainter galaxies, that evolves from $M_{\rm UV, max} = -15.5$ at $z = 5.5$ to $M_{\rm UV,max} = -17.3$ at $z = 10.6$ (see their Figure 1 for the values at each redshift), while Shuntov et al. (2025) work with brighter sources corresponding to a fixed threshold of $M_{\rm UV,max} = -19.1$.

\end{enumerate}

We use a Bayesian analysis to constrain the free parameters of our model by comparing the theoretical predictions with all or a subset of the observational constraints mentioned above. This involves computing the conditional probability distribution or the posterior $\mathcal {P}(\boldsymbol \theta \vert \mathcal{D})$ of the model parameters $\boldsymbol \theta$ given the observational data $\mathcal{D}$, using the Bayes theorem, as follows
\begin{equation}
    \mathcal{P}(\boldsymbol \theta \vert \mathcal{D})=\frac{\mathcal {L}(\mathcal {D} \vert \boldsymbol \theta) ~\pi (\boldsymbol \theta)}{\mathcal{P}(\mathcal{D})},
\end{equation}
where $\mathcal {L}(\mathcal{D} \vert \boldsymbol \theta)$ is the likelihood, i.e., the conditional probability distribution of the data $\mathcal {D}$ given the model parameters $\boldsymbol \theta$, $\pi(\boldsymbol \theta)$ is the prior distribution of the parameters of the model, and $\mathcal {P}(\mathcal{D})$ is the model evidence, which is redundant in our work. Assuming the datasets to be independent, the joint likelihood is calculated as
\begin{equation}
\label{eqn:joint_likelihood_expression}
\mathcal {L}(\mathcal {D} \vert \boldsymbol \theta) =  \prod_\alpha \mathcal {L}(\mathcal {D}_\alpha \vert \boldsymbol \theta),
\end{equation}
where the index $\alpha$ runs over the datasets (among those mentioned above)  that are used in a particular analysis and the likelihood for any individual dataset $\mathcal{D}_{\alpha}$ is given by
\begin{equation}
\label{eqn:individual_likelihood_expression}
\mathcal {L}(\mathcal {D}_\alpha \vert \boldsymbol \theta) =  \exp \left[-\frac{1}{2} ~ \chi^2(\mathcal {D}_\alpha, \boldsymbol \theta) \right] =  \exp \left[-\frac{1}{2} \sum _{i}\left(\frac{\mathcal {D}_{\alpha,i}-\mathcal {M}_{\alpha,i}(\boldsymbol \theta)}{\sigma _{\alpha,i}}\right)^2 \right],
\end{equation}
where $\mathcal{D}_{\alpha,i}$ are the values of the measured data points, $\sigma_{\alpha,i}$ are the associated observational error bars and $\mathcal{M}_{\alpha,i}(\boldsymbol \theta)$ are the values predicted by the model corresponding to the parameter set $\boldsymbol \theta$. The index $i$ runs over all data points in the dataset $\mathcal{D}_{\alpha}$. In this work, we have $\mathcal {D}_\alpha \subseteq \{ \Phi_{\rm UV}(\MUV,z), \tau_{\rm el},  Q_{\rm HI}(z), b_{\rm gal}(z) \}$.

We use the Monte Carlo Markov Chain (MCMC) method to compute the posterior distribution of the free parameters of the model. To sample the parameter space, we use the publicly available package {\tt{COBAYA}}\footnote{https://cobaya.readthedocs.io/en/latest/} \cite{cobaya}. The samples are drawn using 8 parallel chains and the chains are assumed to have converged when the Gelman–Rubin $R-1$ statistic becomes less than a threshold of 0.01. We discard the first  30$\%$  of the steps in the chains as ‘burn-in’ and use the rest for our analysis.

\section{Results from the Baseline Model}
\label{sec:results_baseline}

In this section, we discuss the results obtained by comparing the theoretical predictions of the \textbf{baseline} model with the available observations. For this purpose, we execute two variants of MCMC runs using different combinations of observational datasets, as mentioned below

\begin{itemize}
    \item \textbf{UVLF+reion}: In this case, we use the first three of the observational datasets (i.e., UVLFs, $Q_\mathrm{HI}(z)$ and $\tau_{\rm el}$)  outlined in \secn{sec:data_and_methods}. This corresponds to the default case in \citetalias{Chakraborty2024} and will enable us to constrain the updated \textbf{baseline} model using the new datasets.
    
    \item \textbf{bias+reion}: In this case, we substitute the observed UVLF dataset with the galaxy bias measurements, while retaining all the reionization observables for the likelihood analysis. The primary motivation behind this run is to understand the galaxy-halo connection required to match the galaxy bias measurements, while also ensuring consistency with the current constraints on the timeline of reionization.
\end{itemize}

\begin{table*}[htbp]
 \caption{Parameter constraints obtained from the MCMC-based analysis. The first nine rows correspond to the free parameters of the \textbf{baseline} model, while the remaining are the derived parameters. The free parameters are assumed to have uniform priors in the range mentioned in the second column. The numbers in the other columns show the mean value with 1$\sigma$ errors for different parameters of the model, as obtained from the two MCMC runs (see \secn{sec:results_baseline}).\\} 
 \label{tab:mcmc_results}
 \begin{tabular*}{\columnwidth}{c@{\hspace*{50pt}}c@{\hspace*{50pt}}c@{\hspace*{50pt}}c@{\hspace*{50pt}}}
 \hline   \hline  \\
    Parameters & Priors &  UVLF+reion & bias+reion  \\  \\
  \hline   \hline \\
{\boldmath$\ell_{\varepsilon,0} + \ell_{\varepsilon,\mathrm{jump}} / 2 $} & [-2.0, 2.0] & $-0.238^{-0.064}_{-0.36} $  & $< -0.267$    \\  \\

{\boldmath$\ell_{\varepsilon,0} - \ell_{\varepsilon,\mathrm{jump}} / 2$} & [-2.0, 1.0] & $-0.910^{+0.069}_{-0.044}  $  &  $-0.43^{+0.12}_{-0.16}$   \\ \\

{\boldmath$z_\ast $} & [8.0, 18.0] & $11.62^{+0.17}_{-1.3}$ & $13.7^{+3.5}_{-2.0}$\\ \\

{\boldmath$\Delta z_\ast $} & [0.5, 6.0] & $1.71^{+0.29}_{-0.85}$  & $3.3^{+2.0}_{-1.5}$\\ \\

{\boldmath$\alpha_0 + \alpha_\mathrm{jump} / 2 $} & [0.0, 7.0] & $1.8844^{-0.0063}_{-1.3}$  &  $< 4.04$   \\ \\

{\boldmath$\alpha_0 - \alpha_\mathrm{jump} / 2 $} & [-1.0, 1.0]  & $0.303^{+0.036}_{-0.050}$ & $-0.23^{+0.16}_{-0.12}$  \\ \\

{\boldmath$\log_{10}~(\varepsilon_{\mathrm{esc,10}})$} & [-3.0, 1.0]  & $-0.813^{+0.042}_{-0.035}$ & $-1.42^{+0.27}_{-0.14}$  \\ \\

{\boldmath$\alpha_{\rm esc}$} & [-3.0, 1.0] & $-0.18^{+0.14}_{-0.11}$ & $0.03^{+0.33}_{-0.17}$   \\ \\

{\boldmath$\log_{10}(M_{\mathrm{crit}}/M_\odot)$} & [9.0, 11.0] &$10.07^{+0.30}_{-0.11}$ & $< 9.82$\\ \\

\hline \\

$\tau_{\rm el}  $ & - &  $0.0543^{+0.0020}_{-0.0024}$ & $0.0537^{+0.0021}_{-0.0029}$ \\ \\

$\ell_{\varepsilon,\mathrm{jump}}$ & -  &$0.673^{-0.028}_{-0.40}$ & $-0.18^{+0.58}_{-1.4}$ \\ \\

$\alpha_\mathrm{jump}  $ & - &$1.58145^{+0.00057}_{-1.3}$ &  $3.2^{+1.3}_{-2.8}$   \\ \\
\hline
\end{tabular*}
\end{table*}


\begin{figure}[htbp]
\centering
\includegraphics[width=\columnwidth]{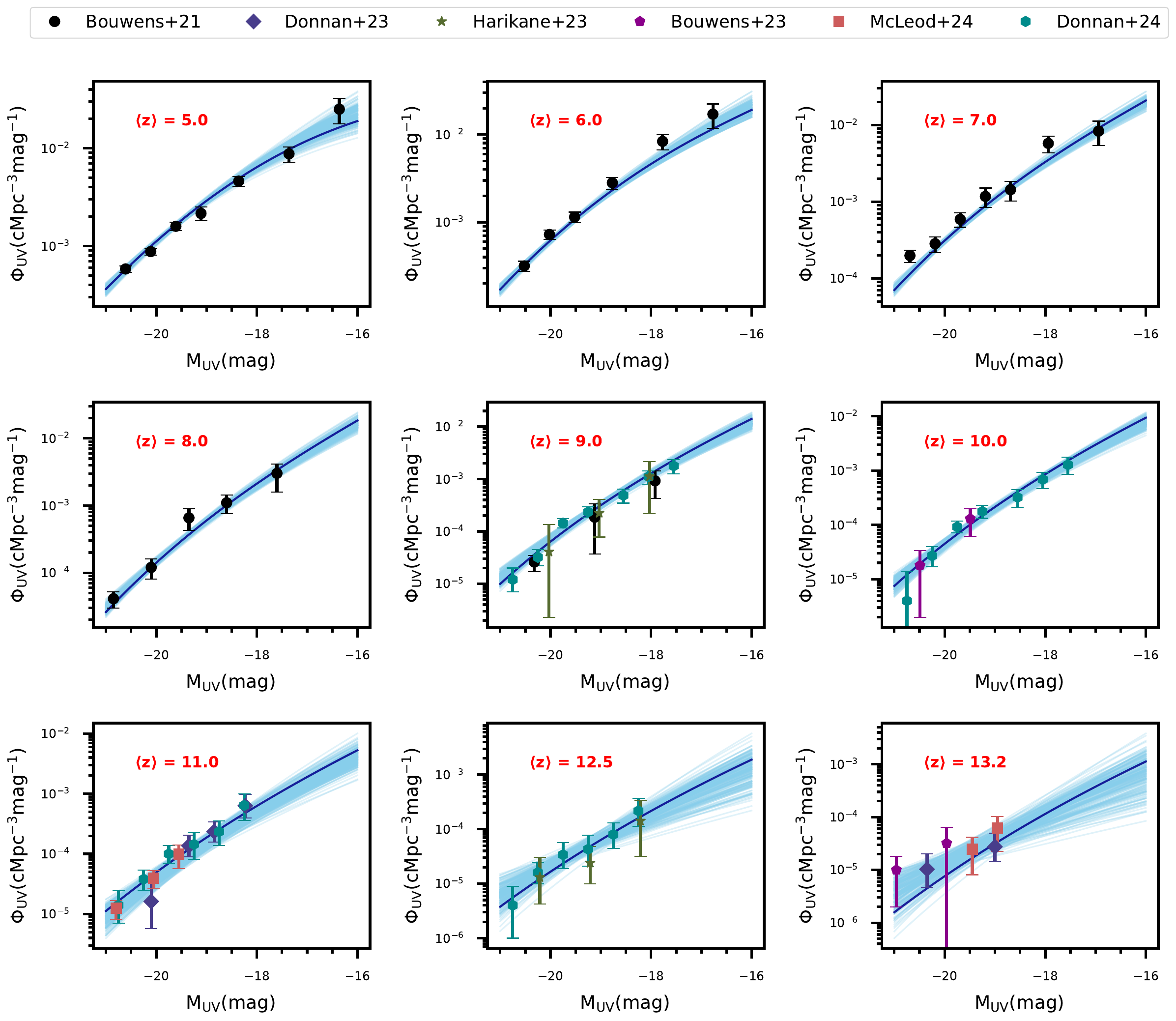}
\caption{The galaxy UV luminosity functions at nine different redshift bins (with their respective mean values $\langle z \rangle$ mentioned in the upper left corner) for 200 random samples drawn from the MCMC chains of the \textbf{UVLF+reion} case, based on the \textbf{baseline} model. In each panel, the solid dark-blue line corresponds to the best-fit model, while the colored data points show the different observational constraints \cite{Bouwens2021, Donnan2023, Harikane2023, Bouwens2023, McLeod2024, Donnan2024} used in the likelihood analysis.}
\label{fig:baseline_case1_UVLF}
\end{figure}

\begin{figure}[htbp]
\centering
\includegraphics[width=\columnwidth]{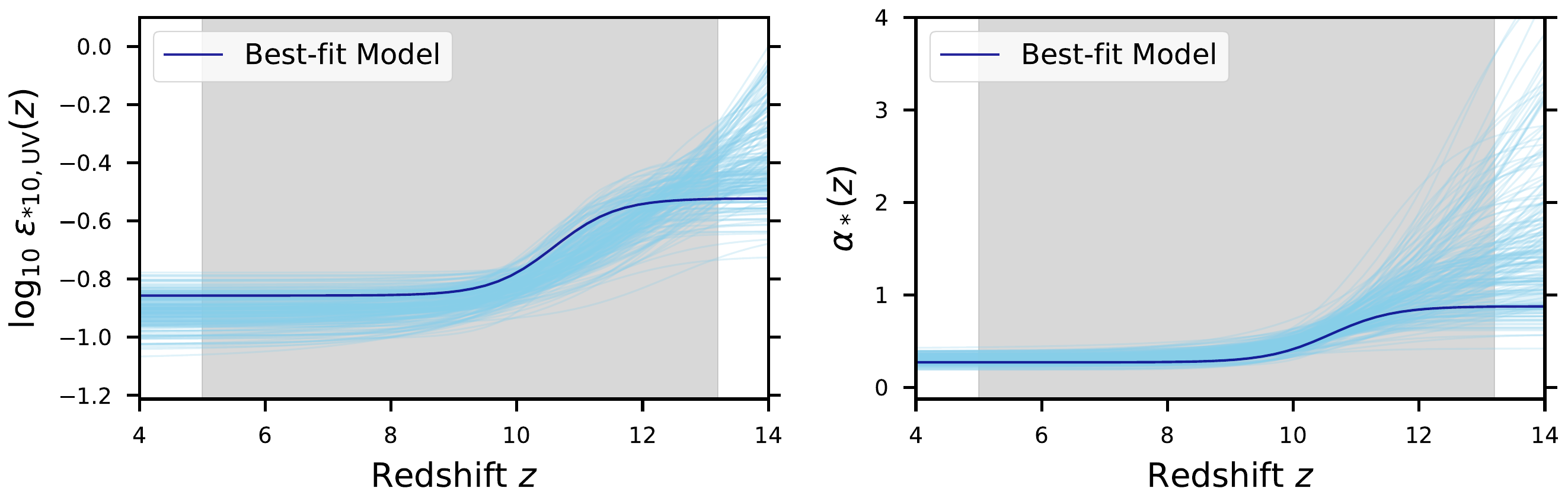}
\caption{The redshift evolution of the normalization (left panel) and power-law (right panel) scaling of the production efficiency of UV radiation with halo mass for 200 random samples drawn from the MCMC chains of the \textbf{UVLF+reion} case, based on the \textbf{baseline} model. The grey box encloses the range of average redshifts (5 $\leq \langle z \rangle \leq$ 13.2) at which UVLF observations have been used for comparison with the model in this work.}
\label{fig:baseline_case1_sfe_params}
\end{figure}

\begin{figure}[htbp]
\centering
\begin{subfigure}[t]{\columnwidth}
    \centering
    \includegraphics[width=0.6\textwidth]{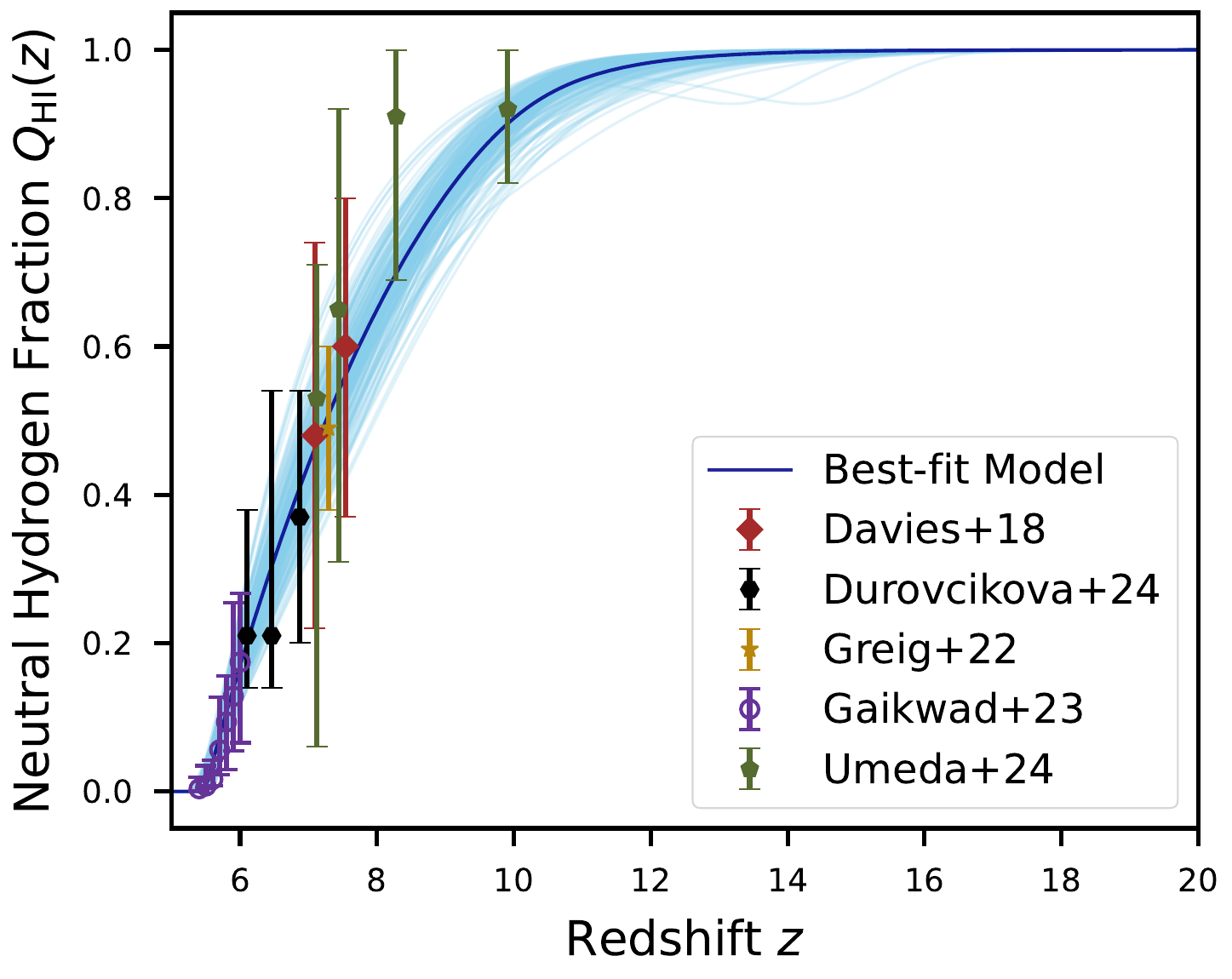}
    \caption{}
    \label{fig:baseline_case1_reion_history}
\end{subfigure}

\vspace{0.5cm} 

\begin{subfigure}[t]{\columnwidth}
    \centering
    \includegraphics[width=0.48\columnwidth]{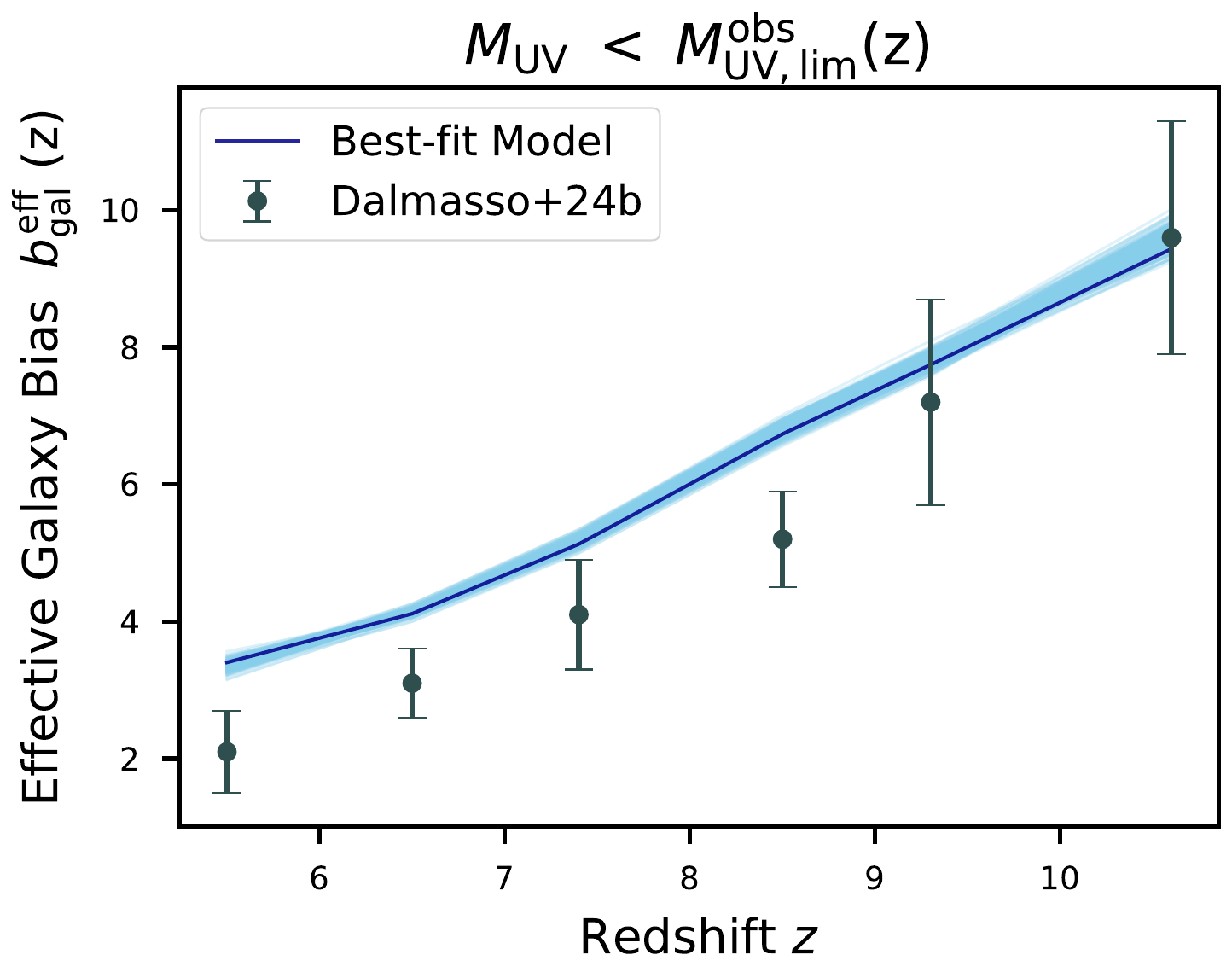}
    \hfill
    \includegraphics[width=0.48\columnwidth]{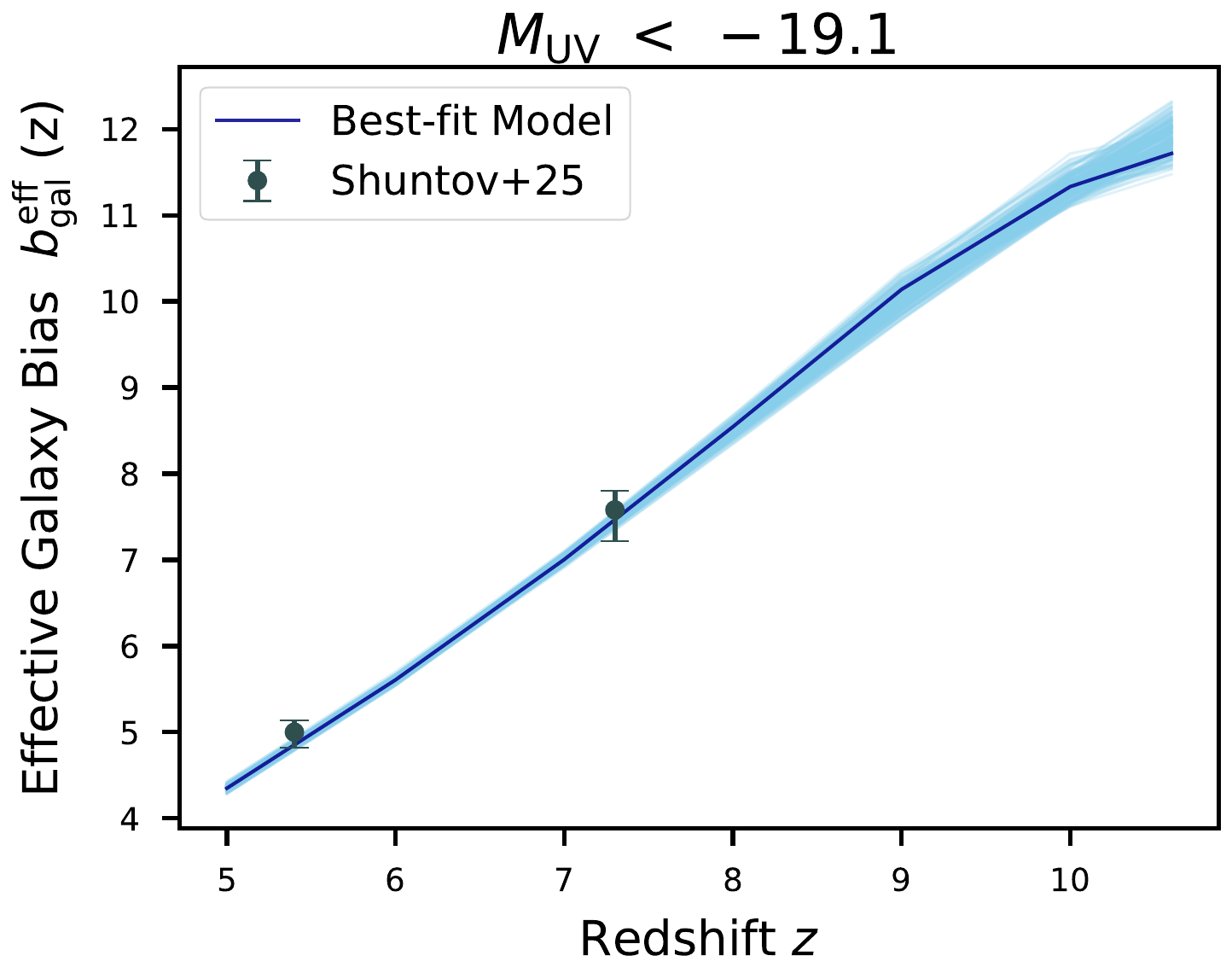 }
    \caption{}
    \label{fig:baseline_case1_gal_bias}
\end{subfigure}
\caption{ The evolution of (a) the globally averaged intergalactic neutral hydrogen fraction and (b) the effective galaxy bias as a function of redshift for 200 random samples drawn from the MCMC chains of the \textbf{UVLF+reion} case, based on the \textbf{baseline} model. The colored data points in each panel represent the respective observational measurements \citep{Davies2018, Greig2022, Gaikwad2023, Umeda2023, Durovcikova2024, DalmassoJWST, Shuntov2025}. Note that, unlike the reionization history, the galaxy bias observations were not included in the likelihood analysis for the \textbf{UVLF+reion} case.}
\label{fig:baseline_case1_reion_history_and_gal_bias}
\end{figure}

We begin with the results obtained from the MCMC runs of the \textbf{UVLF+reion} case. The marginalized constraints on the free and derived parameters are mentioned in the third column of \tab{tab:mcmc_results}. We show the model-predicted UVLFs for 200 random samples from the MCMC chains in \fig{fig:baseline_case1_UVLF}, along with the observational measurements used in the MCMC analysis. The evolution in the efficiency parameters preferred by the data has been plotted in \fig{fig:baseline_case1_sfe_params} for the same 200 random samples. From \tab{tab:mcmc_results}, we find that an increase in the normalization $\varepsilon_{\rm \ast 10,UV}$ and slope $\alpha_\ast$ of the UV efficiency parameter is required at $z \geq$ 10 to match the evolving UVLF observations from JWST, while remaining approximately constant at lower redshifts down to $z = 5$. These findings are qualitatively similar to those reported in our earlier work (\citetalias{Chakraborty2024}), and their interpretation has already been extensively discussed there. From the present analysis, we obtain improved constraints on the timing of this transition, which is expected to occur between $z \approx 10$ and $z \approx 12$ over a redshift interval of $\Delta z \approx 1-2$. Interestingly, the inclusion of the $z = 5$ UVLF observational data in the analysis helps constrain the value of $\log_{10}(\Mcrit/M_\odot)$ to $\approx 10.07$. This is not surprising since the total UVLF at $z = 5$ predicted by our model is exactly equal to the UVLF from feedback-affected regions as hydrogen reionization is complete (i.e., $Q_{\mathrm{HII}} = 1 $) by then in our models. This higher value of $M_{\rm crit}$ — about twice the limit reported in \citetalias{Chakraborty2024} ($M_{\rm crit} < 10^{9.7}~M_\odot$ at $68\%$ confidence) — implies a stronger radiative feedback scenario, wherein its impact is now extended to somewhat more massive halos, while the effect on low- and intermediate-mass halos inside ionized regions becomes slightly more pronounced. The global reionization histories for the 200 random samples from the MCMC chains are shown in \fig{fig:baseline_case1_reion_history}. As a result of this slightly stronger feedback, the model now requires a higher escaping ionizing efficiency of $\approx 15\%$ for $10^{10}~M_\odot$ halos (compared to $\varepsilon_{\rm esc,10} \approx 11\%$ obtained in \citetalias{Chakraborty2024}) to remain consistent with the reionization observables. We also find that the inferred power-law index $\alpha_{\rm esc}$ of the halo-mass dependent escaping ionizing efficiency is somewhat flatter compared to our earlier work ($\alpha_{\rm esc} = -0.38^{+0.19}_{-0.15}$), although still preferring negative values (at 68\% confidence). This flattening of the slope can be understood as a direct consequence of the higher value of $M_{\rm crit}$: since radiative feedback now suppresses star formation even in halos with masses equal to or exceeding $10^{10} M_\odot$, their escaping ionizing efficiency must increase to provide enough ionizing photons that can drive cosmic reionization through its final stages. In other words, the ionizing photon output is now more evenly shared among halos of different masses, which manifests as a marginally shallower $\varepsilon_{\rm esc}-M_h$ relation.

In \fig{fig:baseline_case1_gal_bias}, we compare the redshift evolution of the effective galaxy bias predicted by models of the \textbf{UVLF+reion} case with the recent bias estimates obtained from JWST at $z > 5$ \cite{DalmassoJWST, Shuntov2025}. We emphasize that these measurements were not used to constrain the theoretical model, but instead serve as an independent test of its predictive capability. Our models reproduce the general trend of an increasing galaxy bias with redshift seen in observations. They show excellent agreement with the observational estimates obtained by Shuntov et al. (2025) for the very bright galaxies ($M_{\mathrm{UV}} < -19.1$) at $\langle z \rangle = 5.4$ and $\langle z \rangle = 7.3$. However, when accounting for comparatively fainter galaxies at similar redshifts, the model systematically overestimates the average bias relative to the measurements of Dalmasso et al. (2024) and is, in fact, unable to capture the steep evolution in galaxy bias observed over $5 < z < 11$ in their study. Nevertheless, the model predictions remain consistent with their reported measurements at $z > 9$, which are based on moderately luminous galaxies with $M_{\mathrm{UV}} \lesssim -17$.

The discrepancy between our model predictions and clustering data of Dalmasso et al. (2024) at $z < 9$ indicates that the constructed galaxy–halo connection requires modification to accurately describe fainter and moderately bright galaxies ($M_{\mathrm{UV}} > -19.1$) at lower redshifts. Since galaxies residing in low-mass halos are expected to cluster less strongly than those in high-mass halos \cite{Mo1996, ST99, Benson2000, Jenkins2001, Tinker2010, Coil2013}, if galaxies fainter than $\MUV = -19.1$ at $5<z<9$  occupy lower-mass halos than currently predicted by the model, then the smaller observed bias can be naturally explained.

\begin{figure}[htbp]
\centering
\begin{subfigure}[t]{\columnwidth}
    \centering
    \includegraphics[width=0.6\textwidth]{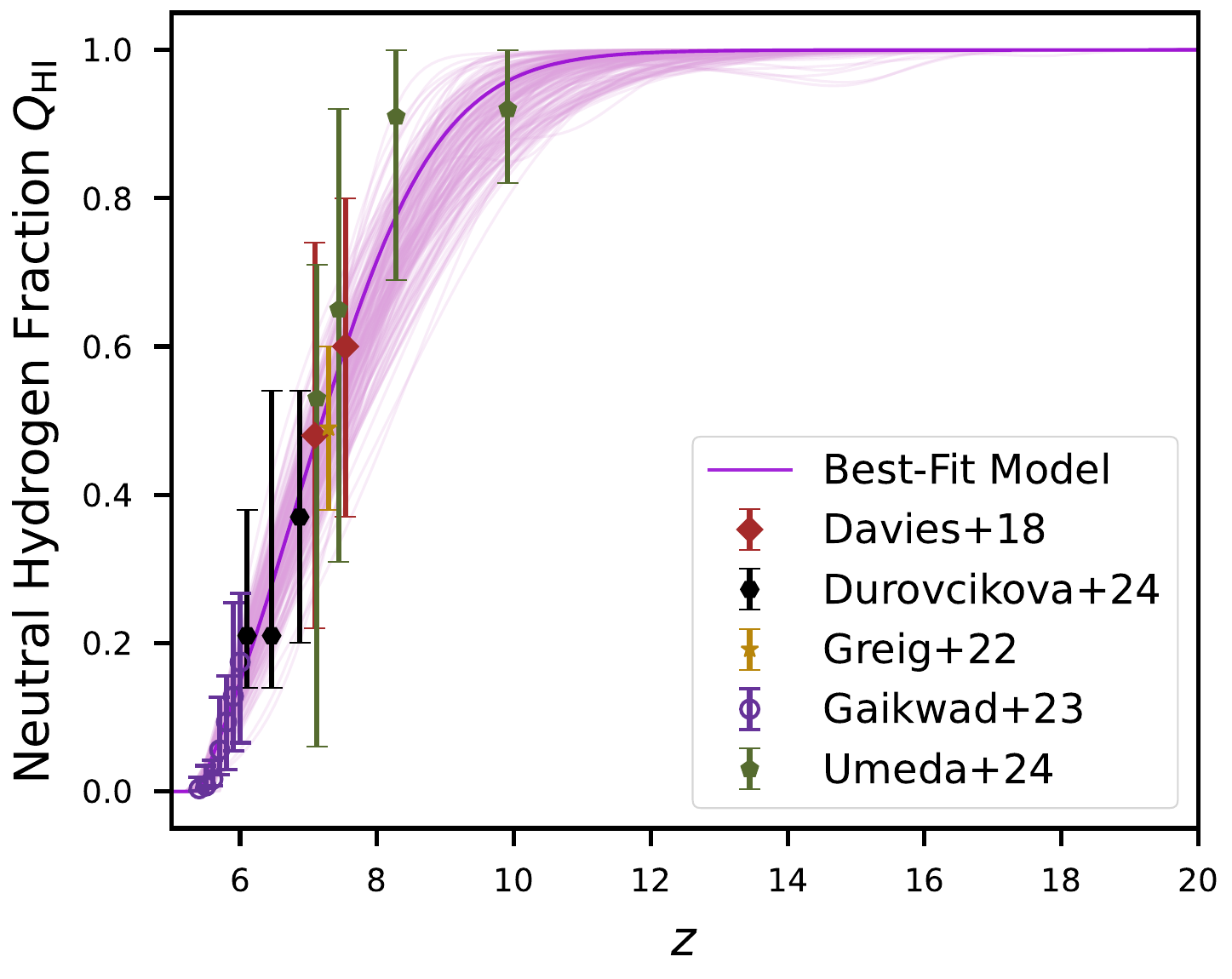}
    \caption{}
    \label{fig:baseline_case2_reion_history}
\end{subfigure}

\vspace{0.5cm} 

\begin{subfigure}[t]{\columnwidth}
    \centering
    \includegraphics[width=0.48\columnwidth]{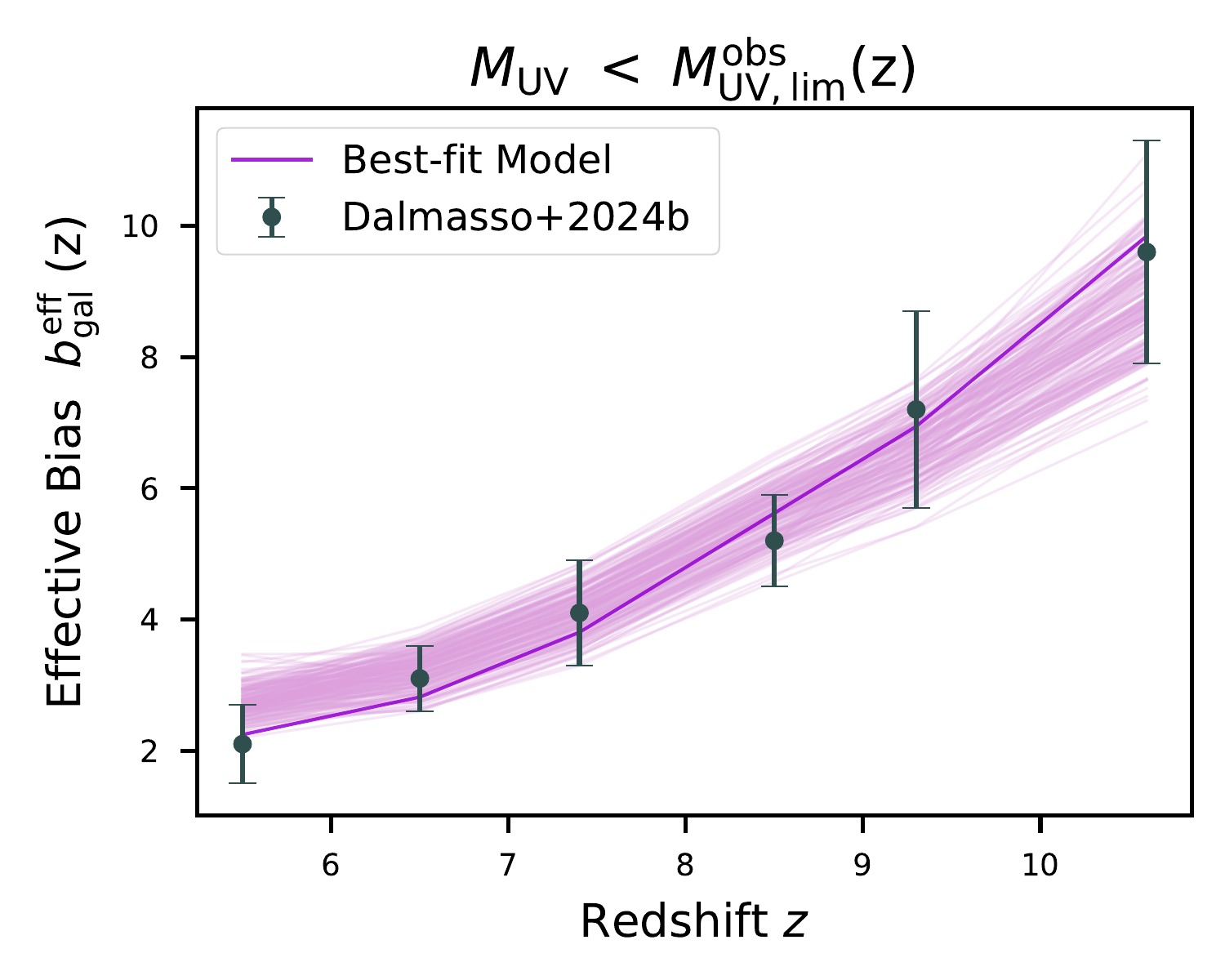}
    \hfill
    \includegraphics[width=0.48\columnwidth]{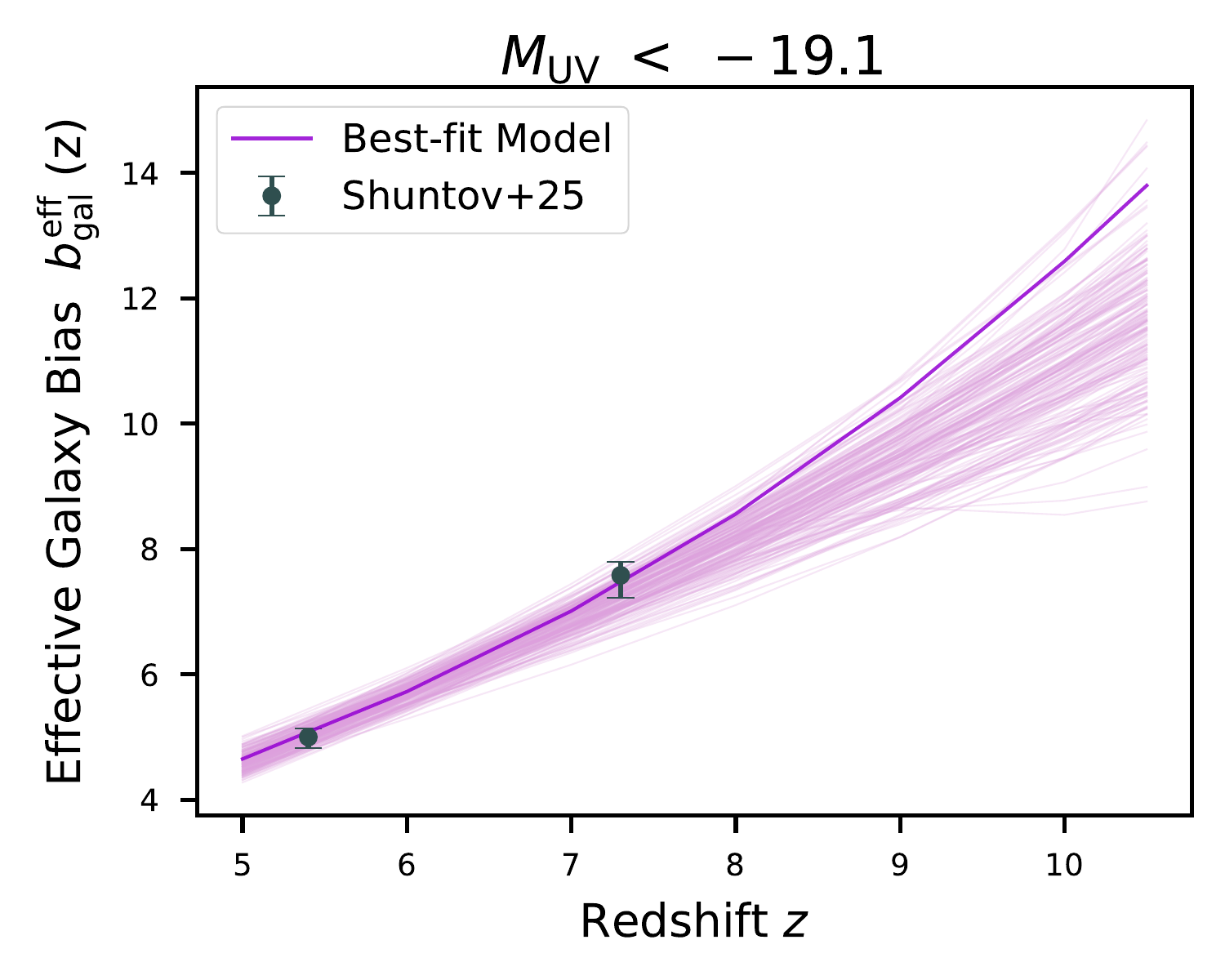}
    \caption{}
    \label{fig:baseline_case2_gal_bias}
\end{subfigure}
\caption{Same as \fig{fig:baseline_case1_reion_history_and_gal_bias} but for 200 random samples drawn from the MCMC chains of the \textbf{bias+reion} case, based on the \textbf{baseline} model. Note that the likelihood analysis for the \textbf{bias+reion} case includes both the reionization history and galaxy bias observations, whereas the UVLF datasets were excluded.
\label{fig:baseline_case2_reion_history_and_gal_bias}}
\end{figure}
\begin{figure}[htbp]
\centering
\includegraphics[width=\columnwidth]{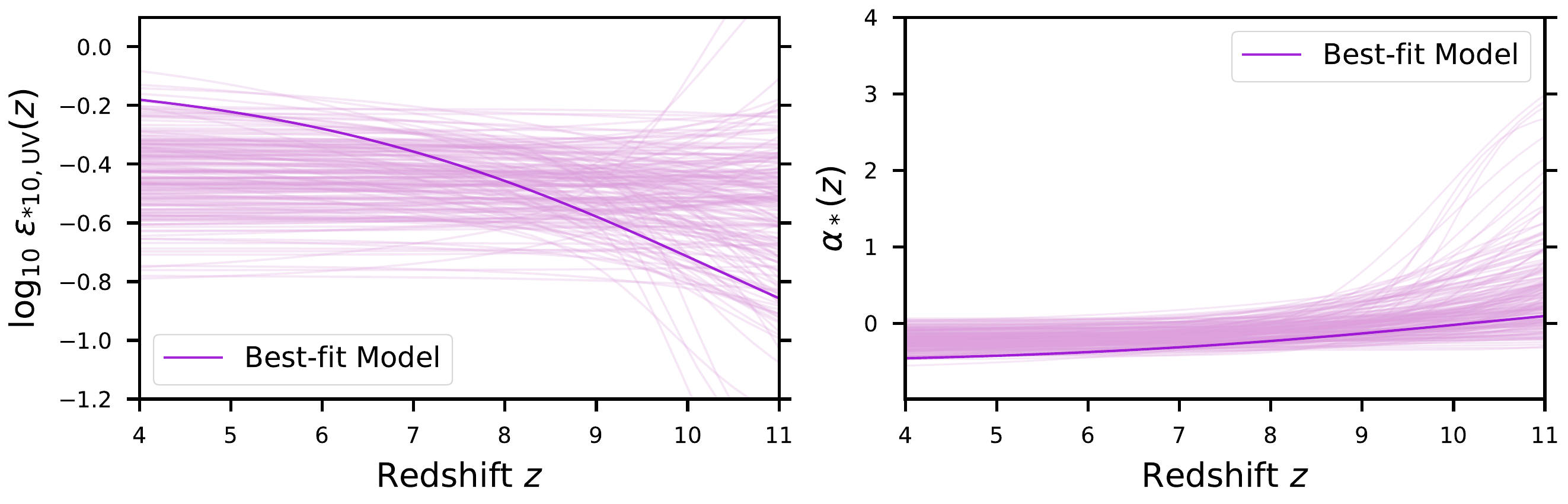}
\caption{ Same as \fig{fig:baseline_case1_sfe_params} but for 200 random samples drawn from the MCMC chains of the \textbf{bias+reion} case, based on the \textbf{baseline} model.}
\label{fig:baseline_case2_sfe_params}
\end{figure}

To investigate the cause of this discrepancy in greater detail, we carried out an analysis, referred to as the \textbf{bias+reion} case, where the requirement for the galaxy-halo connection to match the galaxy UVLFs is relaxed. Instead, we focus on simultaneously fitting the \textbf{baseline} model to both the galaxy bias and the reionization observables ($Q_{\rm HI}$ and $\tau_{\rm el}$). The constraints on the model parameters for this case are mentioned in the fourth column of \tab{tab:mcmc_results}. Additionally, \fig{fig:baseline_case2_reion_history_and_gal_bias} shows the evolution in galaxy bias and the reionization histories for 200 random samples drawn from the MCMC chains of the \textbf{bias+reion} case alongside the corresponding observational measurements.  We also show the redshift evolution in the efficiency parameters for these samples in \fig{fig:baseline_case2_sfe_params}.

\begin{figure}[htbp]
\centering
\includegraphics[width=\columnwidth]{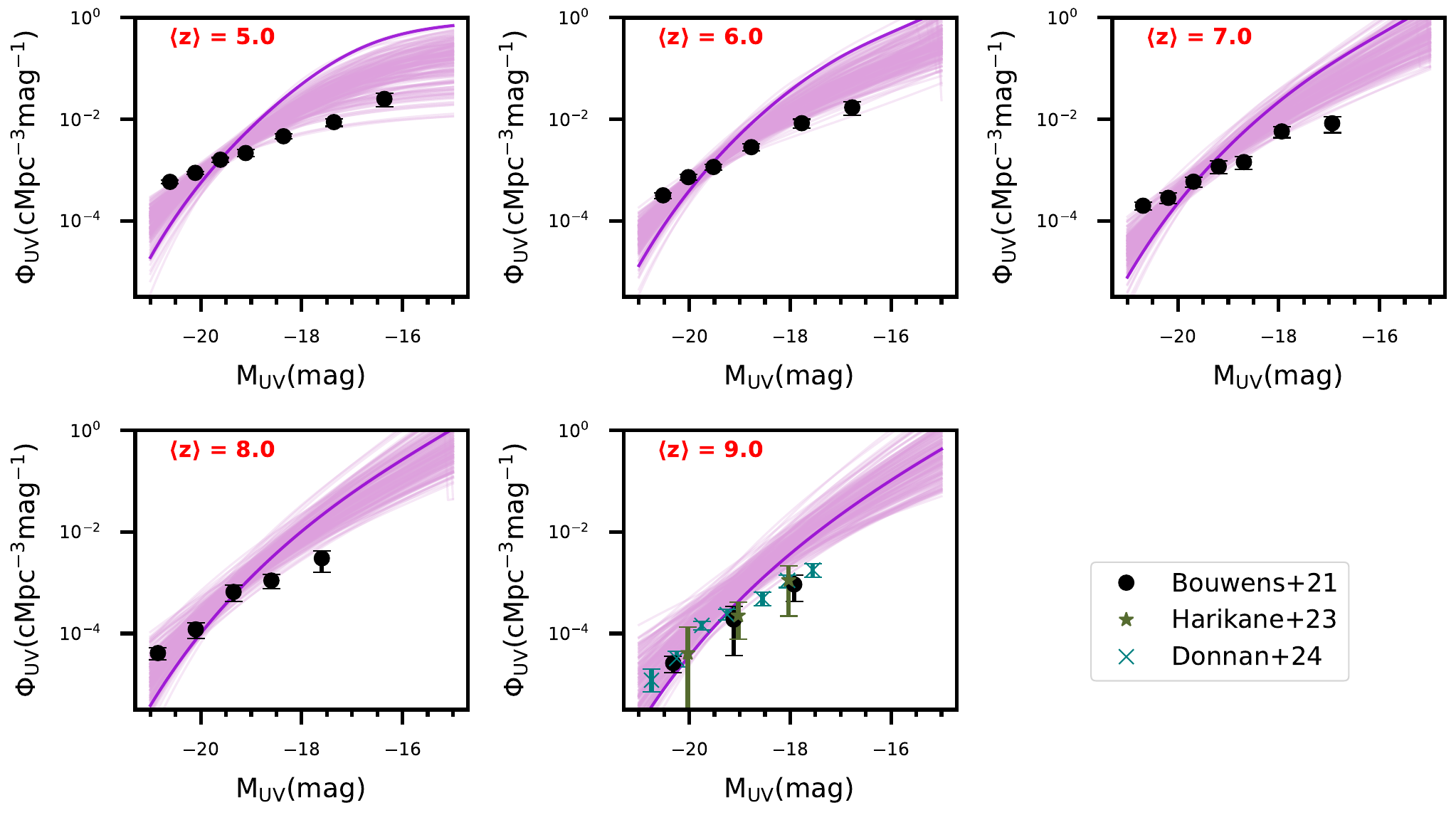}
\caption{The \textbf{\textit{predicted}} galaxy UV luminosity functions at $ 5 \leq z \leq 9$ for 200 random samples drawn from the MCMC chains of the \textbf{bias+reion} case, based on the \textbf{baseline} model. In each panel, the solid dark-violet line corresponds to the best-fit model, while the colored data points show the different observational constraints at $z \leq 9$ \cite{Bouwens2021, Harikane2023, Donnan2024}.}. 
\label{fig:baseline_case2_UVLF}
\end{figure}

 We find the normalization $\varepsilon_{*10,\mathrm{UV}}$ of the UV 
 efficiency parameter takes significantly higher values (compared to the \textbf{UVLF+reion} case) at lower redshifts ($z \lesssim 9$), showing no significant evolution with redshift, as $\ell_{\varepsilon,\mathrm{jump}}$ is statistically consistent with zero.  The power-law index $\alpha_\ast$ is also negative at $z \lesssim 9$, thereby boosting the efficiency of star-formation (and hence, UV radiation) in low-mass halos relative to massive ones. The resulting increase in UV efficiency shifts galaxies of fixed UV luminosity into lower–mass, yet more abundant, halos. Consequently, while this galaxy–halo connection allows the model to reproduce the observed galaxy bias of faint UV-selected galaxies at $z \leq 9$, it over-predicts their number density over the same redshift range, as illustrated in \fig{fig:baseline_case2_UVLF}. In addition, the decreasing slope of the $L_{\mathrm{UV}}$–$M_h$ relation also results in a slight deficit of very bright galaxies ($\MUV < - 19)$ relative to the observations. Furthermore, to satisfy reionization constraints, the model requires these over-luminous galaxies to have a relatively low escaping ionizing efficiency, $\varepsilon_{\rm esc} \sim 3\%$, with no dependence on halo mass.

From the discussions so far, it is evident that the \textbf{baseline} model struggles to simultaneously reproduce the observed galaxy UVLF and bias (see also Gelli et al. (2024) \cite{Gelli2024}). The tension arises because the requirements imposed by each dataset push the model in conflicting directions — for example, reproducing the $z < 9$ galaxy UVLFs overestimates the bias of faint galaxies at these redshifts, whereas matching the observed bias produces an excess of galaxies towards the faint-end of the UVLF.

Extending this analysis, we also made an attempt to simultaneously fit all three observables --- galaxy UVLF, galaxy bias, and reionization history --- using the \textbf{baseline} model, as described in \app{appendix:baseline_all_observations}. We find that matching both the UV luminosity function and galaxy bias at $z < 10$ demands a highly contrived scenario in which star formation or UV production efficiency is extremely high in low-mass halos, while these halos ($M_h < 10^{11} M_\odot$) are also strongly suppressed by radiative feedback. Although such a model predicts an abundance and bias of faint galaxies in good agreement with lower-$z$ observations, it is unable to reproduce the $z>10$ UVLF measurements as successfully as the \textbf{UVLF+reion} case. Moreover, reionization proceeds extremely rapidly in these scenarios, ending much earlier ($z \approx 9$) than permitted by observational constraints. This further illustrates the difficulty of reconciling all three datasets within the \textbf{baseline} model.

Therefore, our next step is to explore extensions or modifications to the \textbf{baseline} model, aiming to reconcile both summary statistics (the galaxy UVLF and bias) and the reionization history within a consistent physical framework. We discuss this in the next section.

\section{Extensions to the Baseline Model: Inclusion of a Duty Cycle}
\label{sec:duty_cycle}

\subsection{Description of the duty cycle prescription}
\label{subsec:duty_cycle_description}

A potential solution to address the surplus of UV-emitting galaxies that arises while tuning our \textbf{baseline} model to match the clustering measurements is to introduce a duty cycle, which would result in only a fraction of the underlying galaxy population being ``observable'' at any given time. We assume that, at any redshift, only those halos that formed within a given preceding time interval $\Delta t$ will be able to host ``detectable'' UV-bright galaxies, thereby naturally yielding a duty cycle that depends on both redshift and halo mass. This is motivated by the fact that a recent increase in mass, particularly due to mergers or accretion, within a halo can correlate with episodes of intense star formation that boost the UV emission from the resident galaxy. In this work, we adopt the duty cycle parameterization of Trenti et al. (2010) \cite{Trenti2010}, wherein the fraction ($\epsilon_{\rm DC}$) of dark-matter halos of a particular mass that host detectable UV-bright galaxies at a given cosmic time is given by

\begin{equation}
\label{eq:dc_definition}
\epsilon_{\rm DC}(\Delta t, M_h,z) = \frac{\displaystyle \int_{M_h}^{+\infty} \bigg[\dfrac{dn}{dM_h}(M'_h,z) - \dfrac{dn}{dM_h}(M'_h,z_{\Delta t}) \bigg] \; dM'_h}{\displaystyle \displaystyle \int_{M_h}^{+\infty} \dfrac{dn}{dM_h}(M'_h,z) \; dM'_h}.
\end{equation}
where $\Delta t = t_H(z) - t_H(z_{\Delta t})$.

\begin{figure}
    \centering
    \includegraphics[width=\columnwidth]{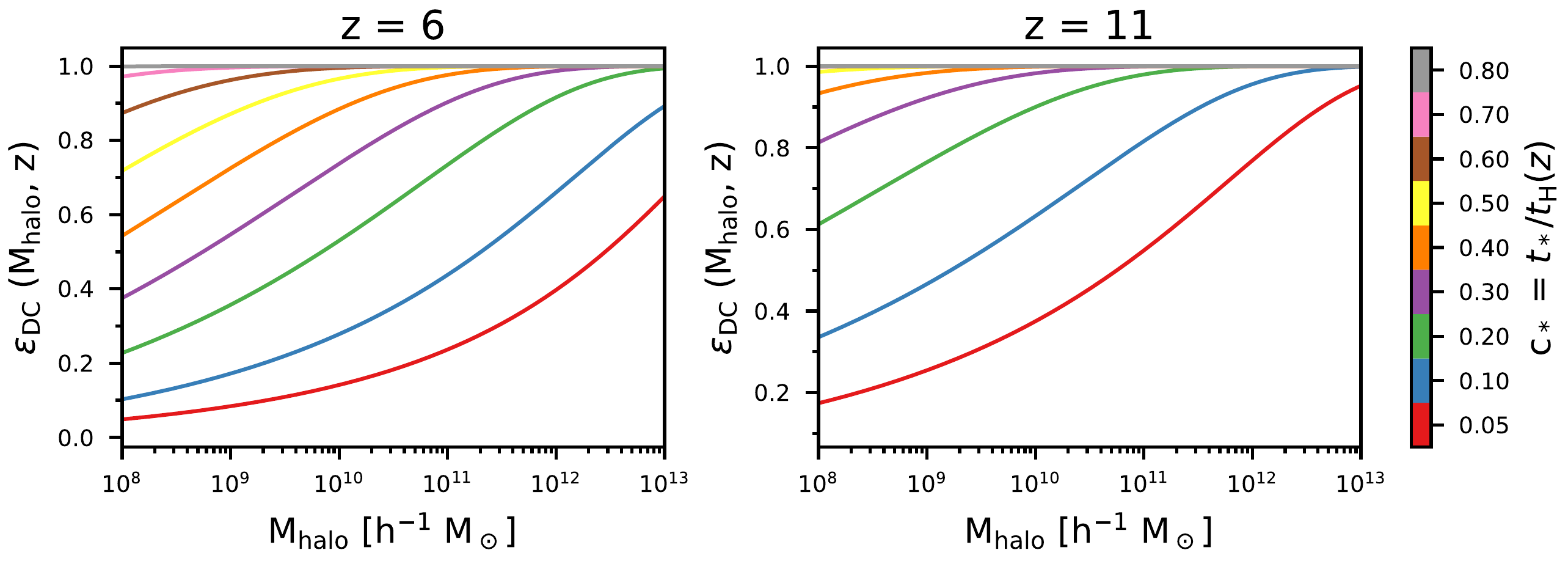}
    \caption{The effective duty cycle as a function of halo mass at two representative redshifts for different \textbf{constant} values of $c_\ast$. Note that to compare the duty cycle at these two redshifts for a fixed value of $t_*$ (rather than a fixed $c_\ast$), one would need to use the following mapping: $c_\ast(z=11) = \left(12/7\right)^{-3/2} \times c_\ast(z=6) \approx0.445 \times c_\ast(z=6)$.} 
    \label{fig:epsilonDC_plots}
\end{figure}

Incorporating an effective duty cycle into the model impacts the calculation of the UV luminosity function (\eqn{eqn:lumfunc_full}) and the ionizing emissivity (\eqns{eq:niondot_nofb}{eq:niondot_fb}) by modifying the occupancy of dark matter halos through the following transformation
\begin{equation}
    \dfrac{dn}{dM_h}(M'_h,z) \rightarrow \epsilon_{\rm DC}~(\Delta t, M'_h,z)~\dfrac{dn}{dM_h}(M'_h,z)
\label{eq:effect_of_DC}
\end{equation}

From \eqn{eq:dc_definition}, it is evident that the limiting case of $\epsilon_{\rm DC}~(M_h,z)$ = 1 is recovered when $\Delta t \rightarrow \infty$. We assume that this timescale, $\Delta t$, which is associated with the global evolution of the halo mass function, to be equal to the star formation timescale $t_\ast = c_\ast~ t_H(z)$. The resulting theoretical model, involving a non-unity duty cycle, will henceforth be referred to as the \textbf{extended} model.

In \fig{fig:epsilonDC_plots}, we show the variation of the effective duty cycle with halo mass and redshift for different constant values of $c_\ast$, which sets the star-formation timescale through $t_\ast = c_\ast ~t_H(z)$. For a fixed $c_\ast$, the duty cycle increases with halo mass, reflecting the rapid evolution of the high-mass end of the halo mass function, and with redshift, since halos of a given mass assemble more quickly at earlier times. Additionally, at a fixed redshift and fixed halo mass, the halo duty cycle also increases with an increase in the value of $c_\ast$, as the extended time interval allows a larger number of halos to form. Note that $c_\ast$, although assumed to be constant both in the above discussion and in our \textbf{baseline} model, may in principle vary as a function of halo mass at a given redshift. In such cases, the duty cycle would smoothly transition between the different evolutionary tracks shown for fixed values of $c_\ast$ in \fig{fig:epsilonDC_plots}.

In the \textbf{extended} model, the star-formation time scale $t_\ast$ (or equivalently, $c_\ast$) plays two important roles. It not only determines the duty cycle of halos, but also affects the UV luminosity of the galaxy occupying it, as given by \eqns{eq:LUV_nofb}{eq:LUV_fb}. For example, decreasing the value of $c_\ast$ at any redshift results in halos of a given mass housing increasingly brighter galaxies. However, a lower $c_\ast$ also leads to a corresponding decrease in their duty cycle, thereby reducing the number of these galaxies that will eventually be detectable (at that redshift).

Additionally, the adopted duty-cycle prescription helps break the degeneracy between some of the parameters governing the production efficiency of UV radiation, namely between $t_\ast$ and $f_{\ast}~\big(\mathcal {K}_{\rm UV,fid}/\mathcal {K}_{\rm UV}\big)$. These parameters had previously appeared in a degenerate multiplicative combination in the \textbf{baseline} model as the UV efficiency parameter - $\varepsilon_{{\rm *,UV}}$ (see \eqn{eq:epsilonstarUV}).

To allow sufficient flexibility in the model, we assume a power-law dependence of $c_\ast$ on both halo mass and redshift:
\begin{equation}
c_\ast(M_h, z) = c_{\ast,10}^{z=5} \left( \frac{1+z}{6} \right)^{\beta_{c_\ast}^z} 
\left( \frac{M_h}{10^{10} M_\odot} \right)^{\beta_{c_\ast}^M},
\label{eq:cstar_mass_redshift_definition}
\end{equation}
where $c_{\ast,10}^{z=5}$ is the normalization at a halo mass of $10^{10}\,M_\odot$ and redshift $z=5$, $\beta_{c_\ast}^z$ governs the redshift dependence, and $\beta_{c_\ast}^M$ determines the scaling with halo mass. As $c_\ast(M_h, z)$ represents a fraction, we impose the requirement that $c_\ast(M_h, z) \in [0.001, 1)$.

For the redshift evolution of the parameter $f_{\ast,10}\big(\mathcal {K}_{\rm UV,fid}/\mathcal {K}_{\rm UV}\big)$, we retain the tanh parameterization introduced earlier for $\varepsilon_{*10,\mathrm{UV}}$ but use different notations to avoid any confusion.
\begin{equation}
    \log_{10} \bigg[ f_{\ast,10}\bigg( \dfrac{\mathcal {K}_{\rm UV,fid}}{\mathcal {K}_{\rm UV}}\bigg) \bigg] = f_{\kappa,0} + \dfrac{f_{\kappa , \mathrm{jump}}}{2} \tanh\left(\dfrac{z-z_\ast}{\Delta z_\ast}\right)
\end{equation}

The prescription for $c_\ast$, provided in \eqn{eq:cstar_mass_redshift_definition}, impacts the UV efficiency of halos, $\varepsilon_{\rm *,UV}(M_h,z)$, defined in \eqn{eq:epsilonstarUV_reExpressed}, in two ways. Firstly, it introduces an additional redshift dependence into the normalization, $\varepsilon_{\rm *10,UV}$ (\eqn{eq:epsilonstarUV_normalisation}), through the redefinition
\begin{equation*}
    c_\ast ~\rightarrow~ \tilde{c}_\ast(z) = c_{\ast,10}^{z=5} \left( \dfrac{1+z}{6} \right)^{\beta_{c_\ast}^z},
\end{equation*}
which adds to the $\tanh$ redshift evolution already sourced by $f_{\ast,10}~(\mathcal{K}_{\rm UV,fid}/\mathcal{K}_{\rm UV})$. Secondly, the halo-mass dependence of $c_\ast$ changes the power-law slope of the $\varepsilon_{\rm *,UV}$–$M_h$ relation according to
\begin{equation*}
\label{eq:epsStarUV_Mh_slope}
    \alpha_\ast ~\rightarrow~ \alpha_\ast - \beta_{c_\ast}^M
\end{equation*}

These modifications to $\varepsilon_{\rm *,UV}(M_h,z)$ consequently propagate into the calculations of the UV luminosity (\eqns{eq:LUV_nofb}{eq:LUV_fb}) and the ionizing emissivity ( \eqns{eq:niondot_nofb}{eq:niondot_fb}) of galaxies within the \textbf{extended} model. In addition, we remind that the duty cycle, $\epsilon_{\mathrm{DC}}(t_\ast(M_h,z), M_h, z)$, which determines the fraction of halos hosting detectable galaxies (\eqn{eq:effect_of_DC}), independently regulates the computation of the UV luminosity function and the corresponding ionizing photon budget at a given redshift. 

In terms of free parameters, the \textbf{extended} model involves \textbf{\emph{twelve}} free parameters — three more in number ($c_{\ast,10}^{z=5}$, $\beta_{c_\ast}^z$ and $\beta_{c_\ast}^M$) compared to the \textbf{baseline} model. 

\subsection{Results from the extended model}
\label{subsec:results_extended}

In this subsection, we present the results obtained from comparing the theoretical predictions of the \textbf{extended} model with the \textbf{all} observational datasets mentioned in \secn{sec:data_and_methods}. We shall refer to this case as the \textbf{UVLF+bias+reion} case. 
While performing parameter inference for the \textbf{extended} model, we note that the disproportionately large number of data points in the UVLF dataset compared to the galaxy bias dataset causes a standard MCMC analysis using the joint likelihood defined in \eqn{eqn:joint_likelihood_expression} to prioritize reproducing the UVLF measurements and place less emphasis on matching the bias measurements. To circumvent this issue, we assign additional weight ($w$) to the galaxy bias dataset in the likelihood calculation. In this case, the joint likelihood takes the following form
\begin{equation}
\label{eqn:joint_likelihood_expression_with_weight}
\begin{aligned}
\mathcal{L}(\mathcal{D} \vert \boldsymbol{\theta}) 
&= \prod_\alpha \mathcal {L}(\mathcal {D}_\alpha \vert \boldsymbol \theta) \\
&= \exp{\Bigg(-\frac{1}{2} \bigg[ \chi^2(\mathcal{D}_{\Phi_{\rm UV}}, \boldsymbol{\theta}) + w^2~\chi^2(\mathcal{D}_{b_{\rm gal}}, \boldsymbol{\theta}) + \chi^2(\mathcal{D}_{Q_{\rm HI}}, \boldsymbol{\theta}) + \chi^2(\mathcal{D}_{\tau_{\rm el}}, \boldsymbol{\theta})  \bigg]\Bigg)}
\end{aligned}
\end{equation}

We present results for the case where the weight factor is set to $w = 4$, a choice that seems adequate to reasonably match observations of both the galaxy summary statistics, considering the relative proportion of their respective data points (i.e., $N^{\rm UVLF}_{\rm data} / N^{\rm bias}_{\rm data} = 75/8 = 9.4$). We further examine how different choices of $w$ influence the agreement between our model-predicted summary statistics and observational data in \app{appendix:weight_factor_effect}. 
\begin{figure}[htbp]
\centering
\includegraphics[width=\columnwidth]{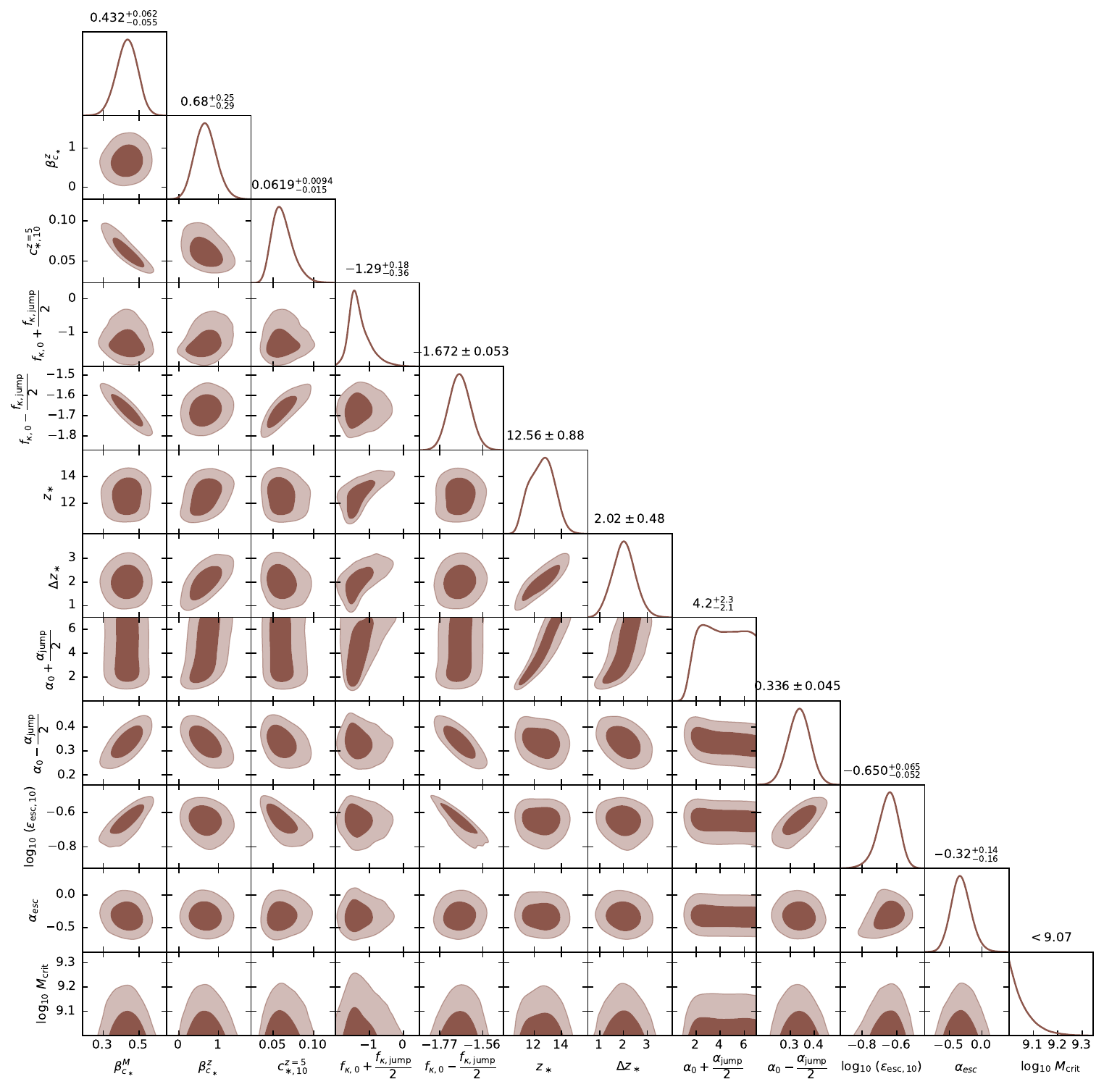}
\caption{Posterior distributions of the free parameters for the \textbf{UVLF+bias+reion} case, based on the \textbf{extended} model. The diagonal panels show the one-dimensional posterior distribution, while the contour plots in the off-diagonal panels represent the two-dimensional joint distribution. The contour levels represent 68\% and 95\% confidence levels. The mean and 68\% confidence intervals are denoted above the one-dimensional posterior distributions of the respective parameters.}
\label{fig:allObs_corner}
\end{figure}

\begin{figure}[htbp]
\centering
\includegraphics[width=\columnwidth]{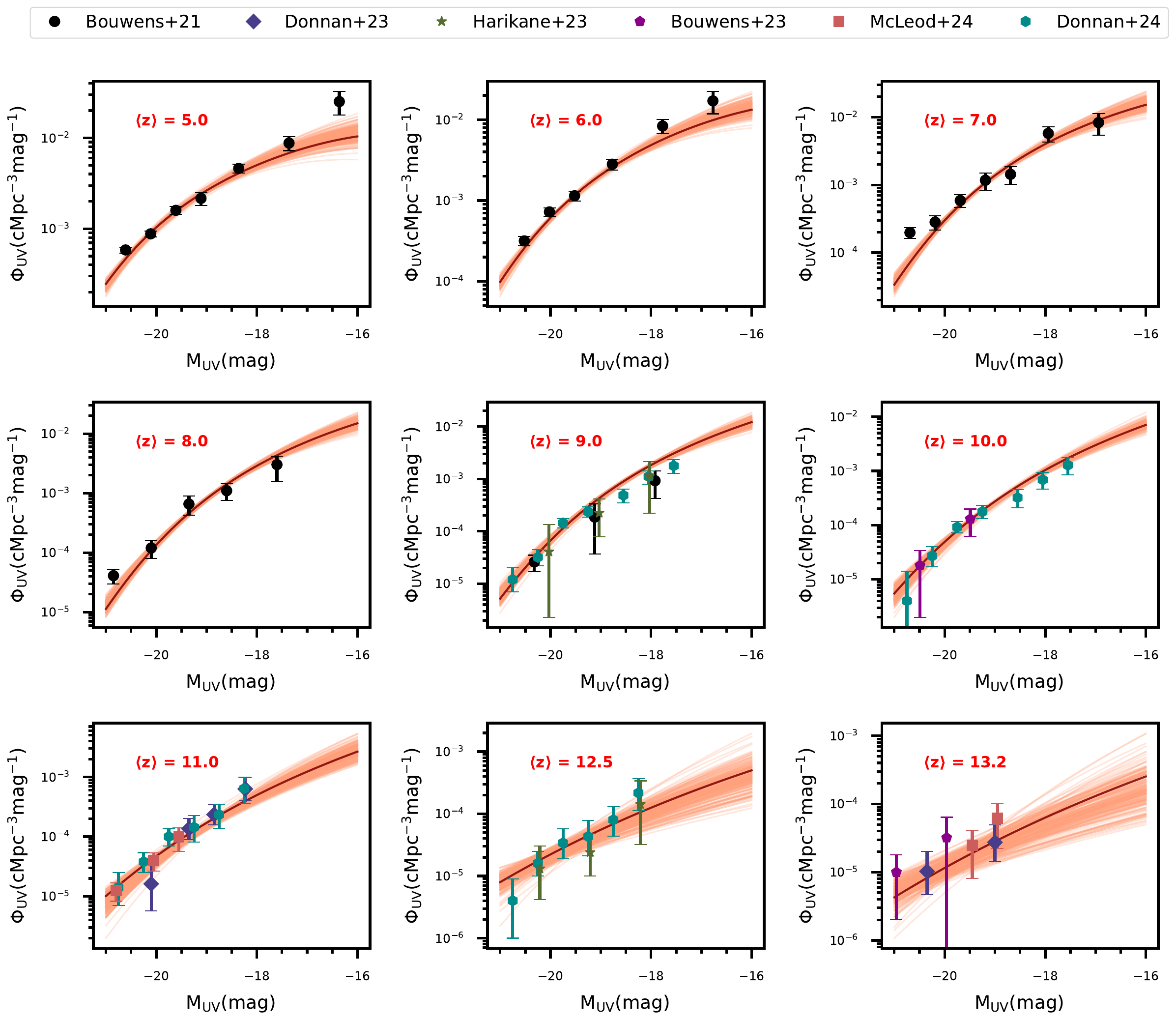}
\caption{Same as \fig{fig:baseline_case1_UVLF} but for 200 random samples drawn from the MCMC chains of the \textbf{UVLF+bias+reion} case, based on the \textbf{extended} model. }
\label{fig:extended_model_UVLF}
\end{figure}

\begin{figure}[htbp]
\centering
\begin{subfigure}[t]{\columnwidth}
    \centering
    \includegraphics[width=0.6\textwidth]{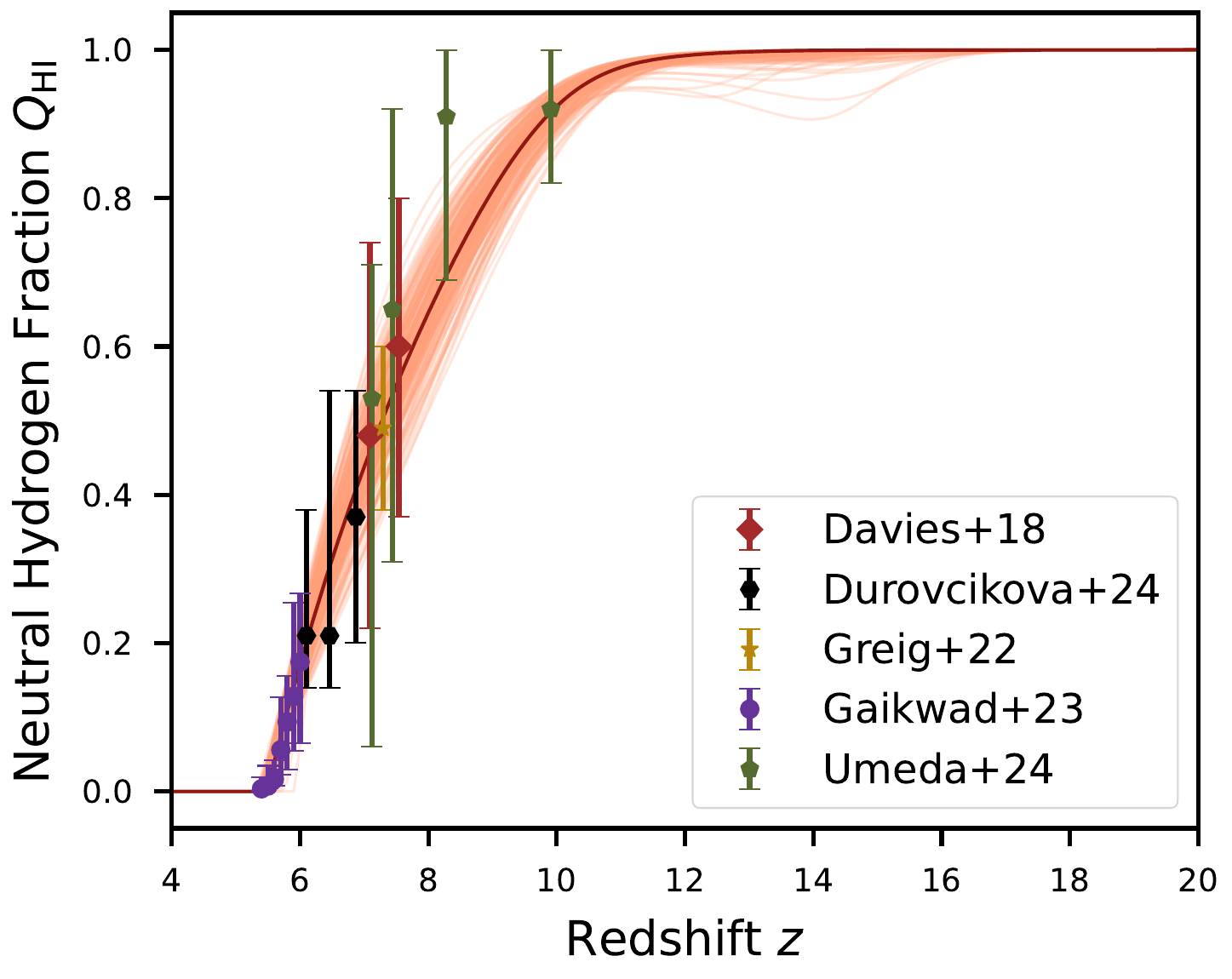}
    \caption{}
    \label{fig:extended_reion_history}
\end{subfigure}

\vspace{0.5cm} 

\begin{subfigure}[t]{\columnwidth}
    \centering
    \includegraphics[width=0.48\columnwidth]{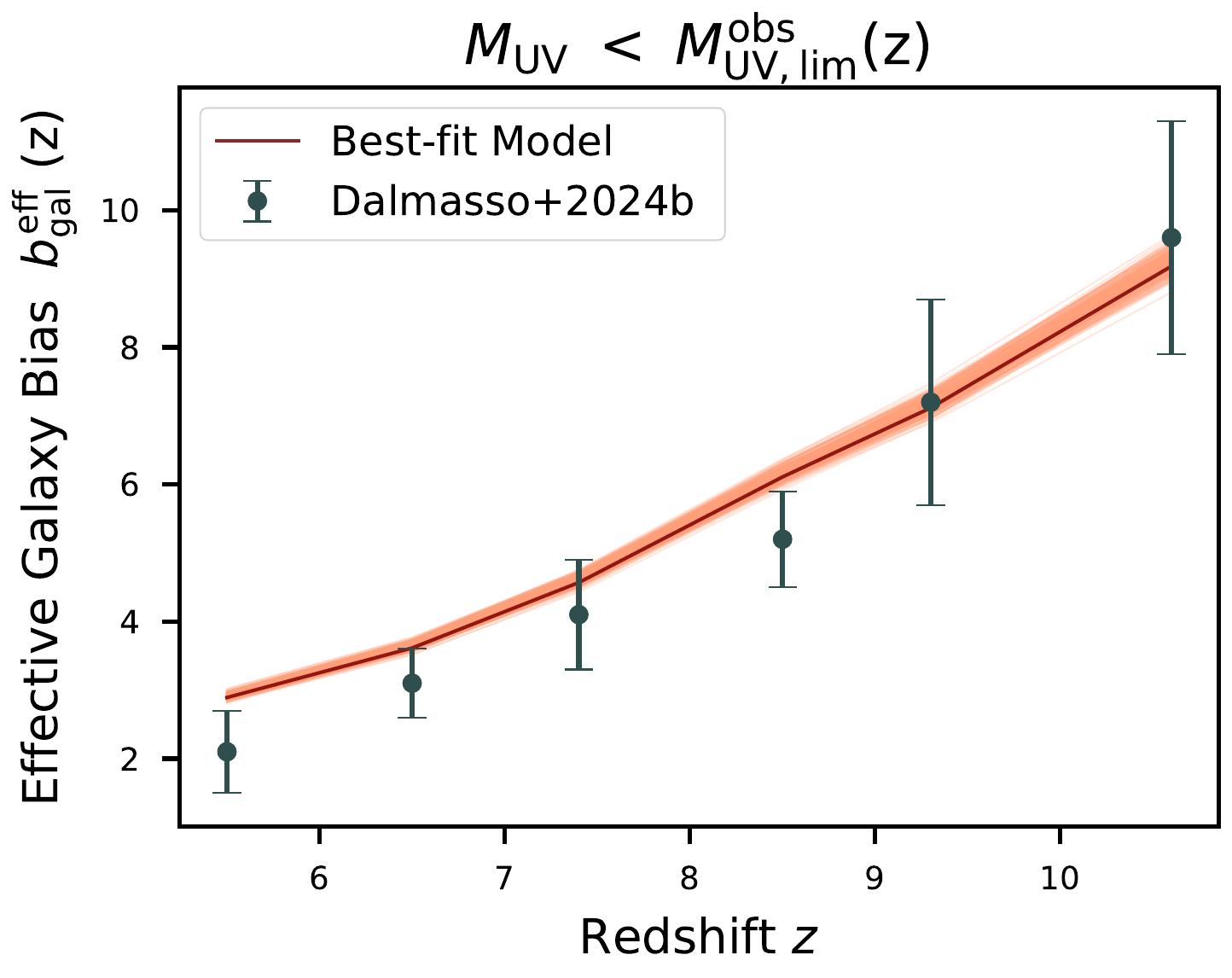}
    \hfill
    \includegraphics[width=0.48\columnwidth]{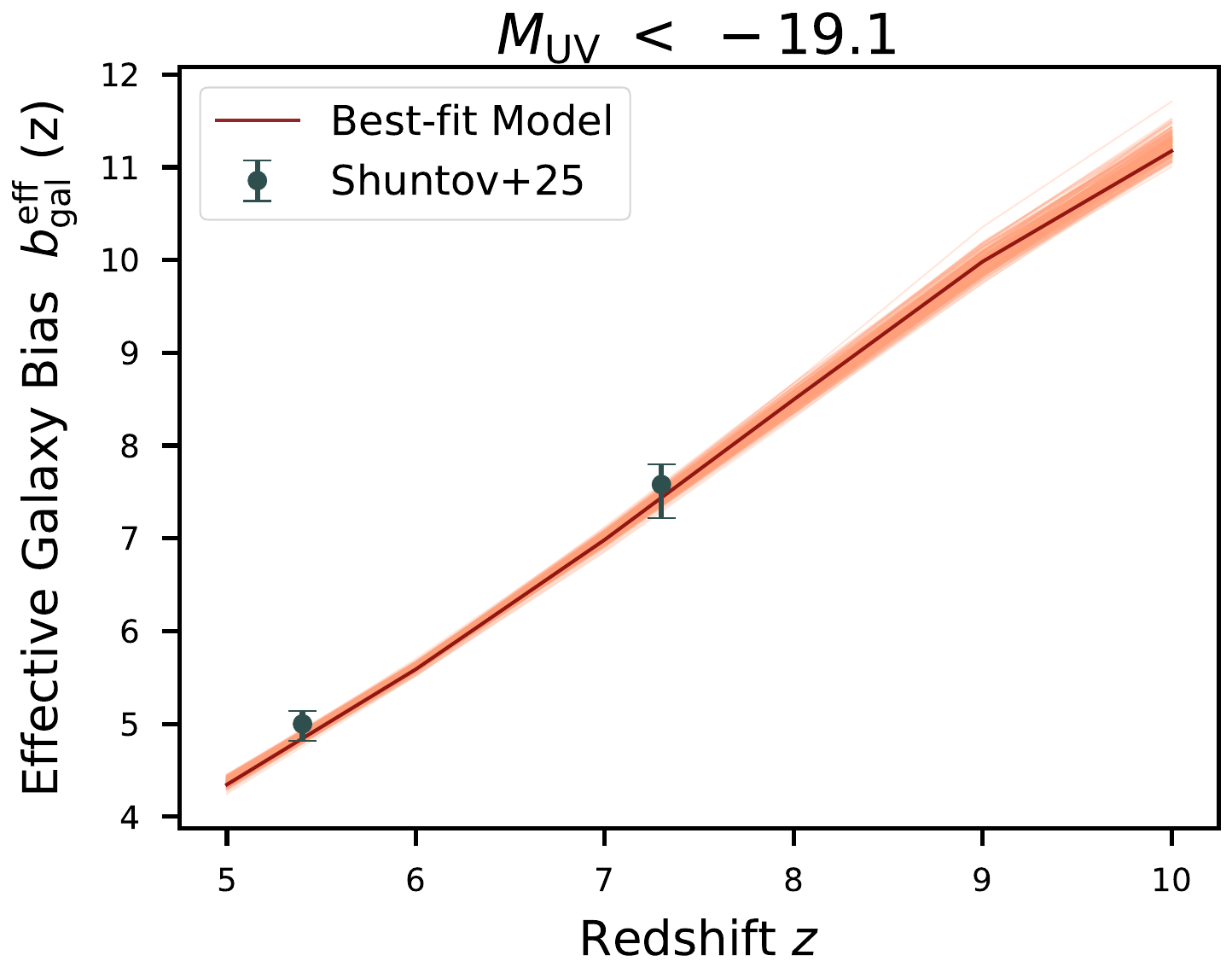}
    \caption{}
    \label{fig:extended_gal_bias}
\end{subfigure}
\caption{Same as \fig{fig:baseline_case1_reion_history_and_gal_bias} but for 200 random samples drawn from the MCMC chains of the \textbf{UVLF+bias+reion} case, based on the \textbf{extended} model.}
\label{fig:extended_reion_history_and gal_bias}
\end{figure}

\begin{figure}[htbp]
\centering
\includegraphics[width=\columnwidth]{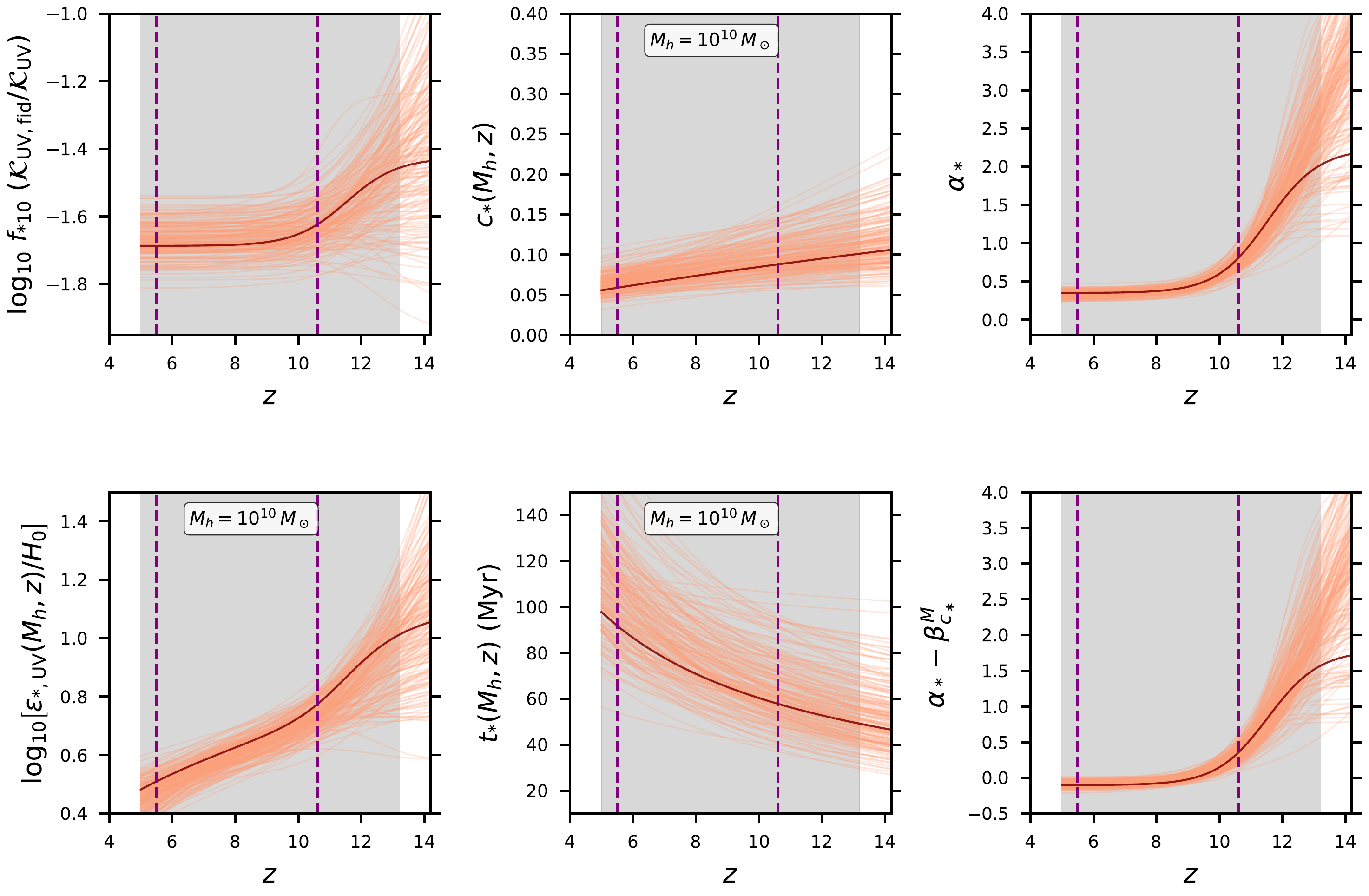}
\caption{ \textbf{Top Row:} The redshift evolution of key astrophysical parameters for 200 random samples drawn from the MCMC chains of the \textbf{UVLF+bias+reion} case, based on the \textbf{extended} model. These parameters include the star-formation efficiency times $\kappaUV^{-1}$  (left panel) of $10^{10} M_\odot$ halos, the star-formation timescale (central panel) of $10^{10} M_\odot$ halos in units of the local Hubble time $t_H(z)$ and the power-law dependence of the star-formation efficiency on halo mass (right panel). In each panel, the grey shaded region and the vertical purple dashed lines indicate the redshift ranges used to compare the model with UVLF observations ($5 \leq \langle z \rangle \leq 13.2$) and galaxy bias measurements ($5.4 \leq \langle z \rangle \leq 10.6$), respectively. \\ \textbf{Bottom Row:} The evolution of other ``derived" parameters in the \textbf{extended} model such as the UV efficiency $\varepsilon_{*,\mathrm{UV}}$ of $10^{10} M_\odot$ halos (expressed in appropriate dimensionless units), the star-formation timescale $t_\ast = c_\ast(M_h,z)~t_H(z)$ of $10^{10} M_\odot$ halos, and the \textit{effective} power-law index $\alpha_\ast - \beta^M_{c_\ast}$, which characterizes the dependence of the UV production efficiency $\varepsilon_{*,\mathrm{UV}}$ on halo mass.}
\label{fig:extended_sfe_params}
\end{figure}

\subsubsection{Constraints on the star formation activity and the duty cycle of high-redshift galaxies}

The posterior distributions of the free parameters of the \textbf{extended} model are shown in \fig{fig:allObs_corner}, along with their mean values and 68\% confidence intervals mentioned above the respective one-dimensional posterior distributions. We show the model-predicted UVLFs, reionization histories, and galaxy bias for 200 random samples drawn from the MCMC chains of \textbf{UVLF+bias+reion} case for the \textbf{extended} model in \figs{fig:extended_model_UVLF}{fig:extended_reion_history_and gal_bias}, along with the presently available observational measurements. The evolution of the different redshift-dependent free parameters of the model is shown in the top row of \fig{fig:extended_sfe_params}.

We observe that the model predictions now exhibit markedly improved agreement with all three key observables — UVLFs, reionization history, and galaxy bias — demonstrating the improved capabilities of the \textbf{extended} model relative to the \textbf{baseline} model. A comparison of the best-fit samples for all the galaxy–halo connection prescriptions explored in this work is further shown in \figs{fig:comparison_UVLF}{fig:comparison_reion_history_and_gal_bias} of \app{appendix:comparison_of_all_models}.

\begin{figure}[htbp]
\centering
\includegraphics[width=0.8\columnwidth]{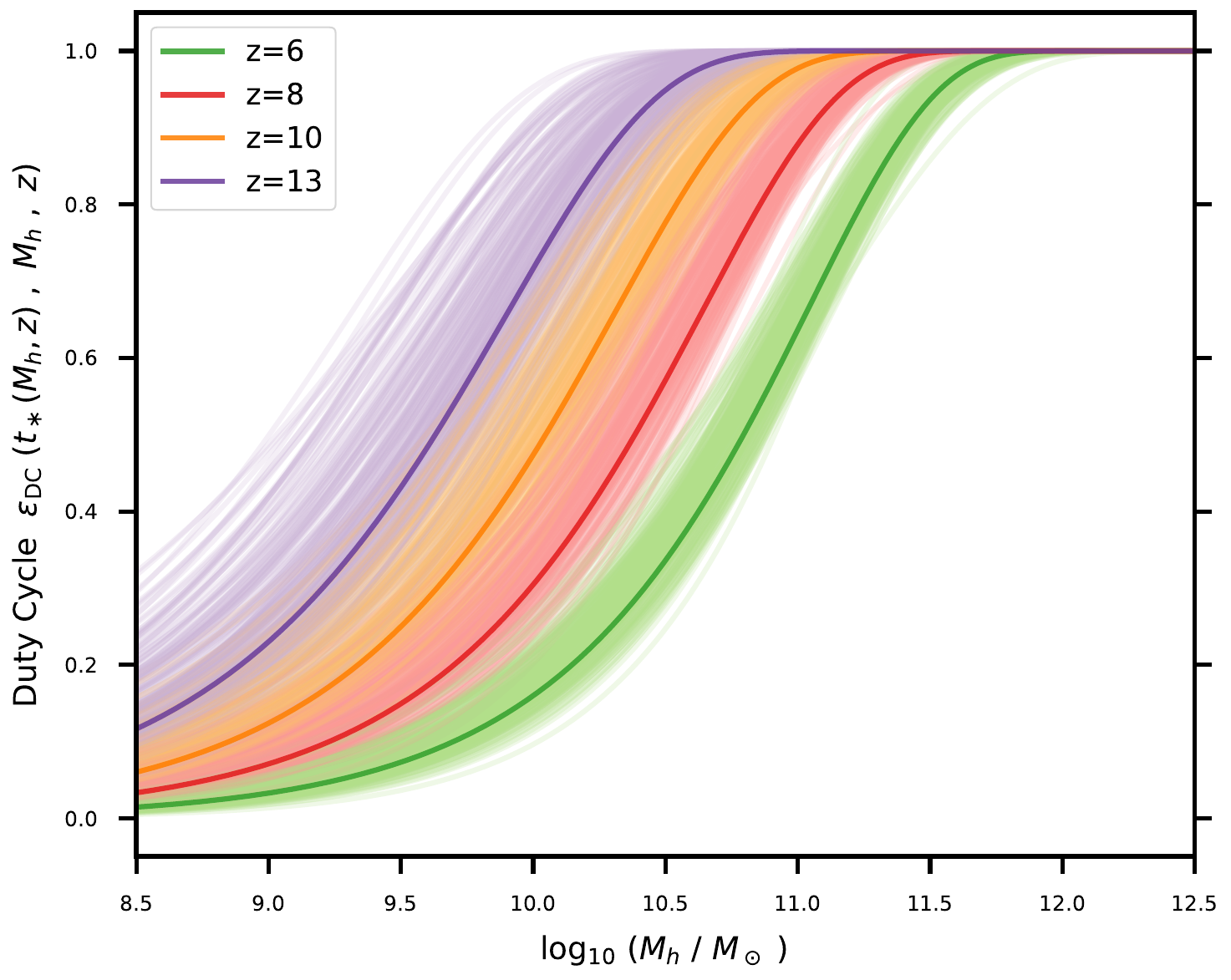}
\caption{The variation in the effective duty cycle with halo mass at four different redshifts for 200 random samples drawn from the MCMC chains of the \textbf{UVLF+bias+reion} case based on the \textbf{extended} model.} 
\label{fig:duty_cycle_for_samples}
\end{figure}

From \fig{fig:allObs_corner}, we see that reproducing all observations with this model requires the star-formation timescale  $t_\ast$ in halos hosting UV-bright galaxies to evolve with redshift, approximately as $t_\ast \propto (1+z)^{\beta^z_{c_\ast} - 3/2}$, which evaluates to
$t_\ast \propto (1+z)^{-0.82^{+0.25}_{-0.29}} ~\mathrm{for}~ \beta^z_{c_\ast} = 0.68^{+0.25}_{-0.29}$, aligning well with the expectation that gas cooling becomes increasingly efficient at high redshift, when halo gas densities are higher \cite{Somerville2025}. The star-formation timescale of $10^{10}~M_\odot$ halos, shown in the bottom-central panel of \fig{fig:extended_sfe_params}, rises from a value of $\approx$ 45 Myr at $z \approx 14$ to around 85 Myr at $z \approx 6$ for the best-fit model. From the evolution of $c_\ast$ shown in the top-central panel of \fig{fig:extended_sfe_params}, we notice that the characteristic timescale $t_\ast$ for the majority of halos --- except for the most massive ones --- remains less than or comparable to the halo dynamical timescale\footnote{The halo dynamical time $t_{\rm dyn} = \sqrt{3\pi/(16G\bar{\rho}_{\rm vir})}$ can be written as $t_{\rm dyn}(z) = (\pi/\sqrt{2\Delta_c})\,t_H(z)$, assuming $\bar{\rho}_{\rm vir}(z) = \Delta_c \rho_c(z)$.} ($t_{\rm dyn} \sim 0.15 - 0.20\,t_H$), across the entire redshift range probed by observations.

At fixed redshift, we also find that the timescale over which gas gets converted into stars increases with halo mass, scaling as $t_\ast \propto M_h^{0.43}$. This trend is consistent with the expectation that although massive halos contain larger gas reservoirs, much of this gas is not readily available in the cold phase and is therefore converted into stars more slowly.

Because the star-formation timescale directly sets the duty cycle in our model, this scaling implies that lower-mass halos have systematically smaller duty cycles. The variation of the effective duty cycle across halo mass at four different redshifts is shown for 200 random samples from the \textbf{extended} model in \fig{fig:duty_cycle_for_samples}. As previously mentioned, while discussing \fig{fig:epsilonDC_plots}, this evolution of the duty cycle as a function of halo mass and redshift in our case closely mirrors the differential growth and assembly of halo populations in the early Universe (e.g., \cite{Trenti2010, Weinberger2019, Maitra2025}). Furthermore, in this duty cycle framework, low-mass halos characterized by shorter star-formation timescales form stars more intermittently, thereby naturally introducing stochasticity among high-redshift galaxies. Although it is distinct from theoretical frameworks that implement stochasticity via a scatter ($\sigma_{\rm UV}$) in the UV luminosity at fixed halo mass \cite{Shen2023, Munoz2023, Gelli2024}, this model nonetheless shares some qualitative similarities with those frameworks, particularly in how its impact varies with halo mass and in the redshift evolution required to reproduce high-$z$ UVLF observations.

Together, these findings suggest that low-mass galaxies undergo short, burst-like episodes of star formation. This behavior is unsurprising, since their shallow gravitational potential wells make them particularly sensitive to gas inflows and outflows, as well as to internal feedback processes, both of which can introduce strong fluctuations in their star-formation activity \cite{Gelli2020, Furlanetto2022, Hopkins2023, Sun2023, Gelli2024}. On the other hand, more massive halos are able to sustain star formation over longer timescales. Moreover, as one moves toward higher redshifts, corresponding to the early stages of galaxy formation, the star-formation activity of galaxies at fixed luminosity becomes increasingly episodic in our model, aligning with inferences drawn from JWST observations of high-$z$ galaxies \cite{Endsley2023, Ciesla2024, Looser2025}.

In scenarios where star formation is spread out over shorter timescales, halos can host exceptionally UV-luminous galaxies, even with moderate star-formation efficiency. Consequently, at a fixed UV luminosity, galaxies will occupy comparatively less massive halos and therefore have lower bias. However, only a very small fraction of these abundant low-mass halos would have undergone recent star formation and thus, be capable of hosting UV-detectable galaxies. The \textbf{extended} model leverages this interplay to reconcile its predictions with both the galaxy UVLF and bias measurements at $z \leq 10$.  

Interestingly, despite star-formation episodes becoming progressively shorter at higher redshifts (bottom-central panel of \fig{fig:extended_sfe_params}), we notice that an increase in the efficiency of star formation or production of UV light per unit star formation is still necessary at $z>10$ for halos with $M_h \geq 10^{10}\,M_\odot$, as indicated by the rapid steepening of the slope, $\alpha_\ast$, and the rise in the normalization, $f_{\ast,10}\big(\mathcal {K}_{\rm UV,fid}/\mathcal {K}_{\rm UV}\big)$, shown in the top-right and top-left panels of \fig{fig:extended_sfe_params}. As a result, the UV efficiency parameter evolves in tandem, increasing towards higher redshift (see bottom-right and bottom-left panels of \fig{fig:extended_sfe_params}).

\begin{figure}[htbp]
\centering
\includegraphics[width=0.8\columnwidth]{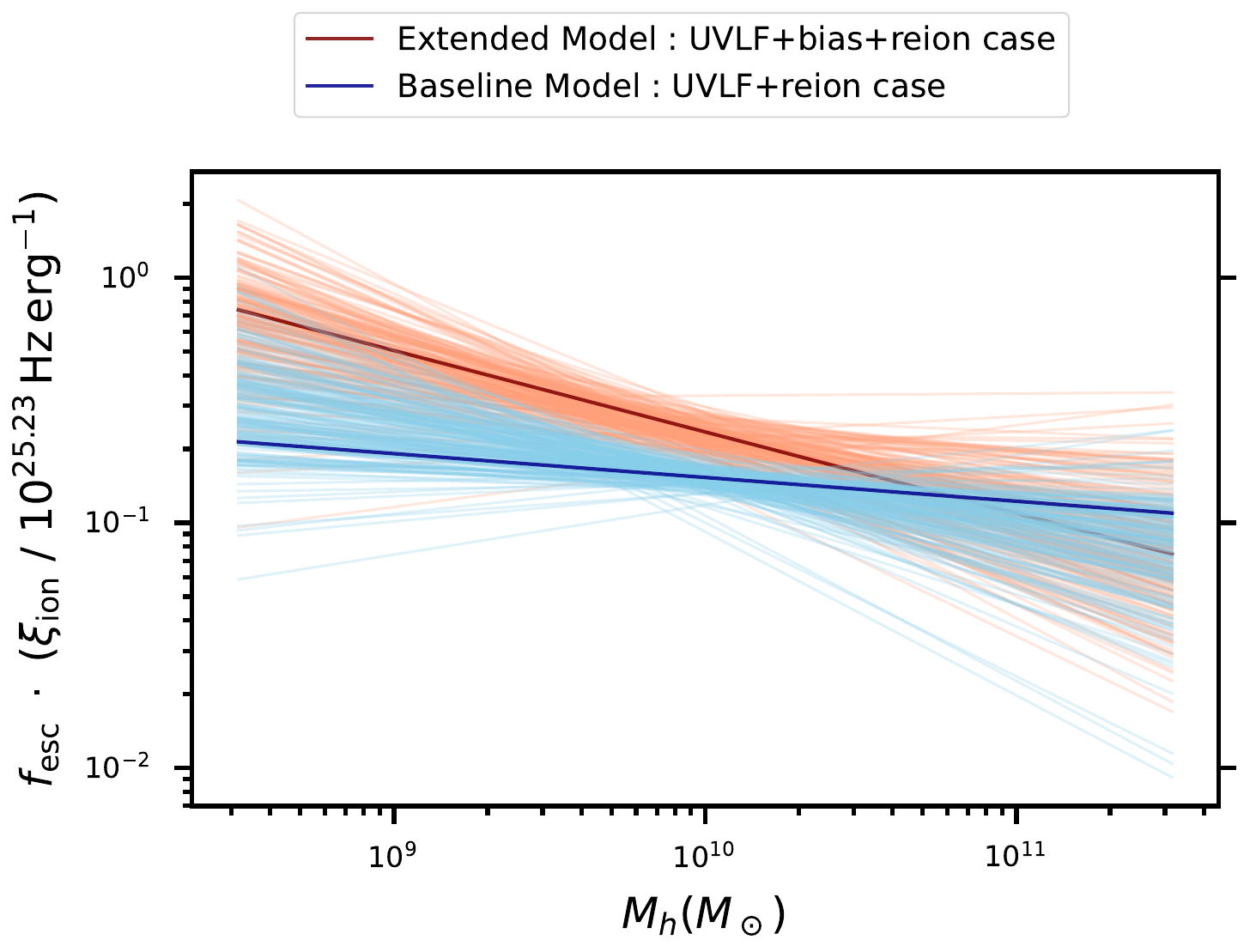}
\caption{The variation in the escaping ionizing efficiency, defined as $\varepsilon_{\rm esc}(M_h) = f_\mathrm{esc}(M_h)\cdot\left(\xi_{\rm ion}~/~\xi_{\rm ion,fid}\right)$, with halo mass for 200 random samples drawn from the MCMC chains of the \textbf{UVLF+bias+reion} case based on the \textbf{extended} model and \textbf{UVLF+reion} case based on the \textbf{baseline} model. The solid lines denote the best-fit model for the respective cases.}
\label{fig:compare_escape_fraction}
\end{figure}

\subsubsection{Constraints on the ionizing properties of high-redshift galaxies}

We now present the constraints obtained on the ionizing properties of high-redshift galaxies from our analysis, and discuss their broader implications.

As the mass-dependent duty cycle by itself already significantly reduces the occupancy of low-mass halos (see \fig{fig:duty_cycle_for_samples}), the critical mass $\Mcrit$ associated with radiative feedback remains unconstrained (within the assumed prior interval) for the \textbf{extended} model, favouring values smaller than $10^{9.1}~M_\odot$.  We further find that an escaping ionizing efficiency decreasing with halo mass yields reionization histories that are in agreement with current constraints. As shown in \fig{fig:compare_escape_fraction}, the slope of $\varepsilon_{\rm esc}-M_h$ relation in the \textbf{UVLF+bias+reion} case of the \textbf{extended} model is relatively steeper than in the \textbf{UVLF+reion} case of the \textbf{baseline} model. This follows from the fact that, in the \textbf{extended} model, low-mass halos with $M_h \lesssim  10^{10} M_\odot$ — despite their comparatively high UV efficiency —  have a low duty cycle, implying that most of these halos are, on average, not actively forming stars. To make up for this, the model requires them to be prolific producers and/or leakers of ionizing photons to satisfy the ionizing photon budget required to match the reionization observations at a given redshift.

\begin{figure}[htbp]
\centering
\includegraphics[width=0.8\columnwidth]{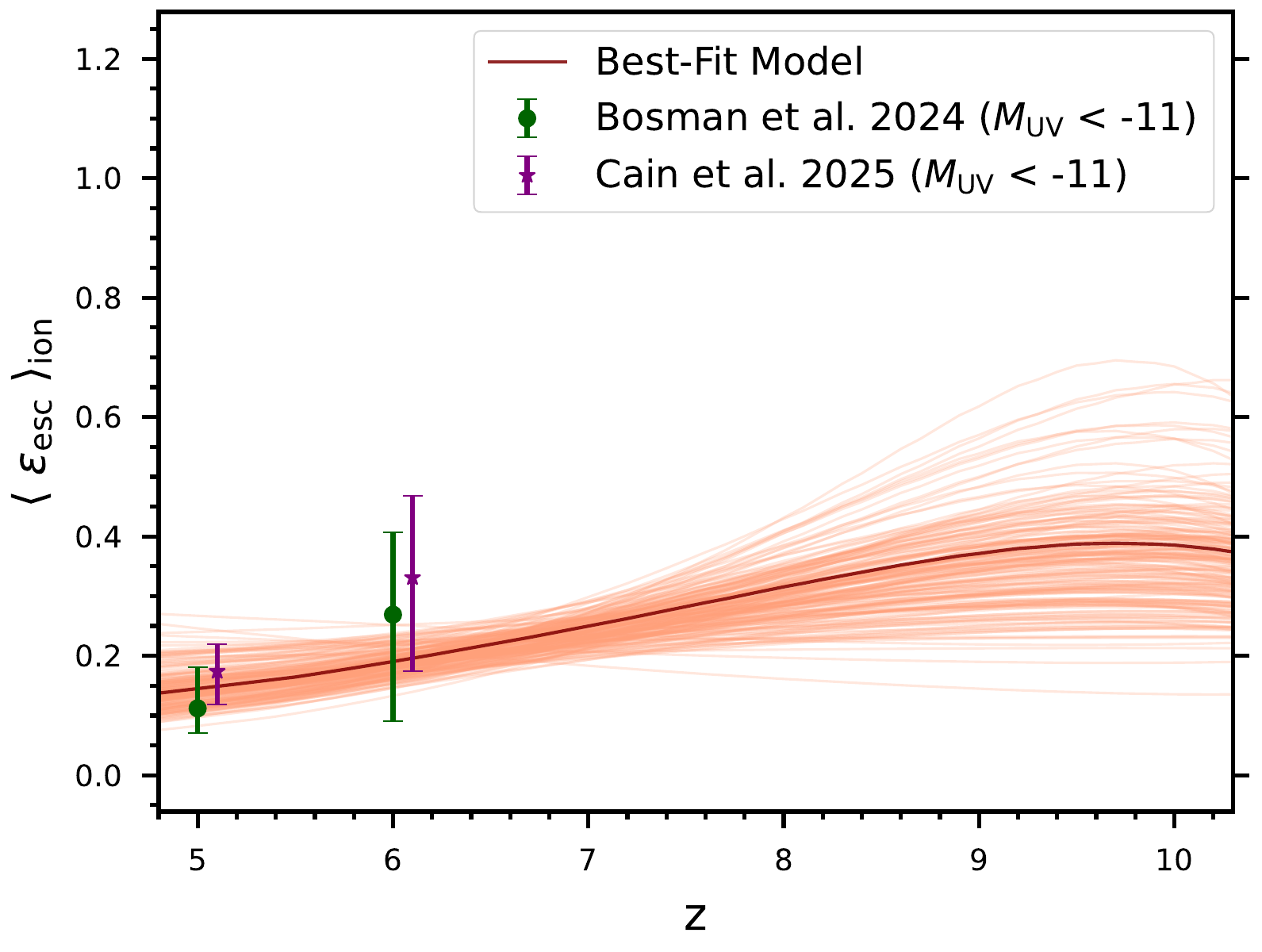}
\caption{The redshift evolution in the population-averaged escaping ionizing efficiency, defined in \eqn{eq:population_averaged_epsilon_escape}, for 200 random samples drawn from the MCMC chains of the \textbf{UVLF+bias+reion} case based on the \textbf{extended} model. The solid line denotes the best-fit model, while colored data points indicate recent measurements at $z = 5$ and $z=6$ inferred using observations of Lyman-$\alpha$ absorption in high-redshift quasar spectra \cite{Bosman2024, Cain2025}. Note that, for clarity, the data point from Cain et al. (2025) \cite{Cain2025} has been slightly shifted along the redshift axis (by $\Delta z = + 0.1$) while plotting to avoid overlap.}
\label{fig:pop_avg_epsilon_escape}
\end{figure}

Therefore, we next examine the redshift evolution of the population-averaged escaping ionizing efficiency, which plays a crucial role in determining the overall progress of cosmic reionization. The average escaping efficiency $\langle \varepsilon_{\rm esc} \rangle_{\rm ion}$, weighted by the intrinsic ionizing photon output, is calculated as (e.g., \cite{Munoz2024}):
\begin{equation}
    \langle \varepsilon_{\rm esc} \rangle_{\rm ion}  
        = \dfrac{\dot{n}_{\rm ion}(z)}
        {\dot{n}^{\rm intrinsic}_{\rm ion}(z)} 
        = \dfrac{\dot{n}_{\rm ion}(z)}
        {\dot{n}_{\rm ion}(z)\big|_{\varepsilon_{\rm esc}(M_h)=1}}
\label{eq:population_averaged_epsilon_escape}
\end{equation}
where $\dot{n}_{\rm ion}(z)$ is the \textit{total} ionizing photon production rate density, defined in \eqn{eq:niondot_total} in terms of \eqns{eq:niondot_nofb}{eq:niondot_fb}, and further modulated by the duty cycle as described in \eqn{eq:effect_of_DC}.

\fig{fig:pop_avg_epsilon_escape} shows the variation of $\langle \varepsilon_{\rm esc} \rangle_{\rm ion}$ with redshift predicted by the \textbf{extended model}. We find that it increases gradually towards high redshifts, from $\approx 15\%$ at $z \sim 5$ and reaching $\approx 40\%$ at $z \sim 9$ in the best-fit model. The predicted values of $\langle \varepsilon_{\rm esc} \rangle_{\rm ion}$ also agree well with the recent measurements based on quasar absorption spectra \cite{Bosman2024, Cain2025}. If the ionizing efficiency $\xi_{\rm ion}$ is assumed to remain constant (i.e., non-evolving) with time across the \textit{entire} galaxy population at our fiducial value, then the evolution of  $\langle \varepsilon_{\rm esc} \rangle_{\rm ion}$ shown in \fig{fig:pop_avg_epsilon_escape} implies that the population-averaged ionizing escape fraction $\langle f_{\rm esc}\rangle$ increases with redshift, in agreement with the qualitative trends inferred in several previous studies \cite{Faisst2016, FaucherGiguere2020, Trebitsch2022, Mitra2023, Ferrara2025}. However, it should be noted that the redshift evolution of $f_{\rm esc}$ remains highly debated, as other works in the literature have reported or predicted alternative behaviors at $z>5$, including a decreasing \cite{Naidu2020} or non-evolving \cite{Khaire2019} escape fraction.

\subsection{Comparison of the extended model's predictions with other observations}

Having constrained the galaxy-halo connection in the \textbf{extended} model to reproduce the observed abundance and bias of high-redshift galaxies, as well as the global reionization history, we now proceed to examine the consistency of the model with other available observations in this subsection.

\begin{figure}[htbp]
\centering
\includegraphics[width=\columnwidth]{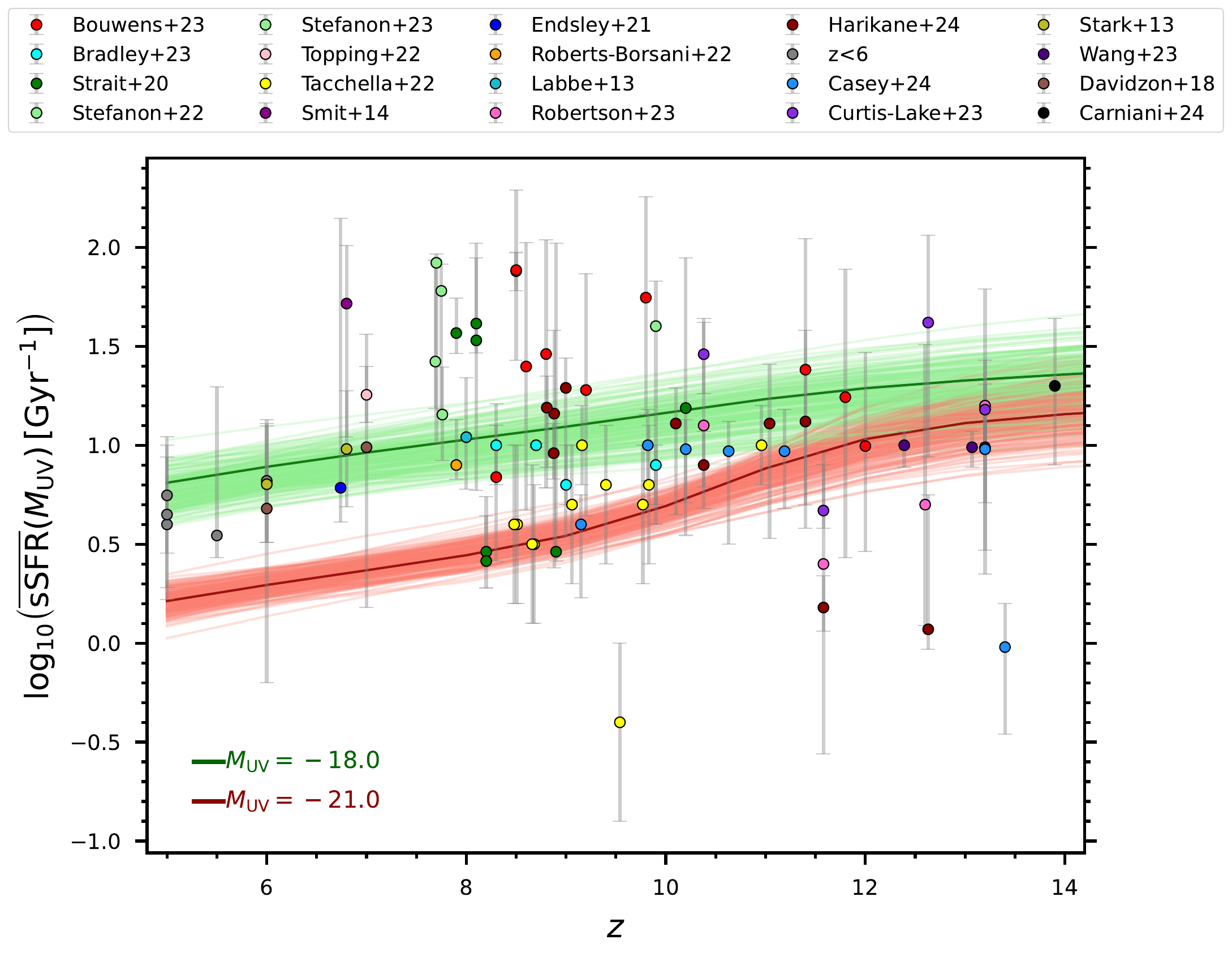}
\caption{The evolution of the specific star formation rate (sSFR = $\dot{M_\ast}/ M_\ast$) at fixed absolute UV magnitudes for 200 random samples drawn from the MCMC chains of the \textbf{UVLF+bias+reion} case, based on the \textbf{extended} model. The colored data points represent various observational estimates of sSFR from the literature \cite{Bouwens2023_XDF_HUDF, Bradley2023, Strait2020, Stefanon2022, Stefanon2023, Topping2022, Tacchella2022, Smit2014, Endsley2021, Roberts-Borsani2022, Labbe2013, Robertson2023, Harikane2024, Santini2017, Salmon2015, Davidzon2018, CurtisLake2023, Stark2013, Wang2023, Casey2024, Carniani2024}.}
\label{fig:extended_model_sSFR}
\end{figure}

Foremost, we investigate the evolution of the specific star formation rate (sSFR) --- defined as the ratio of the star formation rate to stellar mass --- with redshift as implied by this model. In our case, the sSFR depends on both halo mass and redshift, and is given by sSFR($M_h,z$) = $1/t_\ast(M_h,z)$, implying the scaling  --- $\mathrm{sSFR} \propto M_h^{-\beta^M_{c_*}} (1+z)^{3/2 - \beta^z_{c_*}}$. In \fig{fig:extended_model_sSFR}, we compare the model predictions of the `mean' sSFR for galaxies \footnote{This is calculated as $\overline{\rm sSFR}(\MUV, z) = Q_\mathrm{HII}(z)~\mathrm{sSFR}\left(M^{\rm fb}_{h},z\right)$ + $Q_\mathrm{HI}(z)~\mathrm{sSFR}\left( M^{\rm nofb}_{h} \, ,z\right)$, where $M^{\rm fb}_{h}$ and $M^{\rm nofb}_{h}$ denote the masses of the host halo of a galaxy with magnitude $\MUV$. } with fixed absolute UV magnitudes ($\MUV=-18$ and $\MUV=-21$) to observational estimates of sSFR for individual galaxies from the literature \cite{Bouwens2023_XDF_HUDF, Bradley2023, Strait2020, Stefanon2022, Stefanon2023, Topping2022, Tacchella2022, Smit2014, Endsley2021, Roberts-Borsani2022, Labbe2013, Robertson2023, Harikane2024, Santini2017, Salmon2015, Davidzon2018, CurtisLake2023, Stark2013, Wang2023, Casey2024, Carniani2024}. We find that the sSFR predicted by the model increases monotonically with redshift, indicating that galaxies of a given stellar mass are generally more bursty at early times. At $z<10$, the sSFR is lower for brighter galaxies hosted in massive halos, which form stars more steadily, whereas at $z>10$, the evolving slope of the $L_{\rm UV}$–$M_h$ relation shifts these galaxies toward slightly lower-mass halos, resulting in progressively higher sSFRs.
It is worth noting, however, that the sSFR values reported in the literature are usually derived using a variety of tracers that probe star formation over different timescales and thus, the comparison between the observational estimates and our model predictions presented in \fig{fig:extended_model_sSFR} should be regarded as qualitative.

Therefore, the \textbf{extended} model presents a plausible mechanism --- invoking a time-evolving $\epsilon_{\rm \ast UV}(M_h,z)$ parameter along with a redshift- and mass-dependent duty cycle, $\epsilon_{\mathrm{DC}}(t_\ast(M_h,z), M_h, z)$ --- to reconcile the currently available galaxy UVLF and galaxy bias data at $z \gtrsim 5$ while also yielding an sSFR qualitatively consistent with the observations. 

Although our analysis constrains the evolution of the star formation timescale by comparing the predictions of the \textbf{extended} model with observed galaxy UVLFs and clustering, these observations do not individually constrain the parameters, $f_{\ast,10}$ and $\mathcal{K}_{\rm UV,fid}/\mathcal{K}_{\rm UV}$, but only their product. In this regard, other galaxy summary statistics, such as the galaxy stellar mass function (GSMF) that offer a more direct measure of the star-formation efficiency compared to the UVLFs, can help in further breaking the degeneracy between $f_{\ast,10}$ and $\mathcal{K}_{\rm UV,fid}/\mathcal{K}_{\rm UV}$. However, it is important to realize that constructing the GSMF from observations is fraught with several uncertainties and systematic issues associated with estimating stellar masses from UV light. While we refrained from including the observational estimates of the GSMF in our likelihood calculations for these reasons, we now explore the insights that can be obtained by comparing the observed GSMF with that derived from our theoretical model, which already successfully matches other galaxy summary statistics such as the UVLFs and large-scale bias.

In our model, the UV luminosity of a galaxy and its stellar mass are related as follows
\begin{equation}
L_{{\rm UV}}(M_h,z) = \dfrac{1}{\mathcal {K}_{{\rm UV}}} \dfrac{M_*(M_h,z)}{t_{*}(M_h, z)}
\end{equation}
Therefore, since we have already obtained constraints on $c_{*}(M_h,z)$ and $f_{\ast,10}\big(\mathcal {K}_{\rm UV,fid}/\mathcal {K}_{\rm UV}\big)$ by comparing the model predictions to the galaxy UVLF and bias observations, the only input needed for converting the galaxy UV luminosity function into a galaxy stellar mass function is $\mathcal{K}_{\rm UV}$ (more specifically, the ratio $\mathcal{K}_{\rm UV,fid}/\mathcal{K}_{\rm UV}$). As mentioned earlier, this parameter $\mathcal{K}_{\rm UV}$, which depends on the properties of the stellar population and the recent star-formation history, is related to the mass-to-light ratio of galaxies in our model.

\begin{figure}[htbp]
\centering
\includegraphics[width=\columnwidth]{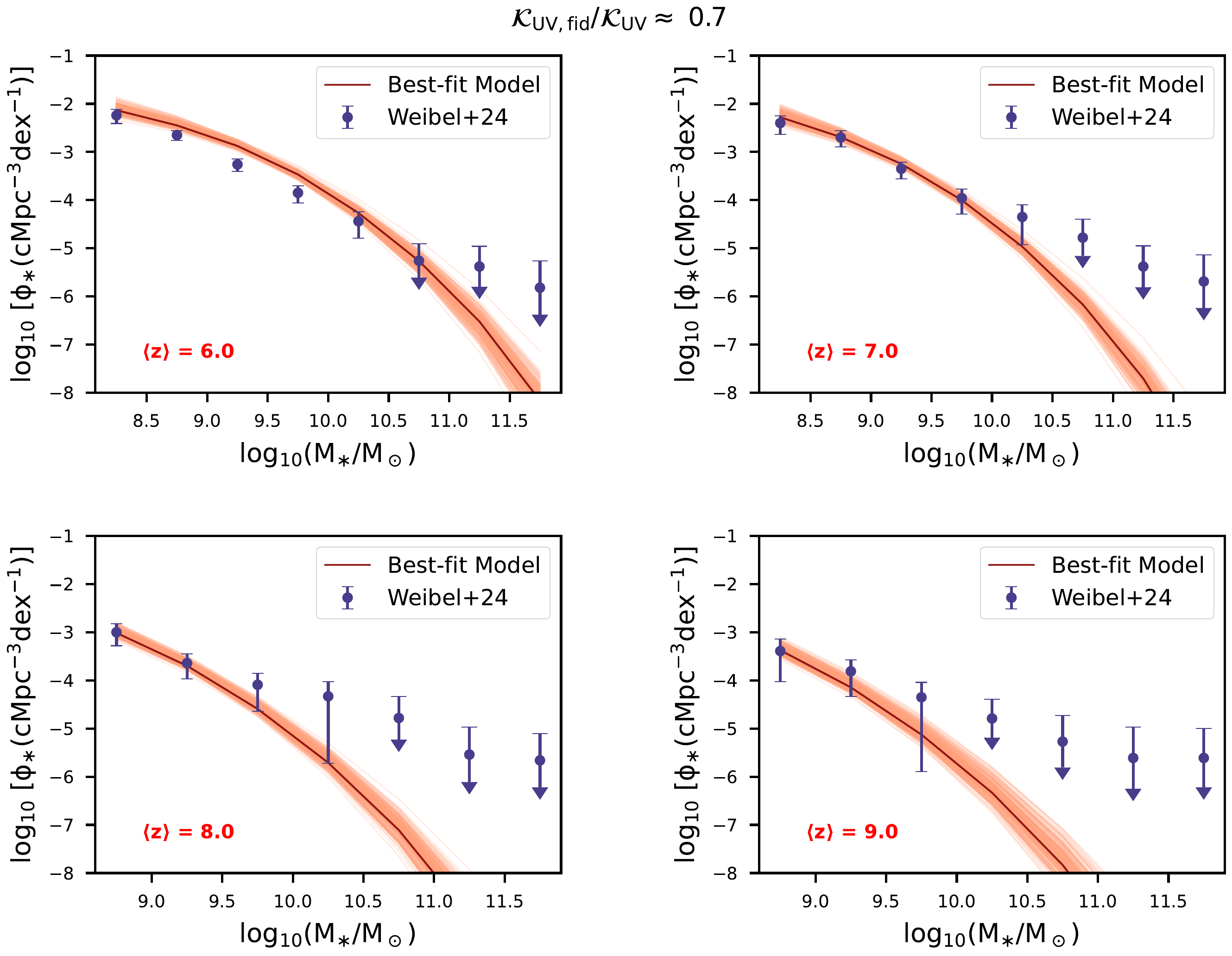}
\caption{The \textbf{\textit{derived}} galaxy stellar mass function (with $\mathcal{K}_{\rm UV,fid}/\mathcal{K}_{\rm UV} \approx 0.7$) corresponding to the best-fit model and 200 random samples drawn from the MCMC chains of the \textbf{UVLF+bias+reion} case with the \textbf{extended} model shown in \fig{fig:extended_model_UVLF}.}
\label{fig:extended_model_SMF}
\end{figure}

For the models shown in \fig{fig:extended_model_UVLF}, we simply vary the ratio $\mathcal{K}_{\rm UV,fid}/\mathcal{K}_{\rm UV}$ until the derived galaxy stellar mass function shows agreement with the observational estimates of the GSMF obtained from JWST by Weibel et al. (2024) \cite{Weibel2024}. We find that setting  $\mathcal{K}_{\rm UV,fid}/\mathcal{K}_{\rm UV} \approx 0.7$ gives a reasonably good visual match to the data, as shown in \fig{fig:extended_model_SMF}. In this regard, it is worth mentioning that this ratio of $\mathcal{K}_{\rm UV,fid}/\mathcal{K}_{\rm UV}$ obtained by us is roughly about $\sim$ 0.5 times the value commonly used when rescaling stellar mass estimates, owing to a different initial mass function adopted in the observational study (Salpeter in our model vs. Kroupa in Weibel et al. (2024) \cite{Weibel2024}), but perhaps remains plausible given additional differences in the choice of the star formation history (continuous vs. delayed-$\tau$) and stellar population synthesis models ({\tt STARBURST99} vs. {\tt BPASS-v2.2.1}) between our study and that assumed in Weibel et al. (2024) \cite{Weibel2024} while deriving the stellar masses from UV observations. From this simple proof-of-principle exercise with the \textbf{extended} model, we find that a scenario in which the star formation efficiency increases with halo mass ($f_\ast \propto M_h^{0.33}$, reaching $\sim 3\%$ for $10^{10}~M_\odot$ halos), while the duration of star formation becomes increasingly shorter towards smaller halos (thereby reducing their duty cycle) provides a reasonable match to observations of all three key galaxy summary statistics — the galaxy stellar mass function, the galaxy UV luminosity function, and the large-scale galaxy bias — \textit{simultaneously} at $z < 10$.

\section{Conclusion}
\label{sec:conclusion}

The unprecedented wealth of observational data from space-based telescopes such as HST and JWST has revolutionized our understanding of galaxy formation and evolution during the first billion years of cosmic history. These observations are also crucial in characterizing the role of early star-forming galaxies in reionizing the Universe, providing stringent tests for theoretical models of high-redshift galaxy populations.

In this study, we leverage the theoretical framework of Chakraborty \& Choudhury (2024) (\citetalias{Chakraborty2024}), which self-consistently links galaxy evolution and reionization, to extract key insights into the astrophysical properties of early galaxies during the Epoch of Reionization. Our main findings can be summarized as follows:

\begin{itemize}

\item Using an updated version of the \citetalias{Chakraborty2024} model with nine free parameters, governing star formation efficiency, radiative feedback–induced UV suppression, and ionizing photon escape fractions in high redshift galaxies, we compare its theoretical predictions against the latest JWST and HST measurements of the UV luminosity function (UVLF) at $z \sim 5 - 15$, as well as constraints from reionization, including the optical depth ($\tau_{\rm el}$) and the globally averaged neutral hydrogen fraction. Our results suggest that reconciling JWST’s UVLF measurements at $z \geq 10$ requires either more efficient star formation or higher UV radiation production per unit stellar mass. Additionally, our analysis supports scenarios in which faint, low-mass galaxies with higher escaping ionizing efficiencies ($\varepsilon_{\rm esc} \equiv f_{\rm esc} \times \xi_{\rm ion}$) are the dominant drivers of cosmic reionization, aligning well with current constraints on the ionization history of the intergalactic medium (IGM).

\item However, while our baseline model successfully connects UV luminosity to halo mass using UVLFs, it fails to reproduce the evolution of large-scale galaxy bias observed by JWST over $5 < z < 11$ for the fainter population of galaxies. Bias measurements at $5 < z < 9$ suggest that faint galaxies at fixed luminosity reside in lower-mass halos than predicted, leading to an overestimation of galaxy abundance in standard models.

\item To address this discrepancy, we introduce a redshift- and mass-dependent duty cycle, assuming that only galaxies in recently assembled halos, i.e., those formed within the characteristic star-formation timescale at each epoch, remain bright in the UV. This naturally leads to a declining duty cycle at $z < 9$, while causing galaxies to appear brighter at fixed halo mass and reconciling the model with both UVLF and bias observations. 

\item Furthermore, adopting a mass-dependent escaping ionizing efficiency, $\varepsilon_{\rm esc}(M_h)$, that decreases with increasing halo mass, this extended model remains consistent with current reionization constraints. In our case, we find that the population-averaged, ionizing-luminosity-weighted escape fraction (assuming a constant fiducial value of $\xi_{\rm ion} = 10^{25.23}~\mathrm{erg^{-1}~Hz}$ across all galaxies) increases with redshift, rising from $\approx 15\%$ at $z \sim 5$ to $\approx 40\%$ at $z \sim 10$.

\end{itemize}

These results highlight the necessity of incorporating higher-order summary statistics, such as large-scale galaxy bias, alongside conventional one-point statistics like the UV luminosity function to refine theoretical models of the galaxy-halo connection at high redshifts. Our study demonstrates that a simple luminosity-based mapping to halo mass is insufficient and that the interplay between star formation timescales, halo assembly, and feedback must be carefully modeled to fully capture the evolution of early galaxies.

Future improvements to our model could further enhance its predictive power. For instance, a more detailed treatment of galaxy star-formation histories would better capture key aspects of galaxy evolution -- such as the stochastic nature of star formation and the evolving stellar populations -- that influence galaxy properties as well as the resulting galaxy-to-galaxy scatter at high redshifts. Additionally, incorporating spatial fluctuations in the ionized hydrogen field would allow direct comparisons with Lyman-$\alpha$ opacity variations at $z \sim 5-6$ and upcoming 21 cm fluctuation experiments, requiring a transition to a semi-numerical framework \cite{Choudhury2025}. Lastly, improved constraints on galaxy clustering at high redshifts would provide further insights into the nature of early galaxy formation, enabling tighter constraints on the astrophysical parameters governing reionization.


\section*{Acknowledgments}

The authors acknowledge support from the Department of Atomic Energy, Government of India, under project no. 12-R\&D-TFR-5.02-0700. The authors thank Andrea Ferrara for useful discussions and the anonymous referee for detailed comments that improved the presentation of the paper.

\section*{Data Availability}

The data generated and presented in this paper will be made available upon reasonable request to the corresponding author.

\newpage

\appendix


\section{Fitting the Galaxy UVLF, Clustering, and Reionization Observations Simultaneously Using the Baseline Model}
\label{appendix:baseline_all_observations}

In this appendix, we present the results obtained from simultaneously fitting the galaxy UVLF, clustering, and reionization observations using the \textbf{baseline} model. We refer to this analysis as the \textbf{UVLF+bias+reion} case. Since the UVLF dataset contains far more measurements than the galaxy bias dataset, it drives the inference of free parameters in a standard MCMC analysis using the joint likelihood defined in \eqn{eqn:joint_likelihood_expression}. To ensure a balanced contribution from all datasets, we explicitly down-weight the contribution of the UVLF data points in the joint likelihood calculation, as follows : 
\begin{equation}
\label{eqn:joint_likelihood_expression_with_weight_copy}
\begin{aligned}
\mathcal{L}(\mathcal{D} \vert \boldsymbol{\theta}) 
= \exp{\Bigg(-\frac{1}{2} \bigg[ \chi^2(\mathcal{D}_{\Phi_{\rm UV}}, \boldsymbol{\theta}) + w^2~\chi^2(\mathcal{D}_{b_{\rm gal}}, \boldsymbol{\theta}) + \chi^2(\mathcal{D}_{Q_{\rm HI}}, \boldsymbol{\theta}) + \chi^2(\mathcal{D}_{\tau_{\rm el}}, \boldsymbol{\theta})  \bigg]\Bigg)},
\end{aligned}
\end{equation}
using a weight factor of $w$=4.

\begin{figure}[htbp]
\centering
\includegraphics[width=\columnwidth]{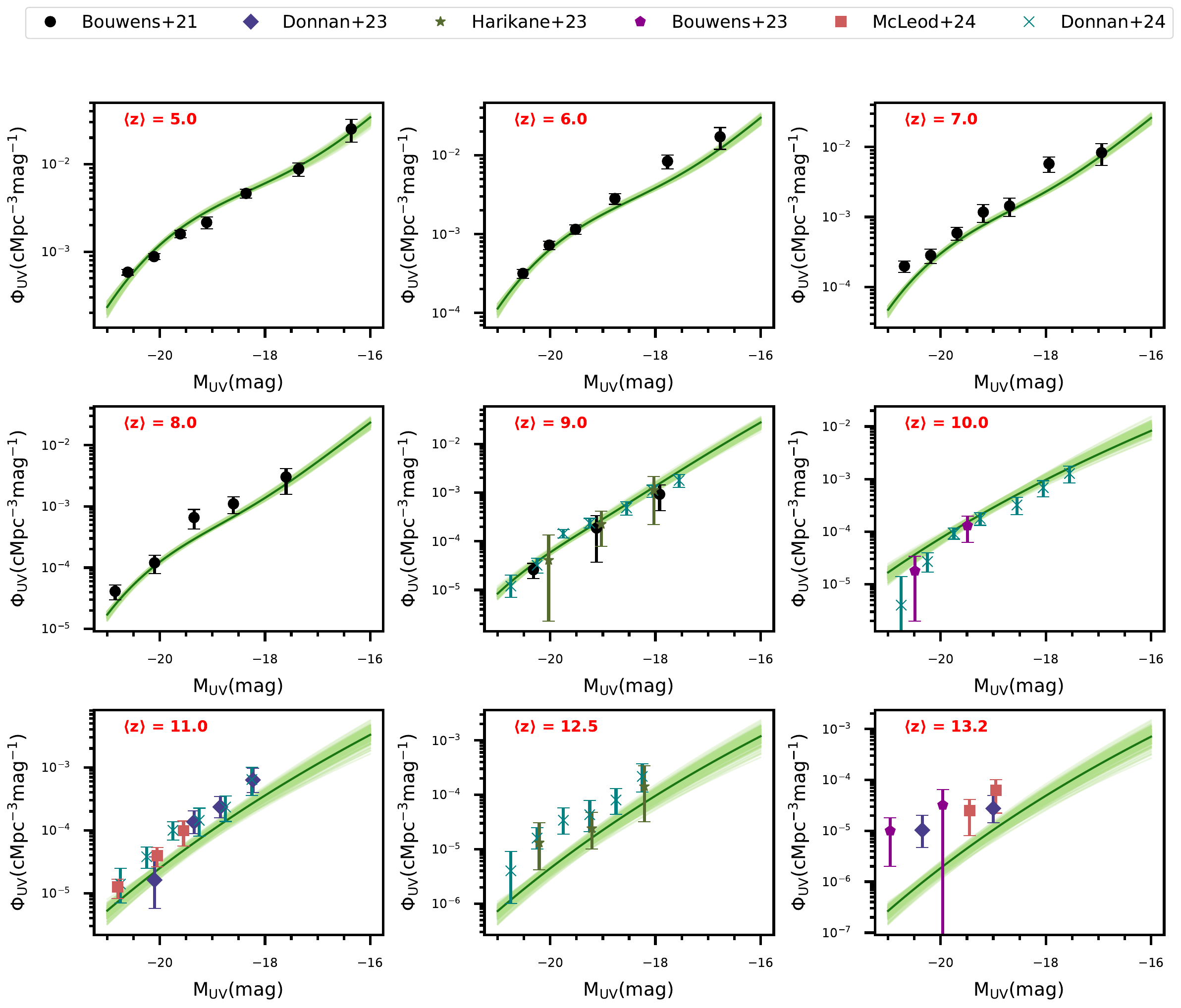}
\caption{Same as \fig{fig:baseline_case1_UVLF} but for 200 random samples drawn from the MCMC chains of the \textbf{UVLF+bias+reion} case, based on the \textbf{baseline} model.}
\label{fig:baseline_case3_UVLF}
\end{figure}

The constraints obtained on the various model parameters for this case are summarized in \tab{tab:appendix_mcmc_results}. We compare the model predictions for the UVLFs, reionization history, and galaxy bias with existing observational measurements in \figs{fig:baseline_case3_UVLF} {fig:baseline_case3_reion_history_and_gal_bias}, showing results for 200 random samples drawn from the MCMC chains of the \textbf{UVLF+bias+reion} case based on the \textbf{baseline} model.  

\begin{figure}[htbp]
\centering
\begin{subfigure}[t]{\columnwidth}
    \centering
    \includegraphics[width=0.6\textwidth]{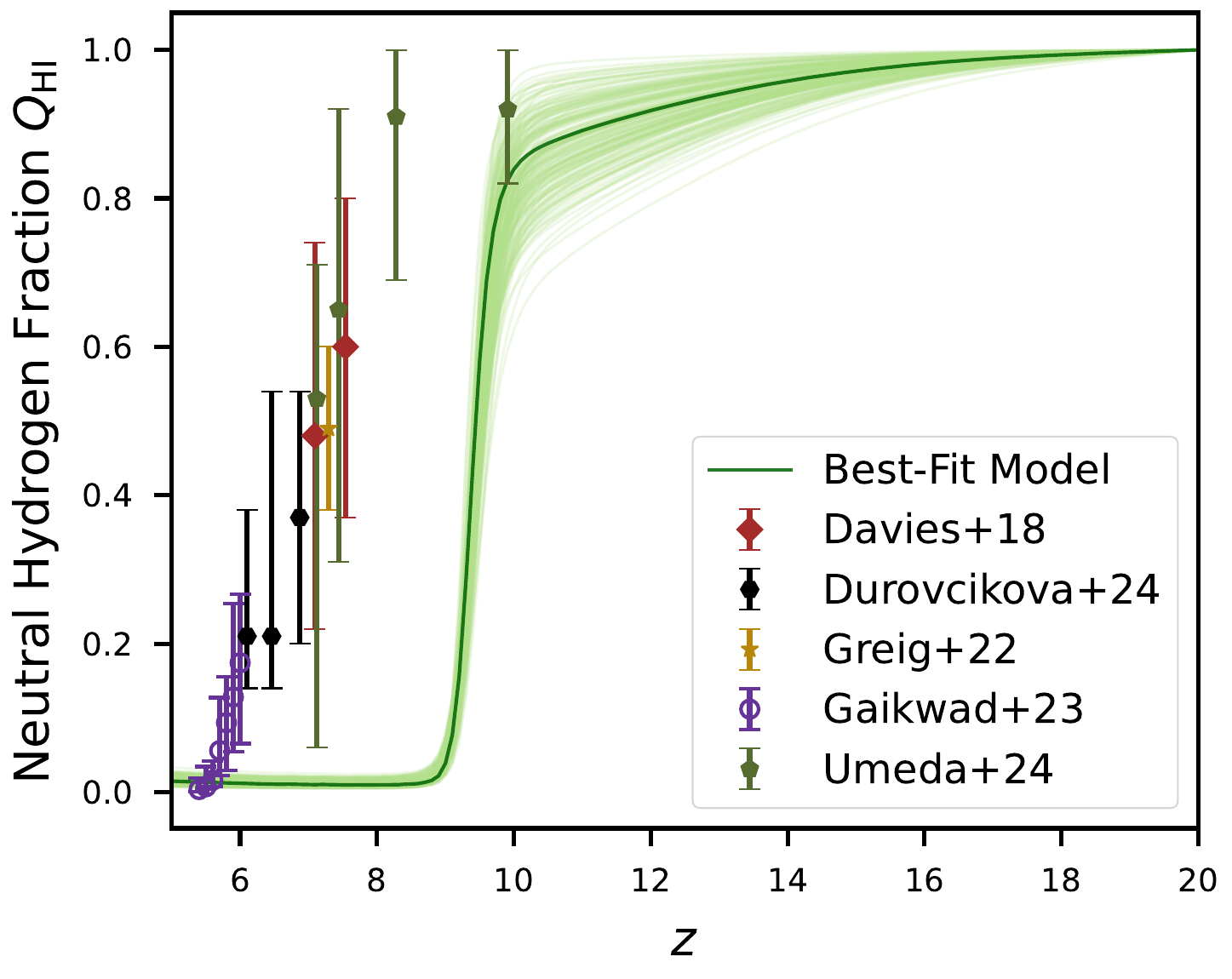}
    \caption{}
    \label{fig:baseline_case3_reion_history}
\end{subfigure}

\vspace{0.5cm} 

\begin{subfigure}[t]{\columnwidth}
    \centering
    \includegraphics[width=0.48\columnwidth]{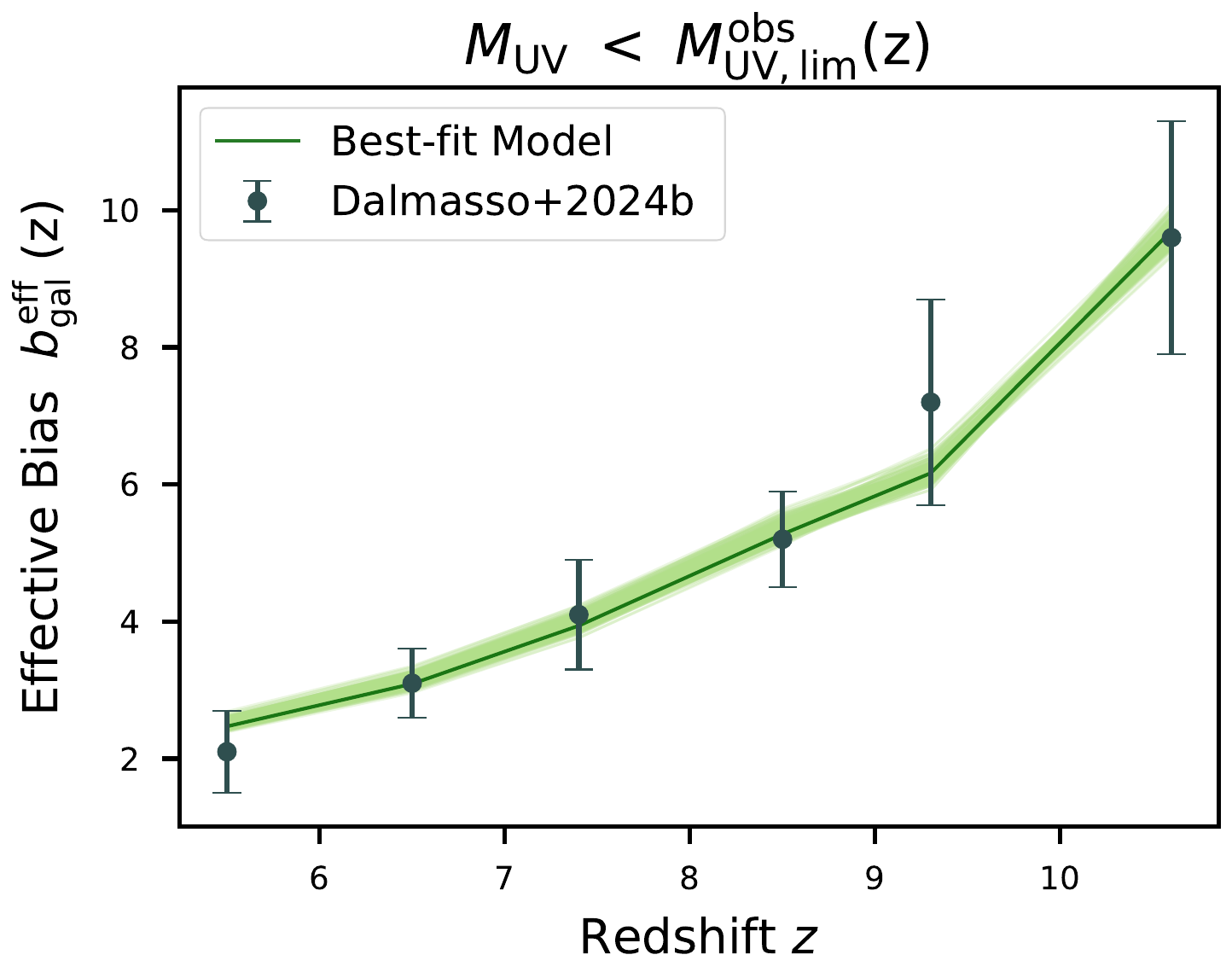}
    \hfill
    \includegraphics[width=0.48\columnwidth]{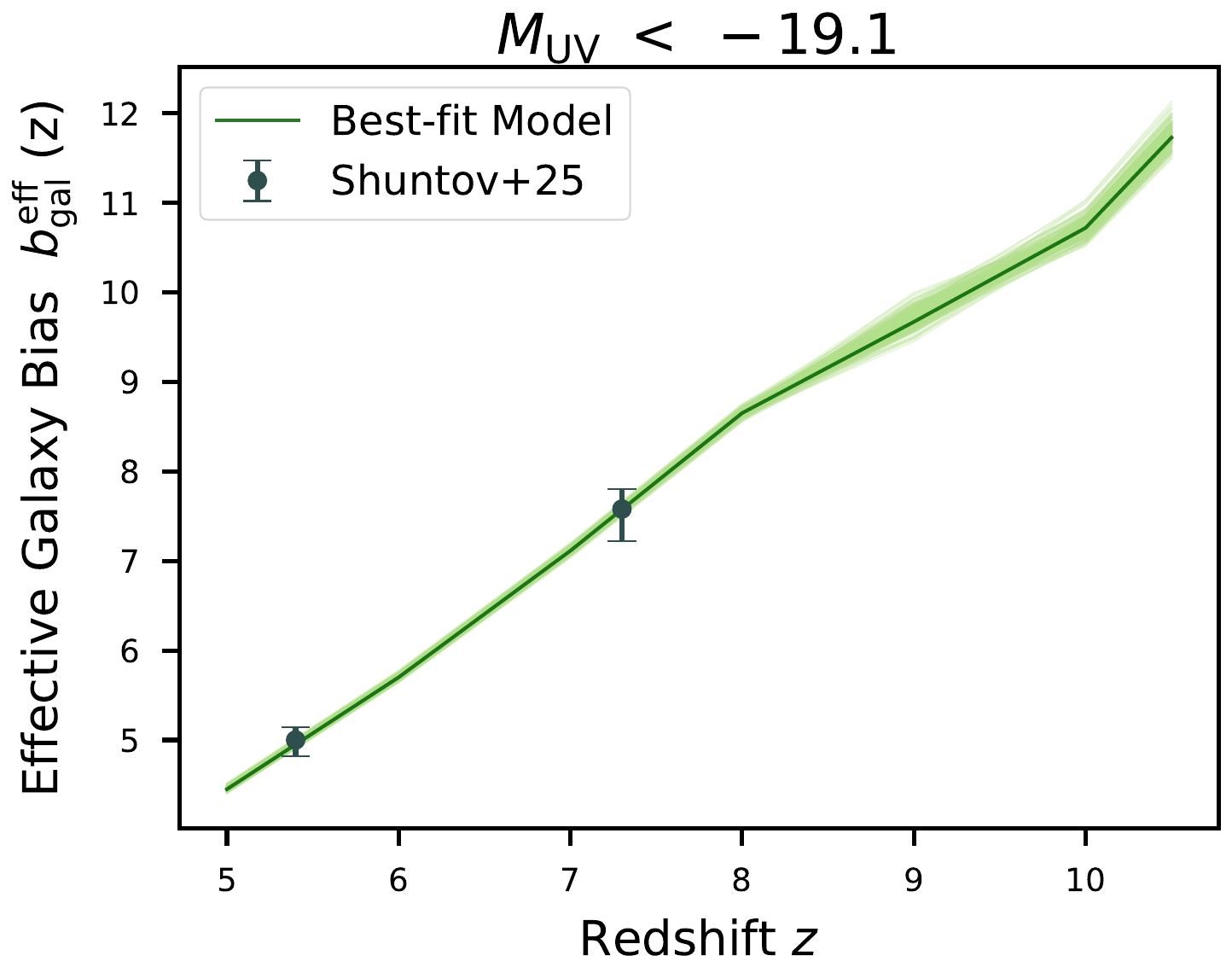 }
    \caption{}
    \label{fig:baseline_case3_gal_bias}
\end{subfigure}
\caption{Same as \fig{fig:baseline_case1_reion_history_and_gal_bias} but for 200 random samples drawn from the MCMC chains of the \textbf{UVLF+bias+reion} case, based on the \textbf{baseline} model.}
\label{fig:baseline_case3_reion_history_and_gal_bias}
\end{figure}

\begin{figure}[htbp]
\centering
\includegraphics[width=\columnwidth] 
{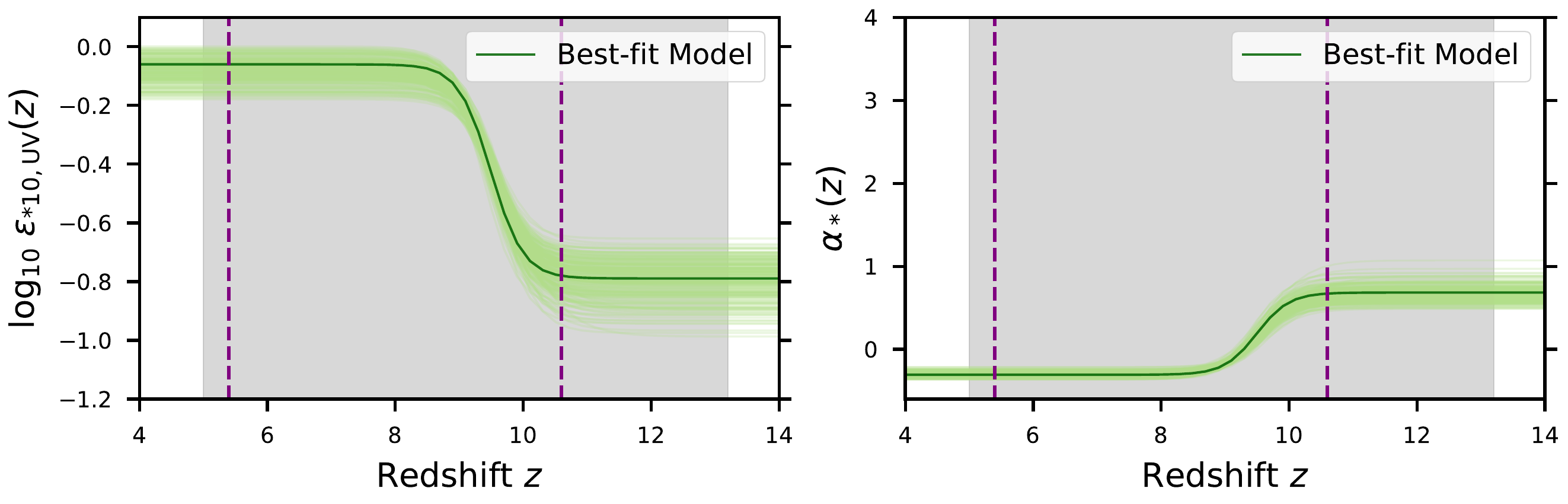}
\caption{Same as \fig{fig:baseline_case1_sfe_params} but for 200 random samples drawn from the MCMC chains of the \textbf{UVLF+bias+reion} case, based on the \textbf{baseline} model. In each panel, the grey box and the vertical purple dashed lines enclose the redshift ranges used for comparing the model with UVLF ($5 \leq \langle z \rangle \leq 13.2$) and various galaxy bias ($5.4 \leq \langle z \rangle \leq 10.6$) observations, respectively.}
\label{fig:baseline_case3_sfe_params}
\end{figure}

From \figs{fig:baseline_case3_UVLF}{fig:baseline_case3_reion_history_and_gal_bias}, we see that the model can reproduce the UVLF and galaxy bias measurements reasonably well, but only under an extreme scenario where low-mass halos exhibit a very high star formation or UV production efficiency at the lower redshifts ($z<9$), as indicated by the constraints obtained on $\alpha_0 - \alpha_\mathrm{jump}/2$ and $\ell_{\varepsilon,0} - \ell_{\varepsilon,\mathrm{jump}}/2$ in \tab{tab:appendix_mcmc_results}. 

As a consequence of enhanced star formation at $z<10$ in this model, reionization occurs extremely rapidly, resembling a nearly step-like transition in $Q_{\rm HI}$, and completes much earlier ($z \approx 9$) compared to the observational constraints on the reionization timeline. Even exceptionally low escaping ionizing efficiencies are unable to delay the process.

We remind that a similar galaxy-halo connection was required in the \textbf{bias+reion} case (\secn{sec:results_baseline}) of the \textbf{baseline} model to explain the evolution of galaxy bias, but it had led to an overestimation of the galaxy abundances. In contrast, in the present scenario, agreement with the observed UVLF at $z \lesssim 9$ — when the intergalactic medium is already fully ionized — is achieved by invoking stronger radiative feedback in halos with masses below $10^{11}~M_\odot$, which suppresses the abundance of faint galaxies.

Moreover, the redshift evolution of the UV efficiency parameter differs from our previous cases, as shown in \fig{fig:baseline_case3_sfe_params}. We find that while the slope ($\alpha_\ast$) of the $\epsilon_{\rm *,UV}-M_h$ relation increases with redshift, enhancing the UV emission from halos at $z>10$, the normalization, $\varepsilon_{\rm *10,UV}$, drops sharply beyond $z \approx 9$ to even lower values. As a result, the UVLF predictions from the model become increasingly discrepant with the available JWST observations at $z>10$.

These results indicate that achieving simultaneous agreement with all three datasets using the \textbf{baseline} model is equally challenging, and may in fact be impossible given its underlying assumptions.

\begin{table*}[htbp]
 \caption{Parameter constraints obtained from the MCMC-based analysis. The first nine rows correspond to the free parameters of the \textbf{baseline} model, while the remaining are the derived parameters. The free parameters are assumed to have uniform priors in the range mentioned in the second column. The numbers in the last column show the mean value with 1$\sigma$ errors for different parameters of the model, as obtained from the \textbf{UVLF+reion+bias} case.\\} 
 \label{tab:appendix_mcmc_results}
 \begin{tabular*}{\columnwidth}{c@{\hspace*{75pt}}c@{\hspace*{75pt}}c@{\hspace*{75pt}}}
 \hline   \hline  \\
    Parameters & Priors &  UVLF+bias+reion  \\  \\
  \hline   \hline \\
{\boldmath$\ell_{\varepsilon,0} + \ell_{\varepsilon,\mathrm{jump}} / 2 $} & [-2.0, 2.0] & $-0.787^{+0.067}_{-0.054}$  \\ \\

{\boldmath$\ell_{\varepsilon,0} - \ell_{\varepsilon,\mathrm{jump}} / 2$} & [-2.0, 1.0] & $-0.080^{+0.043}_{-0.038}$    \\ \\

{\boldmath$z_\ast $} & [8.0, 18.0] & $9.533\pm 0.080$  \\ \\

{\boldmath$\Delta z_\ast $} & [0.5, 6.0] & $< 0.566$  \\ \\

{\boldmath$\alpha_0 + \alpha_\mathrm{jump} / 2 $} & [0.0, 7.0] & $0.667^{+0.069}_{-0.093}$    \\ \\

{\boldmath$\alpha_0 - \alpha_\mathrm{jump} / 2 $} & [-1.0, 1.0]  & $-0.300^{+0.030}_{-0.034}$   \\ \\

{\boldmath$\log_{10}~(\varepsilon_{\mathrm{esc,10}})$} & [-4.0, 1.0]  & $< -3.70 $ \\ \\

{\boldmath$\alpha_{\rm esc}$} & [-3.0, 1.0] & $-2.206^{+0.057}_{-0.15}$  \\ \\

{\boldmath$\log_{10}(M_{\mathrm{crit}}/M_\odot)$} & [9.0, 11.0] &$> 11.0$  \\ \\

\hline \\

$\tau_{\rm el}  $ & - &  $0.0771^{+0.0022}_{-0.0032}$ \\ \\

$\ell_{\varepsilon,\mathrm{jump}}$ & -  &$-0.708\pm 0.079$  \\ \\

$\alpha_\mathrm{jump}  $ & - &$0.967^{+0.086}_{-0.10}$  \\ \\
\hline
\end{tabular*}
\end{table*}


\section{Effect of Different Weighting Choices for Datasets in the Likelihood Analysis}

\label{appendix:weight_factor_effect}

In this appendix, we discuss the sensitivity of our results to the weighting scheme adopted in \secn{subsec:results_extended}. As described in \secn{subsec:results_extended}, the weight factor $w$, which enters in our likelihood calculation (\eqn{eqn:joint_likelihood_expression_with_weight}), controls the relative importance assigned to the galaxy UV luminosity function (UVLF) and the galaxy bias or reionization datasets during the MCMC analysis.  Throughout this work, we used a weight factor of $w=4$, motivated by the relative number of data points from
the galaxy bias datasets and the UVLF datasets, such that $w^2 > N^{\rm UVLF}_{\rm data} / N^{\rm bias}_{\rm data}$.

\begin{figure}[htbp]
    \centering
    \includegraphics[width=\columnwidth]{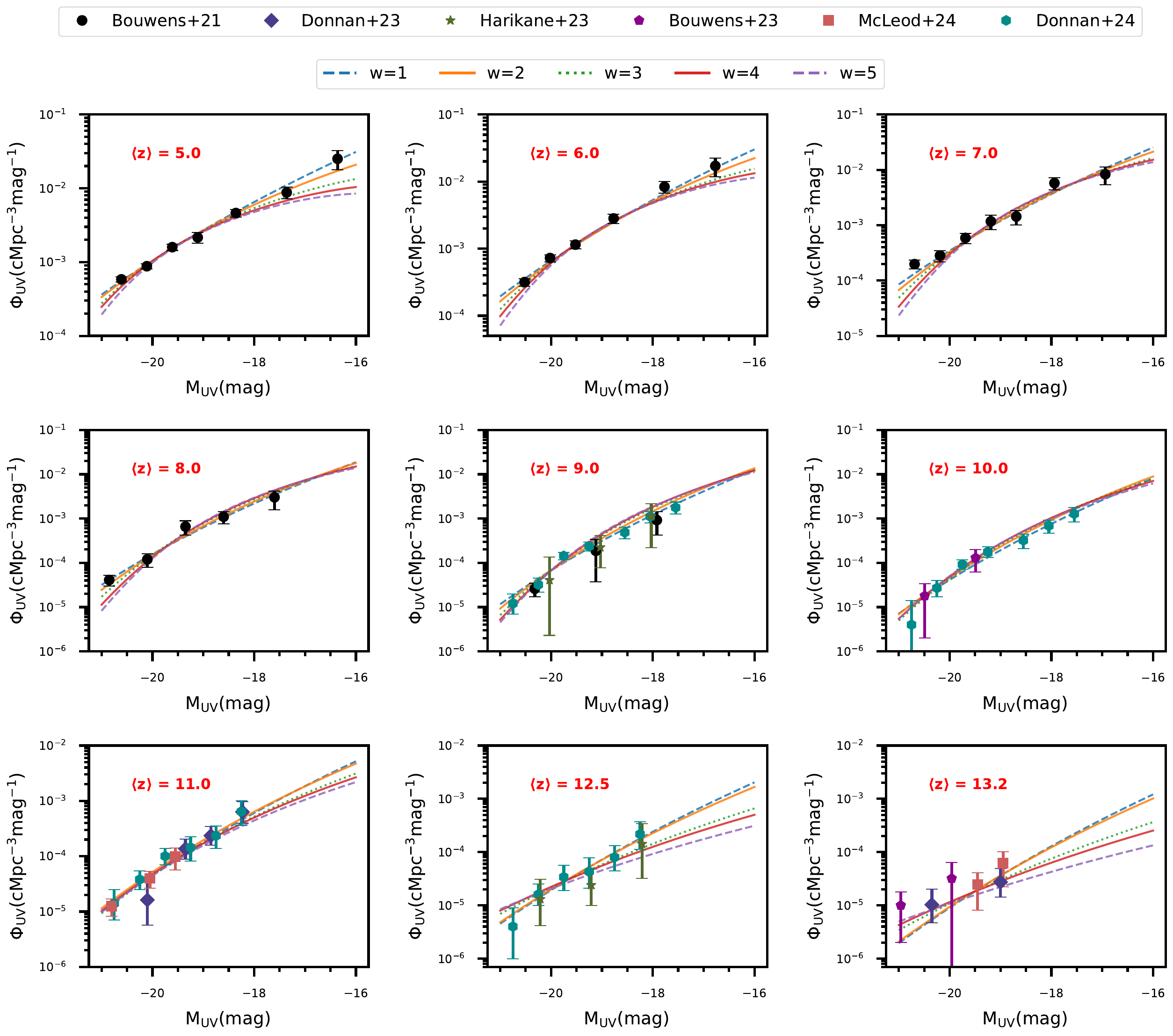}
    \caption{Comparison of the evolution of the galaxy UV luminosity function for the best-fit models corresponding to different choices of the weight factor $w$ (see \eqn{eqn:joint_likelihood_expression_with_weight}) used during the likelihood analysis with the \textbf{extended} model.}
    \label{fig:wt_factor_comparison_UVLF}
\end{figure}

Here, we explore a range of values for $w$, including the case of $w = 1$, which assigns equal weight to all datasets. \figs{fig:wt_factor_comparison_UVLF}{fig:comparison_wtfactor_reionhistory_and_bias} illustrate the resulting evolution of the galaxy UVLF, galaxy bias, and neutral hydrogen fraction, respectively, for the \textbf{best-fit} models corresponding to different weight choices.

\begin{figure}[htbp]
\centering
    \begin{subfigure}[t]{0.95\textwidth}
        \centering
        \includegraphics[width=0.6\textwidth]{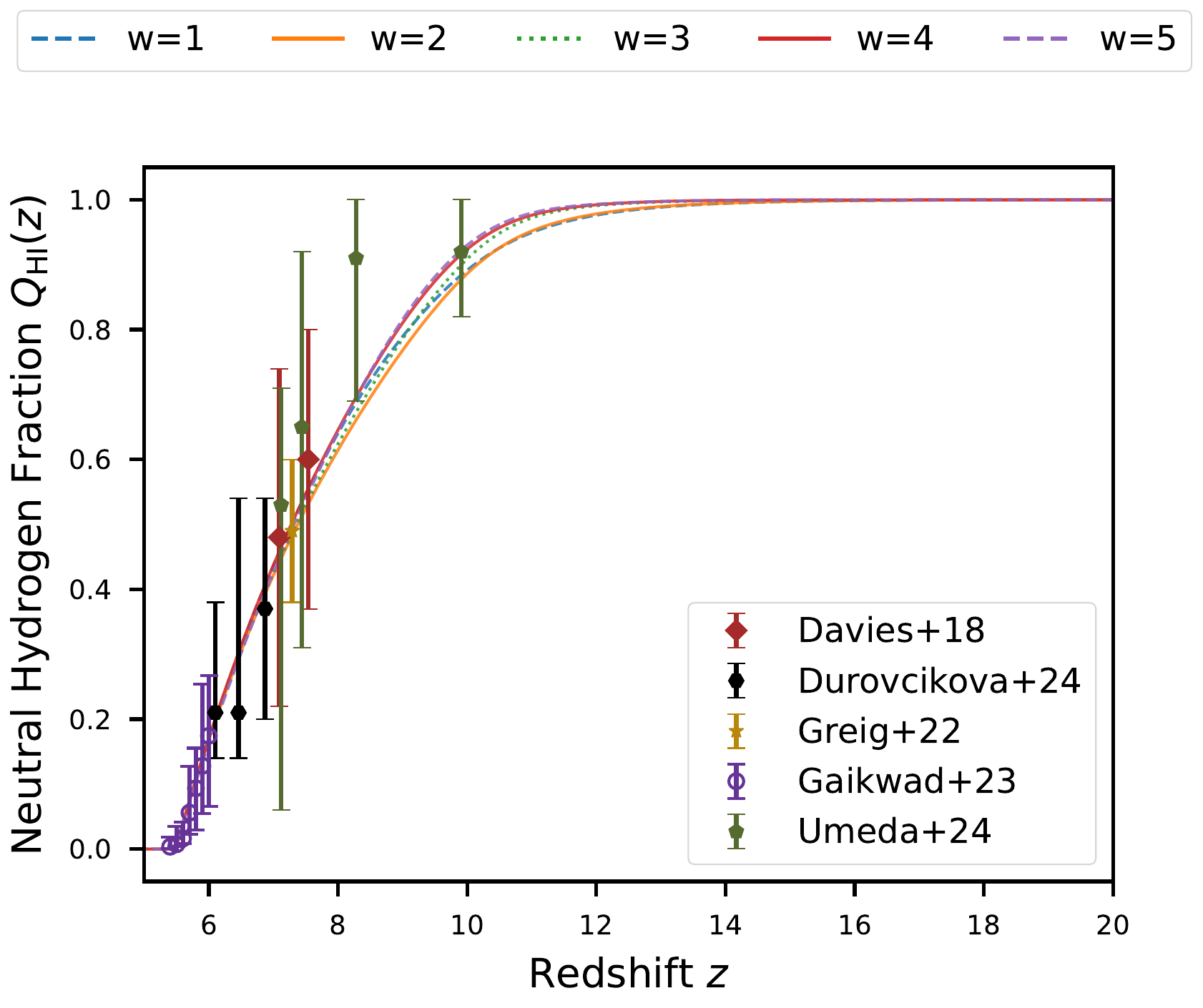}
        \caption{}
        \label{subfig:comparison_wtfactor_reionhist}
    \end{subfigure}
    
    \vspace{0.5cm} 

    \begin{subfigure}[t]{\columnwidth}
        \centering
        \includegraphics[width=0.48\columnwidth]{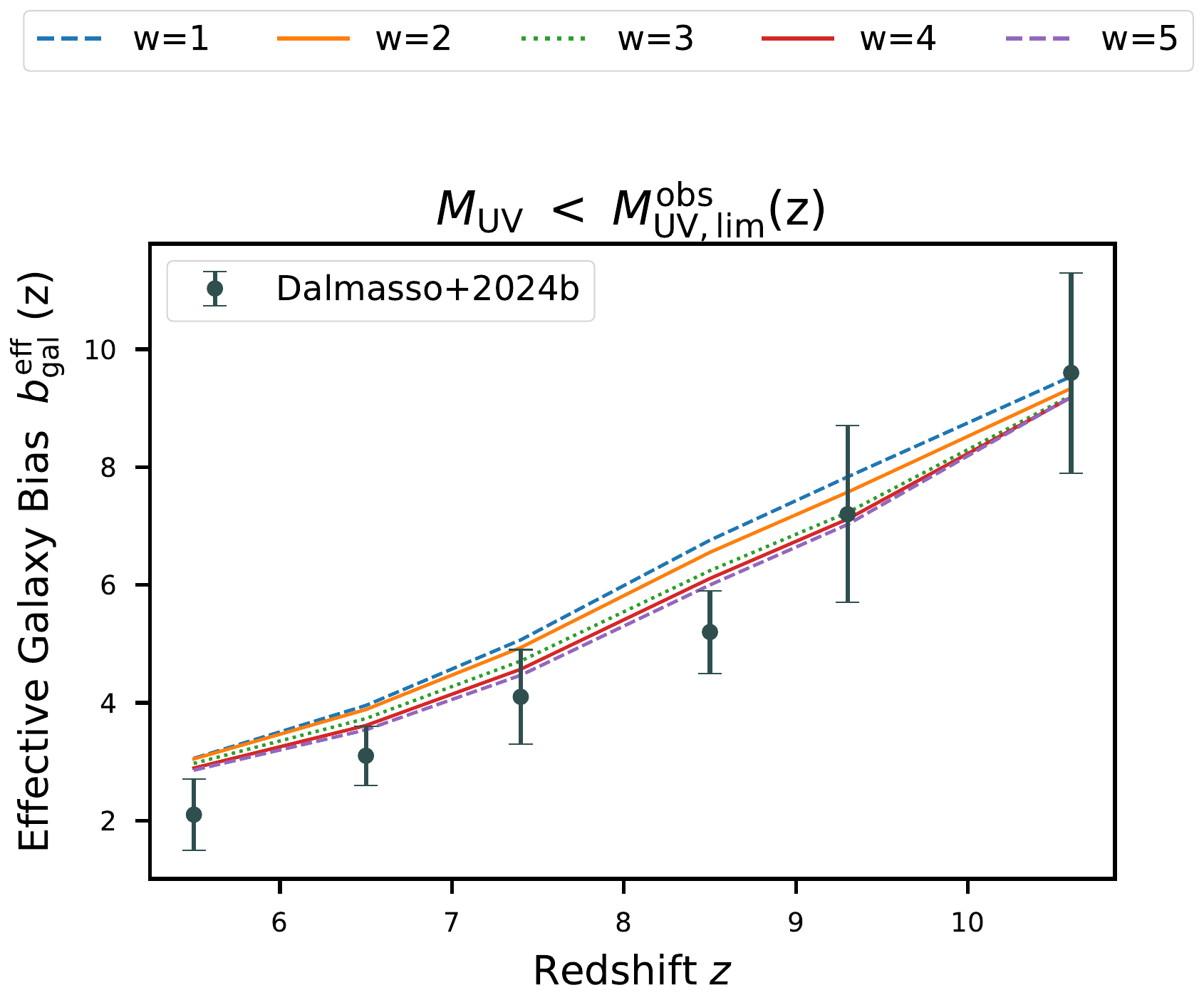}
        \hfill
        \includegraphics[width=0.48\columnwidth]{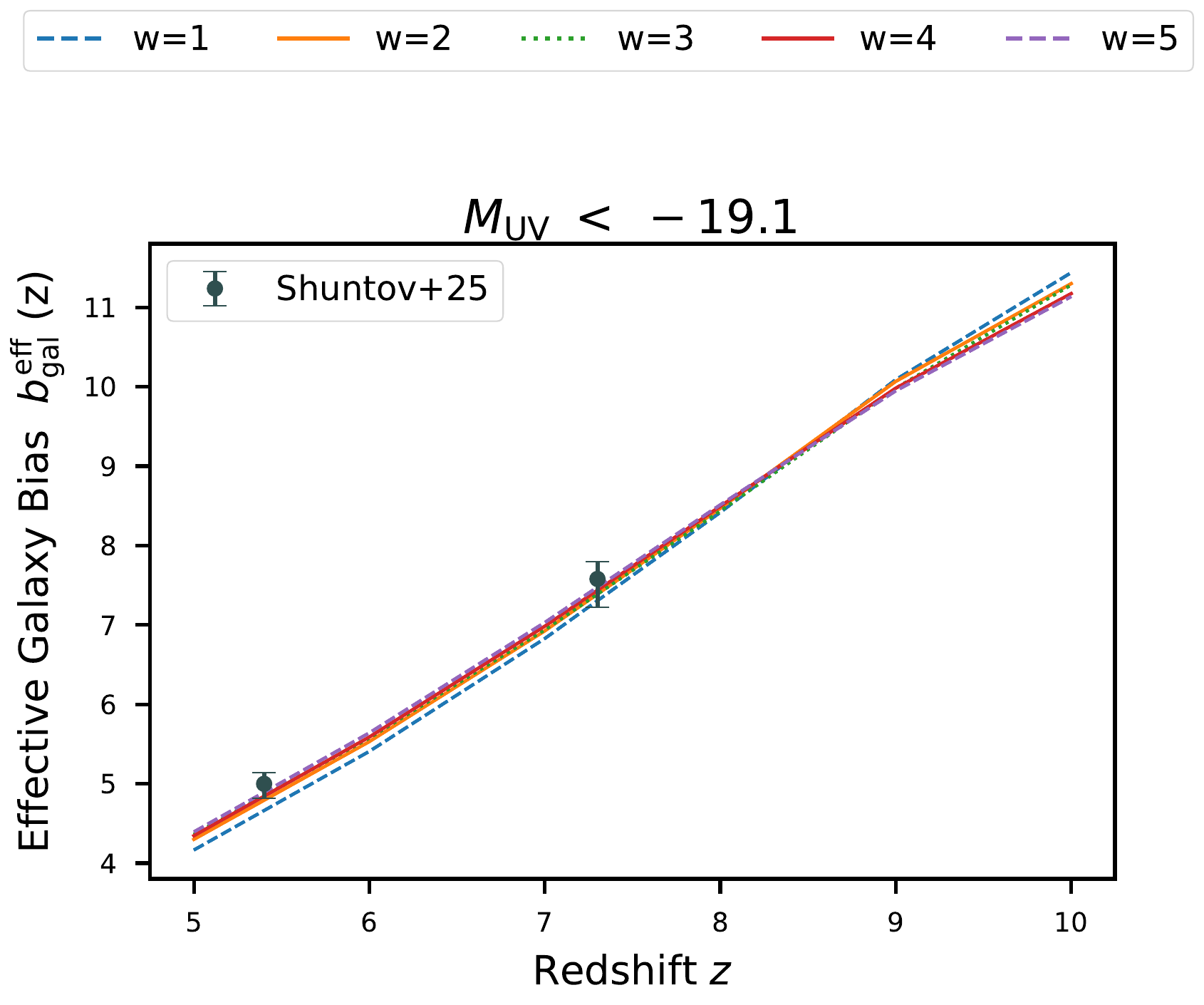}
        \caption{}
        \label{subfig:comparison_wtfactor_bias}
    \end{subfigure}

    \caption{Comparison of the evolution of (a)  intergalactic globally-averaged neutral hydrogen fraction and (b) galaxy bias for the best-fit models corresponding to different choices of the weight factor $w$ (see \eqn{eqn:joint_likelihood_expression_with_weight}) used during the likelihood analysis with the \textbf{extended} model.}
    \label{fig:comparison_wtfactor_reionhistory_and_bias}
\end{figure}

It is evident from \figs{fig:wt_factor_comparison_UVLF}{subfig:comparison_wtfactor_bias} that at lower redshifts, the $w = 1$ case causes the MCMC analysis to prioritize fitting the UVLF data points, which are not only more numerous but also have comparatively smaller observational uncertainties than the bias measurements. Consequently, the case of $w=1$ is then unable to adequately reproduce the galaxy bias datasets at $z \lesssim8.5$.
On the other hand, as shown in \fig{subfig:comparison_wtfactor_reionhist}, the model-predicted reionization history remains largely unaffected across the different weighting schemes considered.

As the value of $w$ is increased, the analysis progressively shifts toward better reproducing the galaxy bias data at lower redshifts ($z \lesssim8.5$) --- however, this comes at the cost of a slightly poorer agreement with the UVLF observations at those same epochs. Therefore, values of $w \sim 4$ help in striking a reasonable balance between the two galaxy datasets in the likelihood analysis.

Nonetheless, as can also be seen from \fig{fig:extended_model_UVLF}, the differences between the model-predicted UVLF for the $w=4$ case and observational measurements at these redshift bins remain modest, especially when considering the intrinsic spread among the model predictions (shown by the light-shaded lines for 200 random samples). This indicates that our main conclusions are robust against reasonable variations in the chosen weight factor.

\newpage
\newpage
\section{Comparison of the Best-Fit Models from Different Galaxy–Halo Connection Prescriptions with Observational Data}

\label{appendix:comparison_of_all_models}

\begin{figure}[htbp]
\centering
\includegraphics[width=\columnwidth]{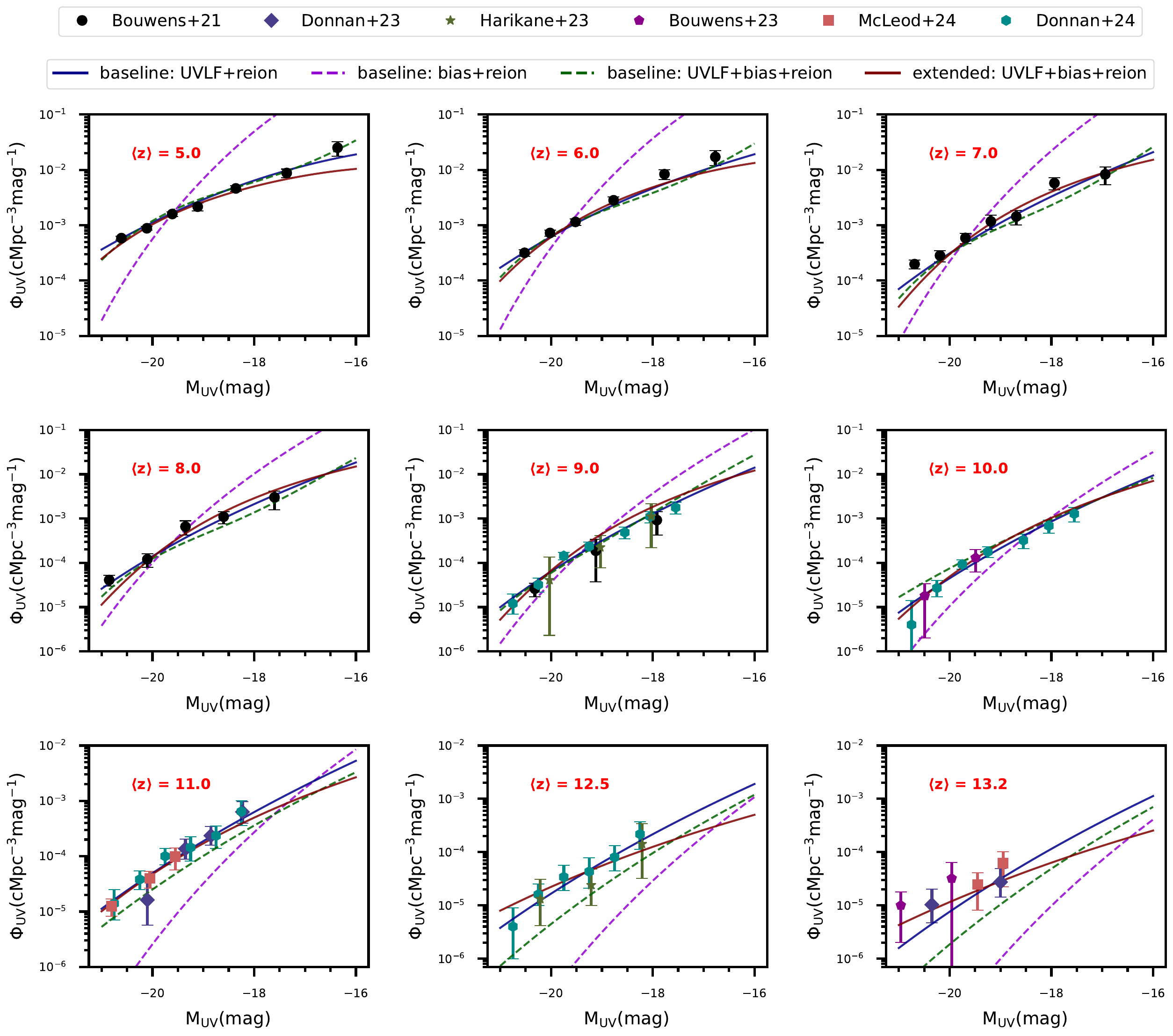}
\caption{ Comparison of the predicted evolution of the galaxy UV luminosity function from the best-fit models for the various cases, based on the \textbf{baseline} and \textbf{extended} models, investigated in \secn{sec:results_baseline}, \secn{sec:duty_cycle} and \app{appendix:baseline_all_observations}.}
\label{fig:comparison_UVLF}
\end{figure}

\begin{figure}[htbp]
\centering
\begin{subfigure}[t]{\columnwidth}
    \centering
    \includegraphics[width=0.6\textwidth]{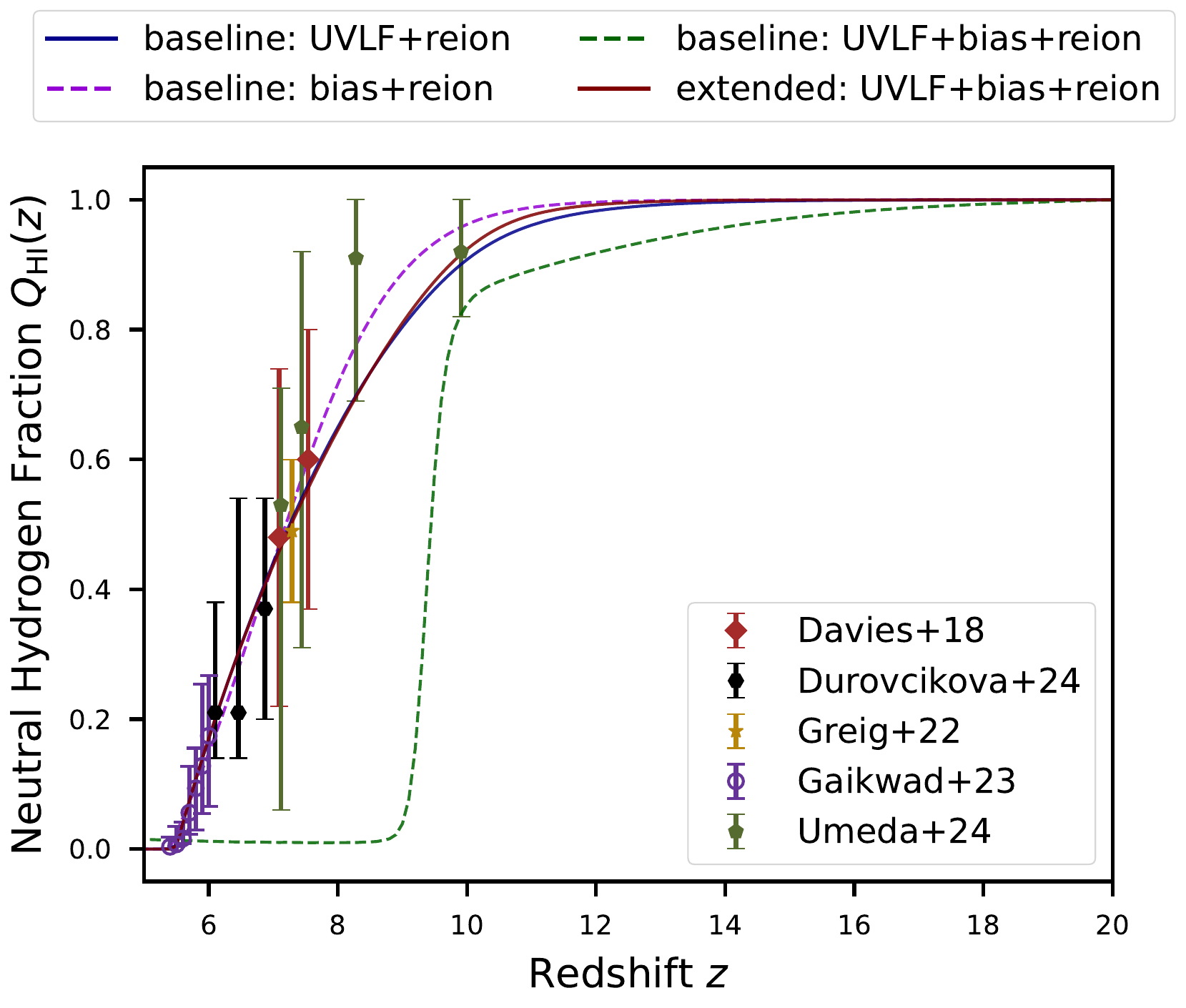}
    \caption{}
    \label{fig:compare_reion_history}
\end{subfigure}

\vspace{0.5cm} 

\begin{subfigure}[t]{\columnwidth}
    \centering
    \includegraphics[width=0.48\columnwidth]{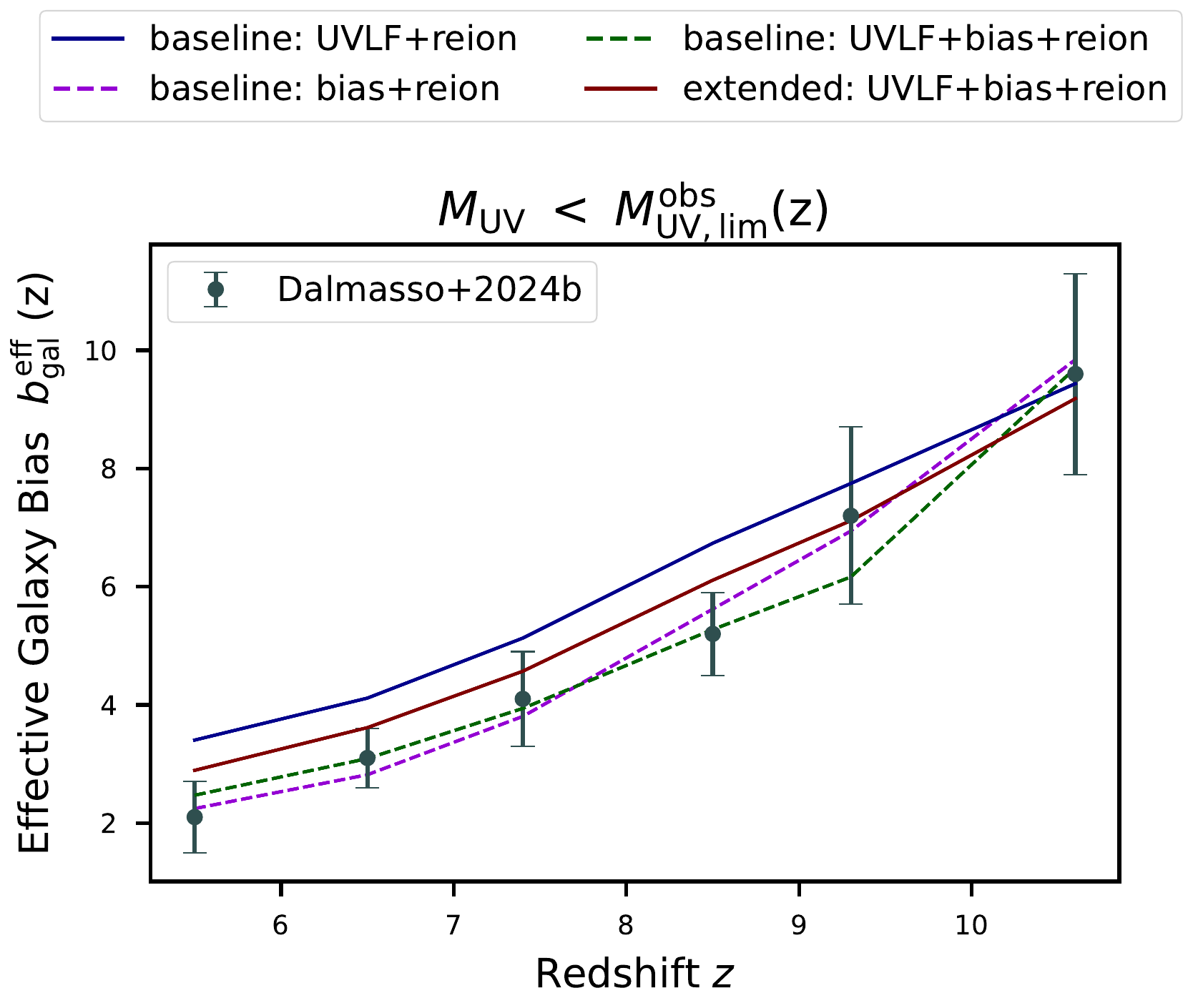}
    \hfill
    \includegraphics[width=0.48\columnwidth]{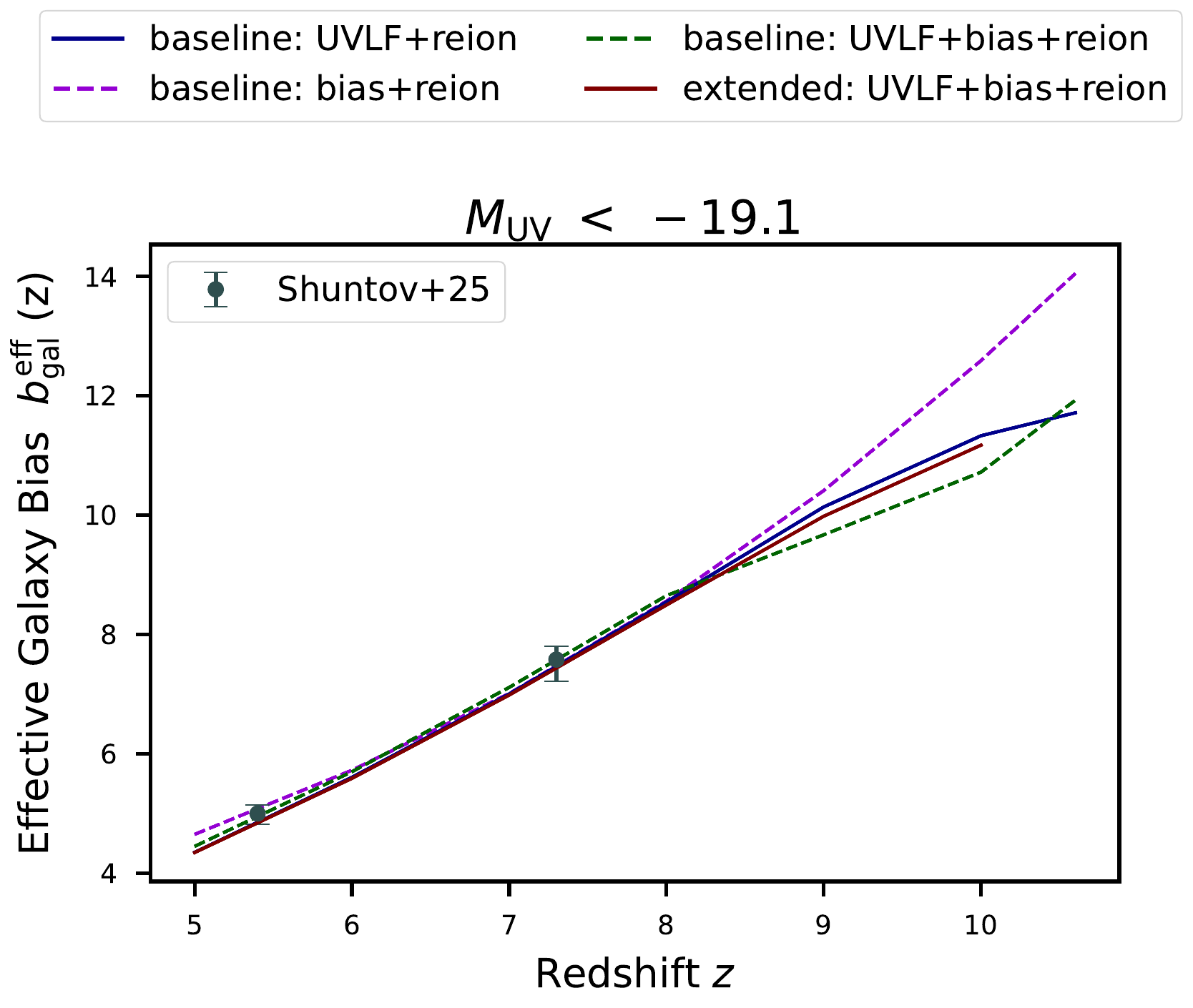 }
    \caption{}
    \label{fig:compare_gal_bias}
\end{subfigure}
\caption{Comparison of the effective galaxy bias and the reionization history predicted by the best-fit model for the various cases based on the \textbf{baseline} and \textbf{extended} models, investigated in \secn{sec:results_baseline}, \secn{sec:duty_cycle} and \app{appendix:baseline_all_observations}.}
\label{fig:comparison_reion_history_and_gal_bias}
\end{figure}

In this appendix, we compare the predictions from the best-fit models of the various cases discussed in \secn{sec:results_baseline}, \secn{sec:duty_cycle} and \app{appendix:baseline_all_observations}, each representing a distinct galaxy–halo connection based on either the \textbf{baseline} or \textbf{extended} model, with available observational data. \fig{fig:comparison_UVLF} shows the predicted evolution of the galaxy UV luminosity function for these models alongside observational measurements from JWST and HST. The evolution of the neutral fraction in the intergalactic medium and effective galaxy bias are shown in \figs{fig:compare_reion_history}{fig:compare_gal_bias} respectively.

\newpage
\newpage

\bibliographystyle{JHEP}
\bibliography{manuscript_final_ver}

\providecommand{\href}[2]{#2}\begingroup\raggedright\begin{thebibliography}{100}

\bibitem{Wechsler2018}
R.H.~{Wechsler} and J.L.~{Tinker}, \emph{{The Connection Between Galaxies and Their Dark Matter Halos}}, \href{https://doi.org/10.1146/annurev-astro-081817-051756}{\emph{\araa} {\bfseries 56} (2018) 435} [\href{https://arxiv.org/abs/1804.03097}{{\ttfamily 1804.03097}}].

\bibitem{Bouwens2015}
R.J.~{Bouwens}, G.D.~{Illingworth}, P.A.~{Oesch}, M.~{Trenti}, I.~{Labb{\'e}}, L.~{Bradley} et~al., \emph{{UV Luminosity Functions at Redshifts z {\ensuremath{\sim}} 4 to z {\ensuremath{\sim}} 10: 10,000 Galaxies from HST Legacy Fields}}, \href{https://doi.org/10.1088/0004-637X/803/1/34}{\emph{\apj} {\bfseries 803} (2015) 34} [\href{https://arxiv.org/abs/1403.4295}{{\ttfamily 1403.4295}}].

\bibitem{Bouwens2017}
R.J.~{Bouwens}, P.A.~{Oesch}, G.D.~{Illingworth}, R.S.~{Ellis} and M.~{Stefanon}, \emph{{The z {\ensuremath{\sim}} 6 Luminosity Function Fainter than -15 mag from the Hubble Frontier Fields: The Impact of Magnification Uncertainties}}, \href{https://doi.org/10.3847/1538-4357/aa70a4}{\emph{\apj} {\bfseries 843} (2017) 129} [\href{https://arxiv.org/abs/1610.00283}{{\ttfamily 1610.00283}}].

\bibitem{Atek2018}
H.~{Atek}, J.~{Richard}, J.-P.~{Kneib} and D.~{Schaerer}, \emph{{The extreme faint end of the UV luminosity function at z {\ensuremath{\sim}} 6 through gravitational telescopes: a comprehensive assessment of strong lensing uncertainties}}, \href{https://doi.org/10.1093/mnras/sty1820}{\emph{\mnras} {\bfseries 479} (2018) 5184} [\href{https://arxiv.org/abs/1803.09747}{{\ttfamily 1803.09747}}].

\bibitem{Ono2018}
Y.~{Ono}, M.~{Ouchi}, Y.~{Harikane}, J.~{Toshikawa}, M.~{Rauch}, S.~{Yuma} et~al., \emph{{Great Optically Luminous Dropout Research Using Subaru HSC (GOLDRUSH). I. UV luminosity functions at z {\ensuremath{\sim}} 4-7 derived with the half-million dropouts on the 100 deg$^{2}$ sky}}, \href{https://doi.org/10.1093/pasj/psx103}{\emph{\pasj} {\bfseries 70} (2018) S10} [\href{https://arxiv.org/abs/1704.06004}{{\ttfamily 1704.06004}}].

\bibitem{Bowler2020}
R.A.A.~{Bowler}, M.J.~{Jarvis}, J.S.~{Dunlop}, R.J.~{McLure}, D.J.~{McLeod}, N.J.~{Adams} et~al., \emph{{A lack of evolution in the very bright end of the galaxy luminosity function from z = 8 to 10}}, \href{https://doi.org/10.1093/mnras/staa313}{\emph{\mnras} {\bfseries 493} (2020) 2059} [\href{https://arxiv.org/abs/1911.12832}{{\ttfamily 1911.12832}}].

\bibitem{harikane2022}
Y.~{Harikane}, Y.~{Ono}, M.~{Ouchi}, C.~{Liu}, M.~{Sawicki}, T.~{Shibuya} et~al., \emph{{GOLDRUSH. IV. Luminosity Functions and Clustering Revealed with 4,000,000 Galaxies at z 2-7: Galaxy-AGN Transition, Star Formation Efficiency, and Implication for Evolution at z > 10}}, \href{https://doi.org/10.3847/1538-4365/ac3dfc}{\emph{\apjs} {\bfseries 259} (2022) 20} [\href{https://arxiv.org/abs/2108.01090}{{\ttfamily 2108.01090}}].

\bibitem{Donnan2023}
C.T.~{Donnan}, D.J.~{McLeod}, J.S.~{Dunlop}, R.J.~{McLure}, A.C.~{Carnall}, R.~{Begley} et~al., \emph{{The evolution of the galaxy UV luminosity function at redshifts z = 8 - 15 from deep JWST and ground-based near-infrared imaging}}, \href{https://doi.org/10.1093/mnras/stac3472}{\emph{\mnras} {\bfseries 518} (2023) 6011} [\href{https://arxiv.org/abs/2207.12356}{{\ttfamily 2207.12356}}].

\bibitem{Harikane2023}
Y.~{Harikane}, M.~{Ouchi}, M.~{Oguri}, Y.~{Ono}, K.~{Nakajima}, Y.~{Isobe} et~al., \emph{{A Comprehensive Study of Galaxies at z 9-16 Found in the Early JWST Data: Ultraviolet Luminosity Functions and Cosmic Star Formation History at the Pre-reionization Epoch}}, \href{https://doi.org/10.3847/1538-4365/acaaa9}{\emph{\apjs} {\bfseries 265} (2023) 5} [\href{https://arxiv.org/abs/2208.01612}{{\ttfamily 2208.01612}}].

\bibitem{Bouwens2023}
R.~{Bouwens}, G.~{Illingworth}, P.~{Oesch}, M.~{Stefanon}, R.~{Naidu}, I.~{van Leeuwen} et~al., \emph{{UV luminosity density results at z > 8 from the first JWST/NIRCam fields: limitations of early data sets and the need for spectroscopy}}, \href{https://doi.org/10.1093/mnras/stad1014}{\emph{\mnras} {\bfseries 523} (2023) 1009} [\href{https://arxiv.org/abs/2212.06683}{{\ttfamily 2212.06683}}].

\bibitem{McLeod2024}
D.J.~{McLeod}, C.T.~{Donnan}, R.J.~{McLure}, J.S.~{Dunlop}, D.~{Magee}, R.~{Begley} et~al., \emph{{The galaxy UV luminosity function at z $\simeq$ 11 from a suite of public JWST ERS, ERO, and Cycle-1 programs}}, \href{https://doi.org/10.1093/mnras/stad3471}{\emph{\mnras} {\bfseries 527} (2024) 5004} [\href{https://arxiv.org/abs/2304.14469}{{\ttfamily 2304.14469}}].

\bibitem{Donnan2024}
C.T.~{Donnan}, R.J.~{McLure}, J.S.~{Dunlop}, D.J.~{McLeod}, D.~{Magee}, K.Z.~{Arellano-C{\'o}rdova} et~al., \emph{{JWST PRIMER: a new multifield determination of the evolving galaxy UV luminosity function at redshifts z = 9 - 15}}, \href{https://doi.org/10.1093/mnras/stae2037}{\emph{\mnras} {\bfseries 533} (2024) 3222} [\href{https://arxiv.org/abs/2403.03171}{{\ttfamily 2403.03171}}].

\bibitem{Whitler2025}
L.~{Whitler}, D.P.~{Stark}, M.W.~{Topping}, B.~{Robertson}, M.~{Rieke}, K.N.~{Hainline} et~al., \emph{{The $z rsim 9$ galaxy UV luminosity function from the JWST Advanced Deep Extragalactic Survey: insights into early galaxy evolution and reionization}}, \href{https://doi.org/10.48550/arXiv.2501.00984}{\emph{arXiv e-prints} (2025) arXiv:2501.00984} [\href{https://arxiv.org/abs/2501.00984}{{\ttfamily 2501.00984}}].

\bibitem{PerezGonzalez2025}
P.G.~{P{\'e}rez-Gonz{\'a}lez}, G.~{{\"O}stlin}, L.~{Costantin}, J.~{Melinder}, S.L.~{Finkelstein}, R.S.~{Somerville} et~al., \emph{{The rise of the galactic empire: luminosity functions at $z\sim17$ and $z\sim25$ estimated with the MIDIS$+$NGDEEP ultra-deep JWST/NIRCam dataset}}, \href{https://doi.org/10.48550/arXiv.2503.15594}{\emph{arXiv e-prints} (2025) arXiv:2503.15594} [\href{https://arxiv.org/abs/2503.15594}{{\ttfamily 2503.15594}}].

\bibitem{Trenti2010}
M.~{Trenti}, M.~{Stiavelli}, R.J.~{Bouwens}, P.~{Oesch}, J.M.~{Shull}, G.D.~{Illingworth} et~al., \emph{{The Galaxy Luminosity Function During the Reionization Epoch}}, \href{https://doi.org/10.1088/2041-8205/714/2/L202}{\emph{\apjl} {\bfseries 714} (2010) L202} [\href{https://arxiv.org/abs/1004.0384}{{\ttfamily 1004.0384}}].

\bibitem{Tacchella2013}
S.~{Tacchella}, M.~{Trenti} and C.M.~{Carollo}, \emph{{A Physical Model for the 0 <\raisebox{-0.5ex}\textasciitilde z <\raisebox{-0.5ex}\textasciitilde 8 Redshift Evolution of the Galaxy Ultraviolet Luminosity and Stellar Mass Functions}}, \href{https://doi.org/10.1088/2041-8205/768/2/L37}{\emph{\apjl} {\bfseries 768} (2013) L37} [\href{https://arxiv.org/abs/1211.2825}{{\ttfamily 1211.2825}}].

\bibitem{Dayal14}
P.~{Dayal}, A.~{Ferrara}, J.S.~{Dunlop} and F.~{Pacucci}, \emph{{Essential physics of early galaxy formation}}, \href{https://doi.org/10.1093/mnras/stu1848}{\emph{\mnras} {\bfseries 445} (2014) 2545} [\href{https://arxiv.org/abs/1405.4862}{{\ttfamily 1405.4862}}].

\bibitem{Mason2015}
C.A.~{Mason}, M.~{Trenti} and T.~{Treu}, \emph{{The Galaxy UV Luminosity Function before the Epoch of Reionization}}, \href{https://doi.org/10.1088/0004-637X/813/1/21}{\emph{\apj} {\bfseries 813} (2015) 21} [\href{https://arxiv.org/abs/1508.01204}{{\ttfamily 1508.01204}}].

\bibitem{Sun&Furlanetto2016}
G.~{Sun} and S.R.~{Furlanetto}, \emph{{Constraints on the star formation efficiency of galaxies during the epoch of reionization}}, \href{https://doi.org/10.1093/mnras/stw980}{\emph{\mnras} {\bfseries 460} (2016) 417} [\href{https://arxiv.org/abs/1512.06219}{{\ttfamily 1512.06219}}].

\bibitem{Tacchella2018}
S.~{Tacchella}, S.~{Bose}, C.~{Conroy}, D.J.~{Eisenstein} and B.D.~{Johnson}, \emph{{A Redshift-independent Efficiency Model: Star Formation and Stellar Masses in Dark Matter Halos at z {\ensuremath{\gtrsim}} 4}}, \href{https://doi.org/10.3847/1538-4357/aae8e0}{\emph{\apj} {\bfseries 868} (2018) 92} [\href{https://arxiv.org/abs/1806.03299}{{\ttfamily 1806.03299}}].

\bibitem{Park++2019}
J.~{Park}, A.~{Mesinger}, B.~{Greig} and N.~{Gillet}, \emph{{Inferring the astrophysics of reionization and cosmic dawn from galaxy luminosity functions and the 21-cm signal}}, \href{https://doi.org/10.1093/mnras/stz032}{\emph{\mnras} {\bfseries 484} (2019) 933} [\href{https://arxiv.org/abs/1809.08995}{{\ttfamily 1809.08995}}].

\bibitem{Ferrara2023}
A.~{Ferrara}, A.~{Pallottini} and P.~{Dayal}, \emph{{On the stunning abundance of super-early, luminous galaxies revealed by JWST}}, \href{https://doi.org/10.1093/mnras/stad1095}{\emph{\mnras} {\bfseries 522} (2023) 3986} [\href{https://arxiv.org/abs/2208.00720}{{\ttfamily 2208.00720}}].

\bibitem{Wang2024}
Y.-Y.~{Wang}, L.~{Lei}, S.-P.~{Tang}, G.-W.~{Yuan} and Y.-Z.~{Fan}, \emph{{Digging into the Ultraviolet Luminosity Functions of Galaxies at High Redshifts: Galaxies Evolution, Reionization, and Cosmological Parameters}}, \href{https://doi.org/10.3847/1538-4357/ad8080}{\emph{\apj} {\bfseries 975} (2024) 285} [\href{https://arxiv.org/abs/2405.09350}{{\ttfamily 2405.09350}}].

\bibitem{Ren2018}
K.~{Ren}, M.~{Trenti} and S.J.~{Mutch}, \emph{{The Cosmic Web around the Brightest Galaxies during the Epoch of Reionization}}, \href{https://doi.org/10.3847/1538-4357/aab094}{\emph{\apj} {\bfseries 856} (2018) 81} [\href{https://arxiv.org/abs/1802.06802}{{\ttfamily 1802.06802}}].

\bibitem{Ren2019}
K.~{Ren}, M.~{Trenti} and C.A.~{Mason}, \emph{{The Brightest Galaxies at Cosmic Dawn from Scatter in the Galaxy Luminosity versus Halo Mass Relation}}, \href{https://doi.org/10.3847/1538-4357/ab2117}{\emph{\apj} {\bfseries 878} (2019) 114} [\href{https://arxiv.org/abs/1905.04848}{{\ttfamily 1905.04848}}].

\bibitem{Munoz2023}
J.B.~{Mu{\~n}oz}, J.~{Mirocha}, S.~{Furlanetto} and N.~{Sabti}, \emph{{Breaking degeneracies in the first galaxies with clustering}}, \href{https://doi.org/10.1093/mnrasl/slad115}{\emph{\mnras} {\bfseries 526} (2023) L47} [\href{https://arxiv.org/abs/2306.09403}{{\ttfamily 2306.09403}}].

\bibitem{Mirocha2023}
J.~{Mirocha} and S.R.~{Furlanetto}, \emph{{Balancing the efficiency and stochasticity of star formation with dust extinction in $z > 10$ galaxies observed by JWST}}, \href{https://doi.org/10.1093/mnras/stac3578}{\emph{\mnras} {\bfseries 519} (2023) 843} [\href{https://arxiv.org/abs/2208.12826}{{\ttfamily 2208.12826}}].

\bibitem{Gelli2024}
V.~{Gelli}, C.~{Mason} and C.C.~{Hayward}, \emph{{The Impact of Mass-dependent Stochasticity at Cosmic Dawn}}, \href{https://doi.org/10.3847/1538-4357/ad7b36}{\emph{\apj} {\bfseries 975} (2024) 192} [\href{https://arxiv.org/abs/2405.13108}{{\ttfamily 2405.13108}}].

\bibitem{Lee2009}
K.-S.~{Lee}, M.~{Giavalisco}, C.~{Conroy}, R.H.~{Wechsler}, H.C.~{Ferguson}, R.S.~{Somerville} et~al., \emph{{Mapping the Dark Matter from UV Light at High Redshift: An Empirical Approach to Understand Galaxy Statistics}}, \href{https://doi.org/10.1088/0004-637X/695/1/368}{\emph{\apj} {\bfseries 695} (2009) 368} [\href{https://arxiv.org/abs/0808.1727}{{\ttfamily 0808.1727}}].

\bibitem{Weinberger2019}
L.H.~{Weinberger}, M.G.~{Haehnelt} and G.~{Kulkarni}, \emph{{Modelling the observed luminosity function and clustering evolution of Ly {\ensuremath{\alpha}} emitters: growing evidence for late reionization}}, \href{https://doi.org/10.1093/mnras/stz481}{\emph{\mnras} {\bfseries 485} (2019) 1350} [\href{https://arxiv.org/abs/1902.05077}{{\ttfamily 1902.05077}}].

\bibitem{Mirocha2020}
J.~{Mirocha}, \emph{{Prospects for distinguishing galaxy evolution models with surveys at redshifts z {\ensuremath{\gtrsim}} 4}}, \href{https://doi.org/10.1093/mnras/staa3150}{\emph{\mnras} {\bfseries 499} (2020) 4534} [\href{https://arxiv.org/abs/2008.04322}{{\ttfamily 2008.04322}}].

\bibitem{Sun2024}
G.~{Sun}, J.B.~{Mu{\~n}oz}, J.~{Mirocha} and C.-A.~{Faucher-Gigu{\`e}re}, \emph{{Constraining bursty star formation histories with galaxy UV and H$\alpha$ luminosity functions and clustering}}, \href{https://doi.org/10.48550/arXiv.2410.21409}{\emph{arXiv e-prints} (2024) arXiv:2410.21409} [\href{https://arxiv.org/abs/2410.21409}{{\ttfamily 2410.21409}}].

\bibitem{Mo1996}
H.J.~{Mo} and S.D.M.~{White}, \emph{{An analytic model for the spatial clustering of dark matter haloes}}, \href{https://doi.org/10.1093/mnras/282.2.347}{\emph{\mnras} {\bfseries 282} (1996) 347} [\href{https://arxiv.org/abs/astro-ph/9512127}{{\ttfamily astro-ph/9512127}}].

\bibitem{ST99}
R.K.~{Sheth} and G.~{Tormen}, \emph{{Large-scale bias and the peak background split}}, \href{https://doi.org/10.1046/j.1365-8711.1999.02692.x}{\emph{\mnras} {\bfseries 308} (1999) 119} [\href{https://arxiv.org/abs/astro-ph/9901122}{{\ttfamily astro-ph/9901122}}].

\bibitem{Jenkins2001}
A.~{Jenkins}, C.S.~{Frenk}, S.D.M.~{White}, J.M.~{Colberg}, S.~{Cole}, A.E.~{Evrard} et~al., \emph{{The mass function of dark matter haloes}}, \href{https://doi.org/10.1046/j.1365-8711.2001.04029.x}{\emph{\mnras} {\bfseries 321} (2001) 372} [\href{https://arxiv.org/abs/astro-ph/0005260}{{\ttfamily astro-ph/0005260}}].

\bibitem{Tinker2010}
J.L.~{Tinker}, B.E.~{Robertson}, A.V.~{Kravtsov}, A.~{Klypin}, M.S.~{Warren}, G.~{Yepes} et~al., \emph{{The Large-scale Bias of Dark Matter Halos: Numerical Calibration and Model Tests}}, \href{https://doi.org/10.1088/0004-637X/724/2/878}{\emph{\apj} {\bfseries 724} (2010) 878} [\href{https://arxiv.org/abs/1001.3162}{{\ttfamily 1001.3162}}].

\bibitem{Park2016}
J.~{Park}, H.-S.~{Kim}, J.S.B.~{Wyithe}, C.G.~{Lacey}, C.M.~{Baugh}, R.L.~{Barone-Nugent} et~al., \emph{{The clustering and halo occupation distribution of Lyman-break galaxies at z {\ensuremath{\sim}} 4}}, \href{https://doi.org/10.1093/mnras/stw1316}{\emph{\mnras} {\bfseries 461} (2016) 176} [\href{https://arxiv.org/abs/1511.01983}{{\ttfamily 1511.01983}}].

\bibitem{Hatfield2018}
P.W.~{Hatfield}, R.A.A.~{Bowler}, M.J.~{Jarvis} and C.L.~{Hale}, \emph{{The environment and host haloes of the brightest z {\ensuremath{\sim}} 6 Lyman-break galaxies}}, \href{https://doi.org/10.1093/mnras/sty856}{\emph{\mnras} {\bfseries 477} (2018) 3760} [\href{https://arxiv.org/abs/1702.03309}{{\ttfamily 1702.03309}}].

\bibitem{Harikane2016}
Y.~{Harikane}, M.~{Ouchi}, Y.~{Ono}, S.~{More}, S.~{Saito}, Y.-T.~{Lin} et~al., \emph{{Evolution of Stellar-to-Halo Mass Ratio at z = 0 - 7 Identified by Clustering Analysis with the Hubble Legacy Imaging and Early Subaru/Hyper Suprime-Cam Survey Data}}, \href{https://doi.org/10.3847/0004-637X/821/2/123}{\emph{\apj} {\bfseries 821} (2016) 123} [\href{https://arxiv.org/abs/1511.07873}{{\ttfamily 1511.07873}}].

\bibitem{Jose2013}
C.~{Jose}, K.~{Subramanian}, R.~{Srianand} and S.~{Samui}, \emph{{Spatial clustering of high-redshift Lyman-break galaxies}}, \href{https://doi.org/10.1093/mnras/sts503}{\emph{\mnras} {\bfseries 429} (2013) 2333} [\href{https://arxiv.org/abs/1208.2097}{{\ttfamily 1208.2097}}].

\bibitem{Bielby2013}
R.~{Bielby}, M.D.~{Hill}, T.~{Shanks}, N.H.M.~{Crighton}, L.~{Infante}, C.G.~{Bornancini} et~al., \emph{{The VLT LBG Redshift Survey - III. The clustering and dynamics of Lyman-break galaxies at z {\ensuremath{\sim}} 3}}, \href{https://doi.org/10.1093/mnras/sts639}{\emph{\mnras} {\bfseries 430} (2013) 425} [\href{https://arxiv.org/abs/1204.3635}{{\ttfamily 1204.3635}}].

\bibitem{Barone-Nugent2014}
R.L.~{Barone-Nugent}, M.~{Trenti}, J.S.B.~{Wyithe}, R.J.~{Bouwens}, P.A.~{Oesch}, G.D.~{Illingworth} et~al., \emph{{Measurement of Galaxy Clustering at z \raisebox{-0.5ex}\textasciitilde 7.2 and the Evolution of Galaxy Bias from 3.8 < z < 8 in the XDF, GOODS-S, and GOODS-N}}, \href{https://doi.org/10.1088/0004-637X/793/1/17}{\emph{\apj} {\bfseries 793} (2014) 17} [\href{https://arxiv.org/abs/1407.7316}{{\ttfamily 1407.7316}}].

\bibitem{Qiu2018}
Y.~{Qiu}, J.S.B.~{Wyithe}, P.A.~{Oesch}, S.J.~{Mutch}, Y.~{Qin}, I.~{Labb{\'e}} et~al., \emph{{Dependence of galaxy clustering on UV luminosity and stellar mass at z {\ensuremath{\sim}} 4-7}}, \href{https://doi.org/10.1093/mnras/sty2633}{\emph{\mnras} {\bfseries 481} (2018) 4885} [\href{https://arxiv.org/abs/1809.10161}{{\ttfamily 1809.10161}}].

\bibitem{DalmassoHST}
N.~{Dalmasso}, M.~{Trenti} and N.~{Leethochawalit}, \emph{{Galaxy clustering measurements out to redshift z {\ensuremath{\sim}} 8 from Hubble Legacy Fields}}, \href{https://doi.org/10.1093/mnras/stad3901}{\emph{\mnras} {\bfseries 528} (2024) 898} [\href{https://arxiv.org/abs/2312.12329}{{\ttfamily 2312.12329}}].

\bibitem{Shuntov2025}
M.~{Shuntov}, P.A.~{Oesch}, S.~{Toft}, R.A.~{Meyer}, A.~{Covelo-Paz}, L.~{Paquereau} et~al., \emph{{Constraints on the early Universe star formation efficiency from galaxy clustering and halo modeling of H{\ensuremath{\alpha}} and [O III] emitters}}, \href{https://doi.org/10.1051/0004-6361/202554618}{\emph{\aap} {\bfseries 699} (2025) A231} [\href{https://arxiv.org/abs/2503.14280}{{\ttfamily 2503.14280}}].

\bibitem{Paquereau2025}
L.~{Paquereau}, C.~{Laigle}, H.J.~{McCracken}, M.~{Shuntov}, O.~{Ilbert}, H.B.~{Akins} et~al., \emph{{Tracing the galaxy-halo connection with galaxy clustering in COSMOS-Web from z = 0.1 to z \raisebox{-0.5ex}\textasciitilde 12}}, \href{https://doi.org/10.48550/arXiv.2501.11674}{\emph{arXiv e-prints} (2025) arXiv:2501.11674} [\href{https://arxiv.org/abs/2501.11674}{{\ttfamily 2501.11674}}].

\bibitem{DalmassoJWST}
N.~{Dalmasso}, N.~{Leethochawalit}, M.~{Trenti} and K.~{Boyett}, \emph{{Galaxy clustering at cosmic dawn from JWST/NIRCam observations to redshift z 11}}, \href{https://doi.org/10.1093/mnras/stae2006}{\emph{\mnras} {\bfseries 533} (2024) 2391} [\href{https://arxiv.org/abs/2402.18052}{{\ttfamily 2402.18052}}].

\bibitem{Dekel2023}
A.~{Dekel}, K.S.~{Sarkar}, Y.~{Birnboim}, N.~{Mandelker} and Z.~{Li}, \emph{{Efficient Formation of Massive Galaxies at Cosmic Dawn by Feedback-Free Starbursts}}, \href{https://doi.org/10.48550/arXiv.2303.04827}{\emph{arXiv e-prints} (2023) arXiv:2303.04827} [\href{https://arxiv.org/abs/2303.04827}{{\ttfamily 2303.04827}}].

\bibitem{Renzini2023}
A.~{Renzini}, \emph{{A transient overcooling in the early Universe? Clues from globular clusters formation}}, \href{https://doi.org/10.1093/mnrasl/slad091}{\emph{\mnras} {\bfseries 525} (2023) L117} [\href{https://arxiv.org/abs/2305.14476}{{\ttfamily 2305.14476}}].

\bibitem{Shen2023}
X.~{Shen}, M.~{Vogelsberger}, M.~{Boylan-Kolchin}, S.~{Tacchella} and R.~{Kannan}, \emph{{The impact of UV variability on the abundance of bright galaxies at $z \geq 9$}}, \href{https://doi.org/10.48550/arXiv.2305.05679}{\emph{arXiv e-prints} (2023) arXiv:2305.05679} [\href{https://arxiv.org/abs/2305.05679}{{\ttfamily 2305.05679}}].

\bibitem{Pallottini2023}
A.~{Pallottini} and A.~{Ferrara}, \emph{{Stochastic star formation in early galaxies: JWST implications}}, \href{https://doi.org/10.48550/arXiv.2307.03219}{\emph{arXiv e-prints} (2023) arXiv:2307.03219} [\href{https://arxiv.org/abs/2307.03219}{{\ttfamily 2307.03219}}].

\bibitem{Inayoshi2022}
K.~{Inayoshi}, Y.~{Harikane}, A.K.~{Inoue}, W.~{Li} and L.C.~{Ho}, \emph{{A Lower Bound of Star Formation Activity in Ultra-high-redshift Galaxies Detected with JWST: Implications for Stellar Populations and Radiation Sources}}, \href{https://doi.org/10.3847/2041-8213/ac9310}{\emph{\apjl} {\bfseries 938} (2022) L10} [\href{https://arxiv.org/abs/2208.06872}{{\ttfamily 2208.06872}}].

\bibitem{Chakraborty2024}
A.~{Chakraborty} and T.R.~{Choudhury}, \emph{{Modelling the star-formation activity and ionizing properties of high-redshift galaxies}}, \href{https://doi.org/10.1088/1475-7516/2024/07/078}{\emph{\jcap} {\bfseries 2024} (2024) 078} [\href{https://arxiv.org/abs/2404.02879}{{\ttfamily 2404.02879}}].

\bibitem{Hutter2024}
A.~{Hutter}, E.R.~{Cueto}, P.~{Dayal}, S.~{Gottl{\"o}ber}, M.~{Trebitsch} and G.~{Yepes}, \emph{{Astraeus X: Indications of a top-heavy initial mass function in highly star-forming galaxies from JWST observations at z>10}}, \href{https://doi.org/10.48550/arXiv.2410.00730}{\emph{arXiv e-prints} (2024) arXiv:2410.00730} [\href{https://arxiv.org/abs/2410.00730}{{\ttfamily 2410.00730}}].

\bibitem{Planck2014}
{Planck Collaboration}, P.A.R.~{Ade}, N.~{Aghanim}, C.~{Armitage-Caplan}, M.~{Arnaud}, M.~{Ashdown} et~al., \emph{{Planck 2013 results. XVI. Cosmological parameters}}, \href{https://doi.org/10.1051/0004-6361/201321591}{\emph{\aap} {\bfseries 571} (2014) A16} [\href{https://arxiv.org/abs/1303.5076}{{\ttfamily 1303.5076}}].

\bibitem{Sobacchi&Mesinger_1Dsims_2013}
E.~{Sobacchi} and A.~{Mesinger}, \emph{{The depletion of gas in high-redshift dwarf galaxies from an inhomogeneous reionization.}}, \href{https://doi.org/10.1093/mnrasl/slt035}{\emph{\mnras} {\bfseries 432} (2013) L51} [\href{https://arxiv.org/abs/1301.6776}{{\ttfamily 1301.6776}}].

\bibitem{Choudhury&Dayal2019}
T.R.~{Choudhury} and P.~{Dayal}, \emph{{Probing the fluctuating ultraviolet background using the Hubble Frontier Fields}}, \href{https://doi.org/10.1093/mnrasl/sly186}{\emph{\mnras} {\bfseries 482} (2019) L19} [\href{https://arxiv.org/abs/1809.01798}{{\ttfamily 1809.01798}}].

\bibitem{Hutter2021}
A.~{Hutter}, P.~{Dayal}, G.~{Yepes}, S.~{Gottl{\"o}ber}, L.~{Legrand} and G.~{Ucci}, \emph{{Astraeus I: the interplay between galaxy formation and reionization}}, \href{https://doi.org/10.1093/mnras/stab602}{\emph{\mnras} {\bfseries 503} (2021) 3698} [\href{https://arxiv.org/abs/2004.08401}{{\ttfamily 2004.08401}}].

\bibitem{Trebitsch2022}
M.~{Trebitsch}, P.~{Dayal}, J.~{Chisholm}, S.L.~{Finkelstein}, A.~{Jaskot}, S.~{Flury} et~al., \emph{{Reionization with star-forming galaxies: insights from the Low-z Lyman Continuum Survey}}, \href{https://doi.org/10.48550/arXiv.2212.06177}{\emph{arXiv e-prints} (2022) arXiv:2212.06177} [\href{https://arxiv.org/abs/2212.06177}{{\ttfamily 2212.06177}}].

\bibitem{Ma2018}
X.~{Ma}, P.F.~{Hopkins}, S.~{Garrison-Kimmel}, C.-A.~{Faucher-Gigu{\`e}re}, E.~{Quataert}, M.~{Boylan-Kolchin} et~al., \emph{{Simulating galaxies in the reionization era with FIRE-2: galaxy scaling relations, stellar mass functions, and luminosity functions}}, \href{https://doi.org/10.1093/mnras/sty1024}{\emph{\mnras} {\bfseries 478} (2018) 1694} [\href{https://arxiv.org/abs/1706.06605}{{\ttfamily 1706.06605}}].

\bibitem{Behroozi19}
P.~{Behroozi}, R.H.~{Wechsler}, A.P.~{Hearin} and C.~{Conroy}, \emph{{UNIVERSEMACHINE: The correlation between galaxy growth and dark matter halo assembly from z = 0-10}}, \href{https://doi.org/10.1093/mnras/stz1182}{\emph{\mnras} {\bfseries 488} (2019) 3143} [\href{https://arxiv.org/abs/1806.07893}{{\ttfamily 1806.07893}}].

\bibitem{Zhu2020}
H.~{Zhu}, C.~{Avestruz} and N.Y.~{Gnedin}, \emph{{Cosmic Reionization On Computers: The Galaxy-Halo Connection between 5 {\ensuremath{\leq}} z {\ensuremath{\leq}} 10}}, \href{https://doi.org/10.3847/1538-4357/aba26d}{\emph{\apj} {\bfseries 899} (2020) 137} [\href{https://arxiv.org/abs/2001.02233}{{\ttfamily 2001.02233}}].

\bibitem{Stefanon2021}
M.~{Stefanon}, R.J.~{Bouwens}, I.~{Labb{\'e}}, G.D.~{Illingworth}, V.~{Gonzalez} and P.A.~{Oesch}, \emph{{Galaxy Stellar Mass Functions from z 10 to z 6 using the Deepest Spitzer/Infrared Array Camera Data: No Significant Evolution in the Stellar-to-halo Mass Ratio of Galaxies in the First Gigayear of Cosmic Time}}, \href{https://doi.org/10.3847/1538-4357/ac1bb6}{\emph{\apj} {\bfseries 922} (2021) 29} [\href{https://arxiv.org/abs/2103.16571}{{\ttfamily 2103.16571}}].

\bibitem{Kannan2022}
R.~{Kannan}, E.~{Garaldi}, A.~{Smith}, R.~{Pakmor}, V.~{Springel}, M.~{Vogelsberger} et~al., \emph{{Introducing the THESAN project: radiation-magnetohydrodynamic simulations of the epoch of reionization}}, \href{https://doi.org/10.1093/mnras/stab3710}{\emph{\mnras} {\bfseries 511} (2022) 4005} [\href{https://arxiv.org/abs/2110.00584}{{\ttfamily 2110.00584}}].

\bibitem{DiCesare2023}
C.~{Di Cesare}, L.~{Graziani}, R.~{Schneider}, M.~{Ginolfi}, A.~{Venditti}, P.~{Santini} et~al., \emph{{The assembly of dusty galaxies at z {\ensuremath{\geq}} 4: the build-up of stellar mass and its scaling relations with hints from early JWST data}}, \href{https://doi.org/10.1093/mnras/stac3702}{\emph{\mnras} {\bfseries 519} (2023) 4632} [\href{https://arxiv.org/abs/2209.05496}{{\ttfamily 2209.05496}}].

\bibitem{Shuntov2025_SHMR}
M.~{Shuntov}, O.~{Ilbert}, S.~{Toft}, R.C.~{Arango-Toro}, H.B.~{Akins}, C.M.~{Casey} et~al., \emph{{COSMOS-Web: Stellar mass assembly in relation to dark matter halos across 0.2 < z < 12 of cosmic history}}, \href{https://doi.org/10.1051/0004-6361/202452570}{\emph{\aap} {\bfseries 695} (2025) A20} [\href{https://arxiv.org/abs/2410.08290}{{\ttfamily 2410.08290}}].

\bibitem{Furlanetto2017}
S.R.~{Furlanetto}, J.~{Mirocha}, R.H.~{Mebane} and G.~{Sun}, \emph{{A minimalist feedback-regulated model for galaxy formation during the epoch of reionization}}, \href{https://doi.org/10.1093/mnras/stx2132}{\emph{\mnras} {\bfseries 472} (2017) 1576} [\href{https://arxiv.org/abs/1611.01169}{{\ttfamily 1611.01169}}].

\bibitem{Oke_ABmag}
J.B.~{Oke}, \emph{{Absolute Spectral Energy Distributions for White Dwarfs}}, \href{https://doi.org/10.1086/190287}{\emph{\apjs} {\bfseries 27} (1974) 21}.

\bibitem{Oke&Gunn_ABmag}
J.B.~{Oke} and J.E.~{Gunn}, \emph{{Secondary standard stars for absolute spectrophotometry.}}, \href{https://doi.org/10.1086/160817}{\emph{\apj} {\bfseries 266} (1983) 713}.

\bibitem{Madau2024}
P.~{Madau}, E.~{Giallongo}, A.~{Grazian} and F.~{Haardt}, \emph{{Cosmic Reionization in the JWST Era: Back to AGNs?}}, \href{https://doi.org/10.3847/1538-4357/ad5ce8}{\emph{\apj} {\bfseries 971} (2024) 75} [\href{https://arxiv.org/abs/2406.18697}{{\ttfamily 2406.18697}}].

\bibitem{Grazian2024}
A.~{Grazian}, E.~{Giallongo}, K.~{Boutsia}, S.~{Cristiani}, F.~{Fontanot}, M.~{Bischetti} et~al., \emph{{What Are the Pillars of Reionization? Revising the AGN Luminosity Function at z {\ensuremath{\sim}} 5}}, \href{https://doi.org/10.3847/1538-4357/ad6980}{\emph{\apj} {\bfseries 974} (2024) 84} [\href{https://arxiv.org/abs/2407.20861}{{\ttfamily 2407.20861}}].

\bibitem{Dayal2025}
P.~{Dayal}, M.~{Volonteri}, J.E.~{Greene}, V.~{Kokorev}, A.D.~{Goulding}, C.C.~{Williams} et~al., \emph{{UNCOVERing the contribution of black holes to reionization}}, \href{https://doi.org/10.1051/0004-6361/202449331}{\emph{\aap} {\bfseries 697} (2025) A211} [\href{https://arxiv.org/abs/2401.11242}{{\ttfamily 2401.11242}}].

\bibitem{Asthana2024}
S.~{Asthana}, M.G.~{Haehnelt}, G.~{Kulkarni}, J.S.~{Bolton}, P.~{Gaikwad}, L.C.~{Keating} et~al., \emph{{The impact of faint AGN discovered by JWST on reionization}}, \href{https://doi.org/10.48550/arXiv.2409.15453}{\emph{arXiv e-prints} (2024) arXiv:2409.15453} [\href{https://arxiv.org/abs/2409.15453}{{\ttfamily 2409.15453}}].

\bibitem{Jiang2025}
D.~{Jiang}, L.~{Jiang}, S.~{Sun}, W.~{Liu} and S.~{Fu}, \emph{{Ruling out AGNs as the dominant source of cosmic reionization with JWST}}, \href{https://doi.org/10.48550/arXiv.2502.03683}{\emph{arXiv e-prints} (2025) arXiv:2502.03683} [\href{https://arxiv.org/abs/2502.03683}{{\ttfamily 2502.03683}}].

\bibitem{Nakajima2020}
K.~{Nakajima}, R.S.~{Ellis}, B.E.~{Robertson}, M.~{Tang} and D.P.~{Stark}, \emph{{The Lyman Continuum Escape Survey. II. Ionizing Radiation as a Function of the [O III]/[O II] Line Ratio}}, \href{https://doi.org/10.3847/1538-4357/ab6604}{\emph{\apj} {\bfseries 889} (2020) 161} [\href{https://arxiv.org/abs/1909.07396}{{\ttfamily 1909.07396}}].

\bibitem{Izotov2021}
Y.I.~{Izotov}, G.~{Worseck}, D.~{Schaerer}, N.G.~{Guseva}, J.~{Chisholm}, T.X.~{Thuan} et~al., \emph{{Lyman continuum leakage from low-mass galaxies with M$_{{\ensuremath{\star}}}$ < {}10$^{8}$ M$_{{\ensuremath{\odot}}}$}}, \href{https://doi.org/10.1093/mnras/stab612}{\emph{\mnras} {\bfseries 503} (2021) 1734} [\href{https://arxiv.org/abs/2103.01514}{{\ttfamily 2103.01514}}].

\bibitem{Flury2022}
S.R.~{Flury}, A.E.~{Jaskot}, H.C.~{Ferguson}, G.~{Worseck}, K.~{Makan}, J.~{Chisholm} et~al., \emph{{The Low-redshift Lyman Continuum Survey. II. New Insights into LyC Diagnostics}}, \href{https://doi.org/10.3847/1538-4357/ac61e4}{\emph{\apj} {\bfseries 930} (2022) 126} [\href{https://arxiv.org/abs/2203.15649}{{\ttfamily 2203.15649}}].

\bibitem{Begley2022}
R.~{Begley}, F.~{Cullen}, R.J.~{McLure}, J.S.~{Dunlop}, A.~{Hall}, A.C.~{Carnall} et~al., \emph{{The VANDELS survey: a measurement of the average Lyman-continuum escape fraction of star-forming galaxies at z = 3.5}}, \href{https://doi.org/10.1093/mnras/stac1067}{\emph{\mnras} {\bfseries 513} (2022) 3510} [\href{https://arxiv.org/abs/2202.04088}{{\ttfamily 2202.04088}}].

\bibitem{Saxena2022}
A.~{Saxena}, L.~{Pentericci}, R.S.~{Ellis}, L.~{Guaita}, A.~{Calabr{\`o}}, D.~{Schaerer} et~al., \emph{{No strong dependence of Lyman continuum leakage on physical properties of star-forming galaxies at {\ensuremath{\lesssim}} z {\ensuremath{\lesssim}} 3.5}}, \href{https://doi.org/10.1093/mnras/stab3728}{\emph{\mnras} {\bfseries 511} (2022) 120} [\href{https://arxiv.org/abs/2109.03662}{{\ttfamily 2109.03662}}].

\bibitem{Naidu2022}
R.P.~{Naidu}, P.A.~{Oesch}, P.~{van Dokkum}, E.J.~{Nelson}, K.A.~{Suess}, G.~{Brammer} et~al., \emph{{Two Remarkably Luminous Galaxy Candidates at z {\ensuremath{\approx}} 10-12 Revealed by JWST}}, \href{https://doi.org/10.3847/2041-8213/ac9b22}{\emph{\apjl} {\bfseries 940} (2022) L14} [\href{https://arxiv.org/abs/2207.09434}{{\ttfamily 2207.09434}}].

\bibitem{Chisholm2022}
J.~{Chisholm}, A.~{Saldana-Lopez}, S.~{Flury}, D.~{Schaerer}, A.~{Jaskot}, R.~{Amor{\'\i}n} et~al., \emph{{The far-ultraviolet continuum slope as a Lyman Continuum escape estimator at high redshift}}, \href{https://doi.org/10.1093/mnras/stac2874}{\emph{\mnras} {\bfseries 517} (2022) 5104} [\href{https://arxiv.org/abs/2207.05771}{{\ttfamily 2207.05771}}].

\bibitem{Pahl2023}
A.J.~{Pahl}, A.~{Shapley}, C.C.~{Steidel}, N.A.~{Reddy}, Y.~{Chen}, G.C.~{Rudie} et~al., \emph{{The connection between the escape of ionizing radiation and galaxy properties at z {\ensuremath{\sim}} 3 in the Keck Lyman continuum spectroscopic survey}}, \href{https://doi.org/10.1093/mnras/stad774}{\emph{\mnras} {\bfseries 521} (2023) 3247} [\href{https://arxiv.org/abs/2210.16697}{{\ttfamily 2210.16697}}].

\bibitem{Kimm2014}
T.~{Kimm} and R.~{Cen}, \emph{{Escape Fraction of Ionizing Photons during Reionization: Effects due to Supernova Feedback and Runaway OB Stars}}, \href{https://doi.org/10.1088/0004-637X/788/2/121}{\emph{\apj} {\bfseries 788} (2014) 121} [\href{https://arxiv.org/abs/1405.0552}{{\ttfamily 1405.0552}}].

\bibitem{Paardekooper2015}
J.-P.~{Paardekooper}, S.~{Khochfar} and C.~{Dalla Vecchia}, \emph{{The First Billion Years project: the escape fraction of ionizing photons in the epoch of reionization}}, \href{https://doi.org/10.1093/mnras/stv1114}{\emph{\mnras} {\bfseries 451} (2015) 2544} [\href{https://arxiv.org/abs/1501.01967}{{\ttfamily 1501.01967}}].

\bibitem{Kimm2017}
T.~{Kimm}, H.~{Katz}, M.~{Haehnelt}, J.~{Rosdahl}, J.~{Devriendt} and A.~{Slyz}, \emph{{Feedback-regulated star formation and escape of LyC photons from mini-haloes during reionization}}, \href{https://doi.org/10.1093/mnras/stx052}{\emph{\mnras} {\bfseries 466} (2017) 4826} [\href{https://arxiv.org/abs/1608.04762}{{\ttfamily 1608.04762}}].

\bibitem{Trebitsch2017}
M.~{Trebitsch}, J.~{Blaizot}, J.~{Rosdahl}, J.~{Devriendt} and A.~{Slyz}, \emph{{Fluctuating feedback-regulated escape fraction of ionizing radiation in low-mass, high-redshift galaxies}}, \href{https://doi.org/10.1093/mnras/stx1060}{\emph{\mnras} {\bfseries 470} (2017) 224} [\href{https://arxiv.org/abs/1705.00941}{{\ttfamily 1705.00941}}].

\bibitem{Lewis2020}
J.S.W.~{Lewis}, P.~{Ocvirk}, D.~{Aubert}, J.G.~{Sorce}, P.R.~{Shapiro}, N.~{Deparis} et~al., \emph{{Galactic ionizing photon budget during the epoch of reionization in the Cosmic Dawn II simulation}}, \href{https://doi.org/10.1093/mnras/staa1748}{\emph{\mnras} {\bfseries 496} (2020) 4342} [\href{https://arxiv.org/abs/2001.07785}{{\ttfamily 2001.07785}}].

\bibitem{Rosdahl2022}
J.~{Rosdahl}, J.~{Blaizot}, H.~{Katz}, T.~{Kimm}, T.~{Garel}, M.~{Haehnelt} et~al., \emph{{LyC escape from SPHINX galaxies in the Epoch of Reionization}}, \href{https://doi.org/10.1093/mnras/stac1942}{\emph{\mnras} {\bfseries 515} (2022) 2386} [\href{https://arxiv.org/abs/2207.03232}{{\ttfamily 2207.03232}}].

\bibitem{Yeh2023}
J.Y.C.~{Yeh}, A.~{Smith}, R.~{Kannan}, E.~{Garaldi}, M.~{Vogelsberger}, J.~{Borrow} et~al., \emph{{The THESAN project: ionizing escape fractions of reionization-era galaxies}}, \href{https://doi.org/10.1093/mnras/stad210}{\emph{\mnras} {\bfseries 520} (2023) 2757} [\href{https://arxiv.org/abs/2205.02238}{{\ttfamily 2205.02238}}].

\bibitem{Sharma2016}
M.~{Sharma}, T.~{Theuns}, C.~{Frenk}, R.~{Bower}, R.~{Crain}, M.~{Schaller} et~al., \emph{{The brighter galaxies reionized the Universe}}, \href{https://doi.org/10.1093/mnrasl/slw021}{\emph{\mnras} {\bfseries 458} (2016) L94} [\href{https://arxiv.org/abs/1512.04537}{{\ttfamily 1512.04537}}].

\bibitem{Naidu2020}
R.P.~{Naidu}, S.~{Tacchella}, C.A.~{Mason}, S.~{Bose}, P.A.~{Oesch} and C.~{Conroy}, \emph{{Rapid Reionization by the Oligarchs: The Case for Massive, UV-bright, Star-forming Galaxies with High Escape Fractions}}, \href{https://doi.org/10.3847/1538-4357/ab7cc9}{\emph{\apj} {\bfseries 892} (2020) 109} [\href{https://arxiv.org/abs/1907.13130}{{\ttfamily 1907.13130}}].

\bibitem{Xu2016}
H.~{Xu}, J.H.~{Wise}, M.L.~{Norman}, K.~{Ahn} and B.W.~{O'Shea}, \emph{{Galaxy Properties and UV Escape Fractions during the Epoch of Reionization: Results from the Renaissance Simulations}}, \href{https://doi.org/10.3847/1538-4357/833/1/84}{\emph{\apj} {\bfseries 833} (2016) 84} [\href{https://arxiv.org/abs/1604.07842}{{\ttfamily 1604.07842}}].

\bibitem{Mutch2023}
S.J.~{Mutch}, B.~{Greig}, Y.~{Qin}, G.B.~{Poole} and J.S.B.~{Wyithe}, \emph{{Dark-ages reionization and galaxy formation simulation -- XXI. Constraining the evolution of the ionizing escape fraction}}, \href{https://doi.org/10.48550/arXiv.2303.07378}{\emph{arXiv e-prints} (2023) arXiv:2303.07378} [\href{https://arxiv.org/abs/2303.07378}{{\ttfamily 2303.07378}}].

\bibitem{Ma2020}
X.~{Ma}, E.~{Quataert}, A.~{Wetzel}, P.F.~{Hopkins}, C.-A.~{Faucher-Gigu{\`e}re} and D.~{Kere{\v{s}}}, \emph{{No missing photons for reionization: moderate ionizing photon escape fractions from the FIRE-2 simulations}}, \href{https://doi.org/10.1093/mnras/staa2404}{\emph{\mnras} {\bfseries 498} (2020) 2001} [\href{https://arxiv.org/abs/2003.05945}{{\ttfamily 2003.05945}}].

\bibitem{Kostyuk2023}
I.~{Kostyuk}, D.~{Nelson}, B.~{Ciardi}, M.~{Glatzle} and A.~{Pillepich}, \emph{{Ionizing photon production and escape fractions during cosmic reionization in the TNG50 simulation}}, \href{https://doi.org/10.1093/mnras/stad677}{\emph{\mnras} {\bfseries 521} (2023) 3077} [\href{https://arxiv.org/abs/2207.11278}{{\ttfamily 2207.11278}}].

\bibitem{Dayal2017}
P.~{Dayal}, T.R.~{Choudhury}, V.~{Bromm} and F.~{Pacucci}, \emph{{Reionization and Galaxy Formation in Warm Dark Matter Cosmologies}}, \href{https://doi.org/10.3847/1538-4357/836/1/16}{\emph{\apj} {\bfseries 836} (2017) 16} [\href{https://arxiv.org/abs/1501.02823}{{\ttfamily 1501.02823}}].

\bibitem{Starburst99}
C.~{Leitherer}, D.~{Schaerer}, J.D.~{Goldader}, R.M.G.~{Delgado}, C.~{Robert}, D.F.~{Kune} et~al., \emph{{Starburst99: Synthesis Models for Galaxies with Active Star Formation}}, \href{https://doi.org/10.1086/313233}{\emph{\apjs} {\bfseries 123} (1999) 3} [\href{https://arxiv.org/abs/astro-ph/9902334}{{\ttfamily astro-ph/9902334}}].

\bibitem{Robertson2013}
B.E.~{Robertson}, S.R.~{Furlanetto}, E.~{Schneider}, S.~{Charlot}, R.S.~{Ellis}, D.P.~{Stark} et~al., \emph{{New Constraints on Cosmic Reionization from the 2012 Hubble Ultra Deep Field Campaign}}, \href{https://doi.org/10.1088/0004-637X/768/1/71}{\emph{\apj} {\bfseries 768} (2013) 71} [\href{https://arxiv.org/abs/1301.1228}{{\ttfamily 1301.1228}}].

\bibitem{Robertson2015}
B.E.~{Robertson}, R.S.~{Ellis}, S.R.~{Furlanetto} and J.S.~{Dunlop}, \emph{{Cosmic Reionization and Early Star-forming Galaxies: A Joint Analysis of New Constraints from Planck and the Hubble Space Telescope}}, \href{https://doi.org/10.1088/2041-8205/802/2/L19}{\emph{\apjl} {\bfseries 802} (2015) L19} [\href{https://arxiv.org/abs/1502.02024}{{\ttfamily 1502.02024}}].

\bibitem{Bouwens2016_xi_ion}
R.J.~{Bouwens}, R.~{Smit}, I.~{Labb{\'e}}, M.~{Franx}, J.~{Caruana}, P.~{Oesch} et~al., \emph{{The Lyman-Continuum Photon Production Efficiency {\ensuremath{\xi}} $_{ion}$ of z {\ensuremath{\sim}} 4-5 Galaxies from IRAC-based H{\ensuremath{\alpha}} Measurements: Implications for the Escape Fraction and Cosmic Reionization}}, \href{https://doi.org/10.3847/0004-637X/831/2/176}{\emph{\apj} {\bfseries 831} (2016) 176} [\href{https://arxiv.org/abs/1511.08504}{{\ttfamily 1511.08504}}].

\bibitem{Atek2024_Spectroscopy}
H.~{Atek}, I.~{Labb{\'e}}, L.J.~{Furtak}, I.~{Chemerynska}, S.~{Fujimoto}, D.J.~{Setton} et~al., \emph{{Most of the photons that reionized the Universe came from dwarf galaxies}}, \href{https://doi.org/10.1038/s41586-024-07043-6}{\emph{\nat} {\bfseries 626} (2024) 975}.

\bibitem{Simmonds2024}
C.~{Simmonds}, S.~{Tacchella}, K.~{Hainline}, B.D.~{Johnson}, W.~{McClymont}, B.~{Robertson} et~al., \emph{{Low-mass bursty galaxies in JADES efficiently produce ionizing photons and could represent the main drivers of reionization}}, \href{https://doi.org/10.1093/mnras/stad3605}{\emph{\mnras} {\bfseries 527} (2024) 6139} [\href{https://arxiv.org/abs/2310.01112}{{\ttfamily 2310.01112}}].

\bibitem{Simmonds2024_JADES_xiION}
C.~{Simmonds}, S.~{Tacchella}, K.~{Hainline}, B.D.~{Johnson}, D.~{Pusk{\'a}s}, B.~{Robertson} et~al., \emph{{Ionizing properties of galaxies in JADES for a stellar mass complete sample: resolving the cosmic ionizing photon budget crisis at the Epoch of Reionization}}, \href{https://doi.org/10.1093/mnras/stae2537}{\emph{\mnras} {\bfseries 535} (2024) 2998} [\href{https://arxiv.org/abs/2409.01286}{{\ttfamily 2409.01286}}].

\bibitem{Pahl2025_xiION}
A.~{Pahl}, M.W.~{Topping}, A.~{Shapley}, R.~{Sanders}, N.A.~{Reddy}, L.~{Clarke} et~al., \emph{{A Spectroscopic Analysis of the Ionizing Photon Production Efficiency in JADES and CEERS: Implications for the Ionizing Photon Budget}}, \href{https://doi.org/10.3847/1538-4357/adb1ab}{\emph{\apj} {\bfseries 981} (2025) 134} [\href{https://arxiv.org/abs/2407.03399}{{\ttfamily 2407.03399}}].

\bibitem{Begley2025_xiION}
R.~{Begley}, R.J.~{McLure}, F.~{Cullen}, D.J.~{McLeod}, J.S.~{Dunlop}, A.C.~{Carnall} et~al., \emph{{The evolution of [O III] + H{\ensuremath{\beta}} equivalent width from z = 3-8: implications for the production and escape of ionizing photons during reionization}}, \href{https://doi.org/10.1093/mnras/staf211}{\emph{\mnras} {\bfseries 537} (2025) 3245} [\href{https://arxiv.org/abs/2410.10988}{{\ttfamily 2410.10988}}].

\bibitem{Llerena2025_xiION}
M.~{Llerena}, L.~{Pentericci}, L.~{Napolitano}, S.~{Mascia}, R.~{Amor{\'\i}n}, A.~{Calabr{\`o}} et~al., \emph{{The ionizing photon production efficiency of star-forming galaxies at z {\ensuremath{\sim}} 4{\textendash}10}}, \href{https://doi.org/10.1051/0004-6361/202453251}{\emph{\aap} {\bfseries 698} (2025) A302} [\href{https://arxiv.org/abs/2412.01358}{{\ttfamily 2412.01358}}].

\bibitem{Nikolic2024}
I.~{Nikoli{\'c}}, A.~{Mesinger}, J.E.~{Davies} and D.~{Prelogovi{\'c}}, \emph{{The importance of stochasticity in determining galaxy emissivities and UV LFs during cosmic dawn and reionization}}, \href{https://doi.org/10.1051/0004-6361/202451213}{\emph{\aap} {\bfseries 692} (2024) A142} [\href{https://arxiv.org/abs/2406.15237}{{\ttfamily 2406.15237}}].

\bibitem{Planck2020}
{Planck Collaboration}, N.~{Aghanim}, Y.~{Akrami}, M.~{Ashdown}, J.~{Aumont}, C.~{Baccigalupi} et~al., \emph{{Planck 2018 results. VI. Cosmological parameters}}, \href{https://doi.org/10.1051/0004-6361/201833910}{\emph{\aap} {\bfseries 641} (2020) A6} [\href{https://arxiv.org/abs/1807.06209}{{\ttfamily 1807.06209}}].

\bibitem{Bouwens2021}
R.J.~{Bouwens}, P.A.~{Oesch}, M.~{Stefanon}, G.~{Illingworth}, I.~{Labb{\'e}}, N.~{Reddy} et~al., \emph{{New Determinations of the UV Luminosity Functions from z 9 to 2 Show a Remarkable Consistency with Halo Growth and a Constant Star Formation Efficiency}}, \href{https://doi.org/10.3847/1538-3881/abf83e}{\emph{\aj} {\bfseries 162} (2021) 47} [\href{https://arxiv.org/abs/2102.07775}{{\ttfamily 2102.07775}}].

\bibitem{Mauerhofer&Dayal2023}
V.~{Mauerhofer} and P.~{Dayal}, \emph{{The dust enrichment of early galaxies in the JWST and ALMA era}}, \href{https://doi.org/10.1093/mnras/stad2734}{\emph{\mnras} {\bfseries 526} (2023) 2196} [\href{https://arxiv.org/abs/2305.01681}{{\ttfamily 2305.01681}}].

\bibitem{cobaya}
J.~{Torrado} and A.~{Lewis}, \emph{{Cobaya: code for Bayesian analysis of hierarchical physical models}}, \href{https://doi.org/10.1088/1475-7516/2021/05/057}{\emph{\jcap} {\bfseries 2021} (2021) 057} [\href{https://arxiv.org/abs/2005.05290}{{\ttfamily 2005.05290}}].

\bibitem{Davies2018}
F.B.~{Davies}, J.F.~{Hennawi}, E.~{Ba{\~n}ados}, Z.~{Luki{\'c}}, R.~{Decarli}, X.~{Fan} et~al., \emph{{Quantitative Constraints on the Reionization History from the IGM Damping Wing Signature in Two Quasars at z > 7}}, \href{https://doi.org/10.3847/1538-4357/aad6dc}{\emph{\apj} {\bfseries 864} (2018) 142} [\href{https://arxiv.org/abs/1802.06066}{{\ttfamily 1802.06066}}].

\bibitem{Greig2022}
B.~{Greig}, A.~{Mesinger}, F.B.~{Davies}, F.~{Wang}, J.~{Yang} and J.F.~{Hennawi}, \emph{{IGM damping wing constraints on reionization from covariance reconstruction of two z {\ensuremath{\gtrsim}} 7 QSOs}}, \href{https://doi.org/10.1093/mnras/stac825}{\emph{\mnras} {\bfseries 512} (2022) 5390} [\href{https://arxiv.org/abs/2112.04091}{{\ttfamily 2112.04091}}].

\bibitem{Gaikwad2023}
P.~{Gaikwad}, M.G.~{Haehnelt}, F.B.~{Davies}, S.E.I.~{Bosman}, M.~{Molaro}, G.~{Kulkarni} et~al., \emph{{Measuring the photoionization rate, neutral fraction, and mean free path of H I ionizing photons at 4.9 {\ensuremath{\leq}} z {\ensuremath{\leq}} 6.0 from a large sample of XShooter and ESI spectra}}, \href{https://doi.org/10.1093/mnras/stad2566}{\emph{\mnras} {\bfseries 525} (2023) 4093} [\href{https://arxiv.org/abs/2304.02038}{{\ttfamily 2304.02038}}].

\bibitem{Umeda2023}
H.~{Umeda}, M.~{Ouchi}, K.~{Nakajima}, Y.~{Harikane}, Y.~{Ono}, Y.~{Xu} et~al., \emph{{JWST Measurements of Neutral Hydrogen Fractions and Ionized Bubble Sizes at $z=7-12$ Obtained with Ly$\alpha$ Damping Wing Absorptions in 26 Bright Continuum Galaxies}}, \href{https://doi.org/10.48550/arXiv.2306.00487}{\emph{arXiv e-prints} (2023) arXiv:2306.00487} [\href{https://arxiv.org/abs/2306.00487}{{\ttfamily 2306.00487}}].

\bibitem{Durovcikova2024}
D.~{{\v{D}}urov{\v{c}}{\'\i}kov{\'a}}, A.-C.~{Eilers}, H.~{Chen}, S.~{Satyavolu}, G.~{Kulkarni}, R.A.~{Simcoe} et~al., \emph{{Chronicling the reionization history at $6\lesssim z \lesssim 7$ with emergent quasar damping wings}}, \href{https://doi.org/10.48550/arXiv.2401.10328}{\emph{arXiv e-prints} (2024) arXiv:2401.10328} [\href{https://arxiv.org/abs/2401.10328}{{\ttfamily 2401.10328}}].

\bibitem{Benson2000}
A.J.~{Benson}, S.~{Cole}, C.S.~{Frenk}, C.M.~{Baugh} and C.G.~{Lacey}, \emph{{The nature of galaxy bias and clustering}}, \href{https://doi.org/10.1046/j.1365-8711.2000.03101.x}{\emph{\mnras} {\bfseries 311} (2000) 793} [\href{https://arxiv.org/abs/astro-ph/9903343}{{\ttfamily astro-ph/9903343}}].

\bibitem{Coil2013}
A.L.~{Coil}, \emph{{The Large-Scale Structure of the Universe}},  in \emph{Planets, Stars and Stellar Systems. Volume 6: Extragalactic Astronomy and Cosmology}, T.D.~{Oswalt} and W.C.~{Keel}, eds., vol.~6, p.~387 (2013), \href{https://doi.org/10.1007/978-94-007-5609-0_8}{DOI}.

\bibitem{Somerville2025}
R.S.~{Somerville}, L.Y.A.~{Yung}, L.~{Lancaster}, S.~{Menon}, L.~{Sommovigo} and S.L.~{Finkelstein}, \emph{{Density modulated star formation efficiency: implications for the observed abundance of ultra-violet luminous galaxies at z>10}}, \href{https://doi.org/10.48550/arXiv.2505.05442}{\emph{arXiv e-prints} (2025) arXiv:2505.05442} [\href{https://arxiv.org/abs/2505.05442}{{\ttfamily 2505.05442}}].

\bibitem{Maitra2025}
S.~{Maitra}, G.~{Kulkarni}, S.~{Asthana}, J.S.~{Bolton}, M.G.~{Haehnelt} and L.~{Keating}, \emph{{The Lyman {\ensuremath{\alpha}} emitter bispectrum as a probe of reionization morphology}}, \href{https://doi.org/10.1093/mnras/staf1262}{\emph{\mnras} {\bfseries 542} (2025) 486} [\href{https://arxiv.org/abs/2505.17188}{{\ttfamily 2505.17188}}].

\bibitem{Gelli2020}
V.~{Gelli}, S.~{Salvadori}, A.~{Pallottini} and A.~{Ferrara}, \emph{{The stellar populations of high-redshift dwarf galaxies}}, \href{https://doi.org/10.1093/mnras/staa2410}{\emph{\mnras} {\bfseries 498} (2020) 4134} [\href{https://arxiv.org/abs/2009.03912}{{\ttfamily 2009.03912}}].

\bibitem{Furlanetto2022}
S.R.~{Furlanetto} and J.~{Mirocha}, \emph{{Bursty star formation during the Cosmic Dawn driven by delayed stellar feedback}}, \href{https://doi.org/10.1093/mnras/stac310}{\emph{\mnras} {\bfseries 511} (2022) 3895} [\href{https://arxiv.org/abs/2109.04488}{{\ttfamily 2109.04488}}].

\bibitem{Hopkins2023}
P.F.~{Hopkins}, A.B.~{Gurvich}, X.~{Shen}, Z.~{Hafen}, M.Y.~{Grudi{\'c}}, S.~{Kurinchi-Vendhan} et~al., \emph{{What causes the formation of discs and end of bursty star formation?}}, \href{https://doi.org/10.1093/mnras/stad1902}{\emph{\mnras} {\bfseries 525} (2023) 2241} [\href{https://arxiv.org/abs/2301.08263}{{\ttfamily 2301.08263}}].

\bibitem{Sun2023}
G.~{Sun}, C.-A.~{Faucher-Gigu{\`e}re}, C.C.~{Hayward}, X.~{Shen}, A.~{Wetzel} and R.K.~{Cochrane}, \emph{{Bursty Star Formation Naturally Explains the Abundance of Bright Galaxies at Cosmic Dawn}}, \href{https://doi.org/10.3847/2041-8213/acf85a}{\emph{\apjl} {\bfseries 955} (2023) L35} [\href{https://arxiv.org/abs/2307.15305}{{\ttfamily 2307.15305}}].

\bibitem{Endsley2023}
R.~{Endsley}, D.P.~{Stark}, L.~{Whitler}, M.W.~{Topping}, B.D.~{Johnson}, B.~{Robertson} et~al., \emph{{The Star-forming and Ionizing Properties of Dwarf z\raisebox{-0.5ex}\textasciitilde6-9 Galaxies in JADES: Insights on Bursty Star Formation and Ionized Bubble Growth}}, \href{https://doi.org/10.48550/arXiv.2306.05295}{\emph{arXiv e-prints} (2023) arXiv:2306.05295} [\href{https://arxiv.org/abs/2306.05295}{{\ttfamily 2306.05295}}].

\bibitem{Ciesla2024}
L.~{Ciesla}, D.~{Elbaz}, O.~{Ilbert}, V.~{Buat}, B.~{Magnelli}, D.~{Narayanan} et~al., \emph{{Identification of a transition from stochastic to secular star formation around z = 9 with JWST}}, \href{https://doi.org/10.1051/0004-6361/202348091}{\emph{\aap} {\bfseries 686} (2024) A128} [\href{https://arxiv.org/abs/2309.15720}{{\ttfamily 2309.15720}}].

\bibitem{Looser2025}
T.J.~{Looser}, F.~{D'Eugenio}, R.~{Maiolino}, S.~{Tacchella}, M.~{Curti}, S.~{Arribas} et~al., \emph{{JADES: Differing assembly histories of galaxies: Observational evidence for bursty star formation histories and (mini-)quenching in the first billion years of the Universe}}, \href{https://doi.org/10.1051/0004-6361/202347102}{\emph{\aap} {\bfseries 697} (2025) A88} [\href{https://arxiv.org/abs/2306.02470}{{\ttfamily 2306.02470}}].

\bibitem{Bosman2024}
S.E.I.~{Bosman} and F.B.~{Davies}, \emph{{A measurement of the escaping ionising efficiency of galaxies at redshift 5}}, \href{https://doi.org/10.1051/0004-6361/202451463}{\emph{\aap} {\bfseries 690} (2024) A391} [\href{https://arxiv.org/abs/2409.08315}{{\ttfamily 2409.08315}}].

\bibitem{Cain2025}
C.~{Cain}, A.~{D'Aloisio} and J.~{Mu{\~n}oz}, \emph{{New constraints on the galactic ionising efficiency and escape fraction at 2.5 < z < 6 based on quasar absorption spectra}}, \href{https://doi.org/10.1017/pasa.2025.10071}{\emph{\pasa} {\bfseries 42} (2025) e107} [\href{https://arxiv.org/abs/2503.08778}{{\ttfamily 2503.08778}}].

\bibitem{Munoz2024}
J.B.~{Mu{\~n}oz}, J.~{Mirocha}, J.~{Chisholm}, S.R.~{Furlanetto} and C.~{Mason}, \emph{{Reionization after JWST: a photon budget crisis?}}, \href{https://doi.org/10.48550/arXiv.2404.07250}{\emph{arXiv e-prints} (2024) arXiv:2404.07250} [\href{https://arxiv.org/abs/2404.07250}{{\ttfamily 2404.07250}}].

\bibitem{Faisst2016}
A.L.~{Faisst}, P.~{Capak}, B.C.~{Hsieh}, C.~{Laigle}, M.~{Salvato}, L.~{Tasca} et~al., \emph{{A Coherent Study of Emission Lines from Broadband Photometry: Specific Star Formation Rates and [O III]/H{\ensuremath{\beta}} Ratio at 3 > z > 6}}, \href{https://doi.org/10.3847/0004-637X/821/2/122}{\emph{\apj} {\bfseries 821} (2016) 122} [\href{https://arxiv.org/abs/1601.07173}{{\ttfamily 1601.07173}}].

\bibitem{FaucherGiguere2020}
C.-A.~{Faucher-Gigu{\`e}re}, \emph{{A cosmic UV/X-ray background model update}}, \href{https://doi.org/10.1093/mnras/staa302}{\emph{\mnras} {\bfseries 493} (2020) 1614} [\href{https://arxiv.org/abs/1903.08657}{{\ttfamily 1903.08657}}].

\bibitem{Mitra2023}
S.~{Mitra} and A.~{Chatterjee}, \emph{{Non-parametric reconstruction of photon escape fraction from reionization}}, \href{https://doi.org/10.1093/mnrasl/slad055}{\emph{\mnras} {\bfseries 523} (2023) L35} [\href{https://arxiv.org/abs/2303.02704}{{\ttfamily 2303.02704}}].

\bibitem{Ferrara2025}
A.~{Ferrara}, M.~{Giavalisco}, L.~{Pentericci}, E.~{Vanzella}, A.~{Calabr{\`o}} and M.~{Llerena}, \emph{{Redshift evolution of Lyman continuum escape fraction after JWST}}, \href{https://doi.org/10.48550/arXiv.2505.10619}{\emph{arXiv e-prints} (2025) arXiv:2505.10619} [\href{https://arxiv.org/abs/2505.10619}{{\ttfamily 2505.10619}}].

\bibitem{Khaire2019}
V.~{Khaire} and R.~{Srianand}, \emph{{New synthesis models of consistent extragalactic background light over cosmic time}}, \href{https://doi.org/10.1093/mnras/stz174}{\emph{\mnras} {\bfseries 484} (2019) 4174} [\href{https://arxiv.org/abs/1801.09693}{{\ttfamily 1801.09693}}].

\bibitem{Bouwens2023_XDF_HUDF}
R.J.~{Bouwens}, M.~{Stefanon}, G.~{Brammer}, P.A.~{Oesch}, T.~{Herard-Demanche}, G.D.~{Illingworth} et~al., \emph{{Evolution of the UV LF from z 15 to z 8 using new JWST NIRCam medium-band observations over the HUDF/XDF}}, \href{https://doi.org/10.1093/mnras/stad1145}{\emph{\mnras} {\bfseries 523} (2023) 1036} [\href{https://arxiv.org/abs/2211.02607}{{\ttfamily 2211.02607}}].

\bibitem{Bradley2023}
L.D.~{Bradley}, D.~{Coe}, G.~{Brammer}, L.J.~{Furtak}, R.L.~{Larson}, V.~{Kokorev} et~al., \emph{{High-redshift Galaxy Candidates at z = 9-10 as Revealed by JWST Observations of WHL0137-08}}, \href{https://doi.org/10.3847/1538-4357/acecfe}{\emph{\apj} {\bfseries 955} (2023) 13} [\href{https://arxiv.org/abs/2210.01777}{{\ttfamily 2210.01777}}].

\bibitem{Strait2020}
V.~{Strait}, M.~{Brada{\v{c}}}, D.~{Coe}, L.~{Bradley}, B.~{Salmon}, B.C.~{Lemaux} et~al., \emph{{Stellar Properties of z {\ensuremath{\gtrsim}} 8 Galaxies in the Reionization Lensing Cluster Survey}}, \href{https://doi.org/10.3847/1538-4357/ab5daf}{\emph{\apj} {\bfseries 888} (2020) 124} [\href{https://arxiv.org/abs/1905.09295}{{\ttfamily 1905.09295}}].

\bibitem{Stefanon2022}
M.~{Stefanon}, R.J.~{Bouwens}, I.~{Labb{\'e}}, G.D.~{Illingworth}, P.A.~{Oesch}, P.~{van Dokkum} et~al., \emph{{Blue Rest-frame UV-optical Colors in z 8 Galaxies from GREATS: Very Young Stellar Populations at 650 Myr of Cosmic Time}}, \href{https://doi.org/10.3847/1538-4357/ac3de7}{\emph{\apj} {\bfseries 927} (2022) 48} [\href{https://arxiv.org/abs/2103.06279}{{\ttfamily 2103.06279}}].

\bibitem{Stefanon2023}
M.~{Stefanon}, R.J.~{Bouwens}, I.~{Labb{\'e}}, G.D.~{Illingworth}, V.~{Gonzalez} and P.A.~{Oesch}, \emph{{Deep Spitzer/IRAC Data for z 10 Galaxies Reveal Blue Balmer Break Colors: Young Stellar Populations at 500 Myr of Cosmic Time}}, \href{https://doi.org/10.3847/1538-4357/aca470}{\emph{\apj} {\bfseries 943} (2023) 81} [\href{https://arxiv.org/abs/2206.13525}{{\ttfamily 2206.13525}}].

\bibitem{Topping2022}
M.W.~{Topping}, D.P.~{Stark}, R.~{Endsley}, R.J.~{Bouwens}, S.~{Schouws}, R.~{Smit} et~al., \emph{{The ALMA REBELS Survey: specific star formation rates in the reionization era}}, \href{https://doi.org/10.1093/mnras/stac2291}{\emph{\mnras} {\bfseries 516} (2022) 975} [\href{https://arxiv.org/abs/2203.07392}{{\ttfamily 2203.07392}}].

\bibitem{Tacchella2022}
S.~{Tacchella}, S.L.~{Finkelstein}, M.~{Bagley}, M.~{Dickinson}, H.C.~{Ferguson}, M.~{Giavalisco} et~al., \emph{{On the Stellar Populations of Galaxies at z = 9-11: The Growth of Metals and Stellar Mass at Early Times}}, \href{https://doi.org/10.3847/1538-4357/ac4cad}{\emph{\apj} {\bfseries 927} (2022) 170} [\href{https://arxiv.org/abs/2111.05351}{{\ttfamily 2111.05351}}].

\bibitem{Smit2014}
R.~{Smit}, R.J.~{Bouwens}, I.~{Labb{\'e}}, W.~{Zheng}, L.~{Bradley}, M.~{Donahue} et~al., \emph{{Evidence for Ubiquitous High-equivalent-width Nebular Emission in z \raisebox{-0.5ex}\textasciitilde 7 Galaxies: Toward a Clean Measurement of the Specific Star-formation Rate Using a Sample of Bright, Magnified Galaxies}}, \href{https://doi.org/10.1088/0004-637X/784/1/58}{\emph{\apj} {\bfseries 784} (2014) 58} [\href{https://arxiv.org/abs/1307.5847}{{\ttfamily 1307.5847}}].

\bibitem{Endsley2021}
R.~{Endsley}, D.P.~{Stark}, S.~{Charlot}, J.~{Chevallard}, B.~{Robertson}, R.J.~{Bouwens} et~al., \emph{{MMT spectroscopy of Lyman-alpha at z $\simeq$ 7: evidence for accelerated reionization around massive galaxies}}, \href{https://doi.org/10.1093/mnras/stab432}{\emph{\mnras} {\bfseries 502} (2021) 6044} [\href{https://arxiv.org/abs/2010.03566}{{\ttfamily 2010.03566}}].

\bibitem{Roberts-Borsani2022}
G.~{Roberts-Borsani}, T.~{Morishita}, T.~{Treu}, N.~{Leethochawalit} and M.~{Trenti}, \emph{{The Physical Properties of Luminous z {\ensuremath{\gtrsim}} 8 Galaxies and Implications for the Cosmic Star Formation Rate Density from 0.35 deg$^{2}$ of (Pure-)Parallel HST Observations}}, \href{https://doi.org/10.3847/1538-4357/ac4803}{\emph{\apj} {\bfseries 927} (2022) 236} [\href{https://arxiv.org/abs/2106.06544}{{\ttfamily 2106.06544}}].

\bibitem{Labbe2013}
I.~{Labb{\'e}}, P.A.~{Oesch}, R.J.~{Bouwens}, G.D.~{Illingworth}, D.~{Magee}, V.~{Gonz{\'a}lez} et~al., \emph{{The Spectral Energy Distributions of z \raisebox{-0.5ex}\textasciitilde 8 Galaxies from the IRAC Ultra Deep Fields: Emission Lines, Stellar Masses, and Specific Star Formation Rates at 650 Myr}}, \href{https://doi.org/10.1088/2041-8205/777/2/L19}{\emph{\apjl} {\bfseries 777} (2013) L19} [\href{https://arxiv.org/abs/1209.3037}{{\ttfamily 1209.3037}}].

\bibitem{Robertson2023}
B.E.~{Robertson}, S.~{Tacchella}, B.D.~{Johnson}, K.~{Hainline}, L.~{Whitler}, D.J.~{Eisenstein} et~al., \emph{{Identification and properties of intense star-forming galaxies at redshifts z > 10}}, \href{https://doi.org/10.1038/s41550-023-01921-1}{\emph{Nature Astronomy} {\bfseries 7} (2023) 611} [\href{https://arxiv.org/abs/2212.04480}{{\ttfamily 2212.04480}}].

\bibitem{Harikane2024}
Y.~{Harikane}, K.~{Nakajima}, M.~{Ouchi}, H.~{Umeda}, Y.~{Isobe}, Y.~{Ono} et~al., \emph{{Pure Spectroscopic Constraints on UV Luminosity Functions and Cosmic Star Formation History from 25 Galaxies at z $_{spec}$ = 8.61-13.20 Confirmed with JWST/NIRSpec}}, \href{https://doi.org/10.3847/1538-4357/ad0b7e}{\emph{\apj} {\bfseries 960} (2024) 56} [\href{https://arxiv.org/abs/2304.06658}{{\ttfamily 2304.06658}}].

\bibitem{Santini2017}
P.~{Santini}, A.~{Fontana}, M.~{Castellano}, M.~{Di Criscienzo}, E.~{Merlin}, R.~{Amorin} et~al., \emph{{The Star Formation Main Sequence in the Hubble Space Telescope Frontier Fields}}, \href{https://doi.org/10.3847/1538-4357/aa8874}{\emph{\apj} {\bfseries 847} (2017) 76} [\href{https://arxiv.org/abs/1706.07059}{{\ttfamily 1706.07059}}].

\bibitem{Salmon2015}
B.~{Salmon}, C.~{Papovich}, S.L.~{Finkelstein}, V.~{Tilvi}, K.~{Finlator}, P.~{Behroozi} et~al., \emph{{The Relation between Star Formation Rate and Stellar Mass for Galaxies at 3.5 <= z <= 6.5 in CANDELS}}, \href{https://doi.org/10.1088/0004-637X/799/2/183}{\emph{\apj} {\bfseries 799} (2015) 183} [\href{https://arxiv.org/abs/1407.6012}{{\ttfamily 1407.6012}}].

\bibitem{Davidzon2018}
I.~{Davidzon}, O.~{Ilbert}, A.L.~{Faisst}, M.~{Sparre} and P.L.~{Capak}, \emph{{An Alternate Approach to Measure Specific Star Formation Rates at 2 z 7}}, \href{https://doi.org/10.3847/1538-4357/aaa19e}{\emph{\apj} {\bfseries 852} (2018) 107} [\href{https://arxiv.org/abs/1712.03959}{{\ttfamily 1712.03959}}].

\bibitem{CurtisLake2023}
E.~{Curtis-Lake}, S.~{Carniani}, A.~{Cameron}, S.~{Charlot}, P.~{Jakobsen}, R.~{Maiolino} et~al., \emph{{Spectroscopic confirmation of four metal-poor galaxies at z = 10.3-13.2}}, \href{https://doi.org/10.1038/s41550-023-01918-w}{\emph{Nature Astronomy} {\bfseries 7} (2023) 622} [\href{https://arxiv.org/abs/2212.04568}{{\ttfamily 2212.04568}}].

\bibitem{Stark2013}
D.P.~{Stark}, M.A.~{Schenker}, R.~{Ellis}, B.~{Robertson}, R.~{McLure} and J.~{Dunlop}, \emph{{Keck Spectroscopy of 3 < z < 7 Faint Lyman Break Galaxies: The Importance of Nebular Emission in Understanding the Specific Star Formation Rate and Stellar Mass Density}}, \href{https://doi.org/10.1088/0004-637X/763/2/129}{\emph{\apj} {\bfseries 763} (2013) 129} [\href{https://arxiv.org/abs/1208.3529}{{\ttfamily 1208.3529}}].

\bibitem{Wang2023}
B.~{Wang}, S.~{Fujimoto}, I.~{Labb{\'e}}, L.J.~{Furtak}, T.B.~{Miller}, D.J.~{Setton} et~al., \emph{{UNCOVER: Illuminating the Early Universe-JWST/NIRSpec Confirmation of z > 12 Galaxies}}, \href{https://doi.org/10.3847/2041-8213/acfe07}{\emph{\apjl} {\bfseries 957} (2023) L34} [\href{https://arxiv.org/abs/2308.03745}{{\ttfamily 2308.03745}}].

\bibitem{Casey2024}
C.M.~{Casey}, H.B.~{Akins}, M.~{Shuntov}, O.~{Ilbert}, L.~{Paquereau}, M.~{Franco} et~al., \emph{{COSMOS-Web: Intrinsically Luminous z {\ensuremath{\gtrsim}} 10 Galaxy Candidates Test Early Stellar Mass Assembly}}, \href{https://doi.org/10.3847/1538-4357/ad2075}{\emph{\apj} {\bfseries 965} (2024) 98} [\href{https://arxiv.org/abs/2308.10932}{{\ttfamily 2308.10932}}].

\bibitem{Carniani2024}
S.~{Carniani}, K.~{Hainline}, F.~{D'Eugenio}, D.J.~{Eisenstein}, P.~{Jakobsen}, J.~{Witstok} et~al., \emph{{Spectroscopic confirmation of two luminous galaxies at a redshift of 14}}, \href{https://doi.org/10.1038/s41586-024-07860-9}{\emph{\nat} {\bfseries 633} (2024) 318} [\href{https://arxiv.org/abs/2405.18485}{{\ttfamily 2405.18485}}].

\bibitem{Weibel2024}
A.~{Weibel}, P.A.~{Oesch}, L.~{Barrufet}, R.~{Gottumukkala}, R.S.~{Ellis}, P.~{Santini} et~al., \emph{{Galaxy build-up in the first 1.5 Gyr of cosmic history: insights from the stellar mass function at z 4-9 from JWST NIRCam observations}}, \href{https://doi.org/10.1093/mnras/stae1891}{\emph{\mnras} {\bfseries 533} (2024) 1808} [\href{https://arxiv.org/abs/2403.08872}{{\ttfamily 2403.08872}}].

\bibitem{Choudhury2025}
T.R.~{Choudhury} and A.~{Chakraborty}, \emph{{Capturing Small-Scale Reionization Physics: A Sub-Grid Model for Photon Sinks with SCRIPT}}, \href{https://doi.org/10.48550/arXiv.2504.03384}{\emph{arXiv e-prints} (2025) arXiv:2504.03384} [\href{https://arxiv.org/abs/2504.03384}{{\ttfamily 2504.03384}}].

\end{thebibliography}\endgroup

\end{document}